\documentclass[12pt]{article}
\usepackage{a4wide, amsmath, latexsym, epsfig,
  psfrag,graphicx, rotating, fancyhdr, tabularx, afterpage,booktabs}
\usepackage{subfigure}
\newcommand\nn{\nonumber}                  
\newcommand\ba{\begin{eqnarray}}  
\newcommand\ea{\end{eqnarray}}

\begin{document}
\begin{titlepage}
 
\begin{flushright} 
{ IFJPAN-IV-2011-6 \hskip 1 cm  UAB-FT/695\\   FTUV/2011-0929
\hskip 1 cm IFIC/11-53  \hskip 1 cm  CERN-PH-TH/2012-016 } 
\end{flushright}

\begin{center}
{\bf\huge Resonance chiral Lagrangian currents and $\tau$ decay Monte Carlo}
\end{center}
 
\begin{center}
   {\bf    O. Shekhovtsova$^{a}$, T. Przedzi\'nski$^{b}$, P. Roig$^{c}$ and Z. W\c{a}s$^{d,e}$  }\\ 
{\em $^a$IFIC, Universitat de Val\`encia-CSIC,  Apt. Correus 22085, \\  E-46071, Val\`encia, Spain }\\
       {\em $^b$ The Faculty of Physics, Astronomy and Applied Computer
Science, \\ Jagellonian University, Reymonta 4, 30-059 Cracow, Poland}
\\
{\em $^c$ Grup de F\'{\i}sica Te\`orica, Institut de F\'{\i}sica d'Altes Energies, 
Universitat Aut\`onoma de Barcelona, E-08193 Bellaterra, Barcelona, Spain} \\
       {\em $^d$  Institute of Nuclear Physics, PAN,
        Krak\'ow, ul. Radzikowskiego 152, Poland}\\
{\em $^e$ CERN PH-TH, CH-1211 Geneva 23, Switzerland }
\end{center}
\begin{center}
{\bf   ABSTRACT  }
\end{center} 
   In the present  paper we describe the  set of form factors 
for hadronic 
$\tau$ decays based 
on  {\tt Resonance Chiral Theory}.
The technical  implementation of the form factors 
in {\tt FORTRAN} code is also explained. It is shown 
how it can be installed into {\tt TAUOLA} Monte Carlo program. Then it is
rather easy to implement into software environments of not only  Belle 
and BaBar collaborations but also for FORTRAN and C++ applications of LHC.
The description of the current for each $\tau$  decay mode is complemented with 
technical numerical tests. 
The set is ready for fits, parameters to be used in fits are explained. 
Arrangements to work
with the experimental data not requiring   unfolding  are prepared. 
Hadronic currents, ready for  confrontation with the $\tau$ decay data,
but not yet ready for the general use, 
cover more than 88 \% of hadronic
$\tau$ decay width.

 \vspace{0.1 cm}
{\small
\begin{flushleft}
{   IFJPAN-IV-2011-6 \hskip 1 cm  UAB-FT/695 \hskip 1 cm   FTUV/2011-09-29 \hskip 1 cm IFIC/11-53 \hskip 1 cm CERN-PH-TH/2012-016\\
March, 2012}
\end{flushleft}
}
 
\vspace*{1mm}
\bigskip
\end{titlepage}

\tableofcontents
\newpage


\section{Introduction}\label{sec:introduction}

Measurements of $\tau$ lepton, because of its long lifetime, large mass 
and parity sensitive couplings, 
lead to broad physics interest.  From the perspective of high-energy experiments
such as at LHC, knowledge of $\tau$ lepton properties offers an important ingredient of new 
physics signatures. From the perspective of lower energies,  $\tau$ lepton 
decays constitute 
an excellent laboratory for hadronic interactions. 
In itself,  the  $\tau$ lepton 
decays constitute 
an excellent laboratory for studies of hadronic interactions at the energy scale 
of about 
1 GeV, where neither perturbative QCD methods nor chiral Lagrangians are expected to work 
to a good precision \cite{Braaten:1991qm, Braaten:1988ea, Braaten:1988hc, Narison:1988ni, Pich:1989pq}.
 At present,
hundreds of milions  of $\tau$ decays are amassed by both Belle and BaBar 
experiments. It is of utmost importance to represent such data in a form as 
useful for general applications as possible.

Most of these data samples are not yet analyzed. For example, in Ref.~\cite{Fujikawa:2008ma}
only 10 \% of the collected sample, which means 5.4 M events for $\tau^\pm \to \nu_\tau \pi^\pm \pi^0$,
was used. Future samples at the Belle II or  Frascati Super B facilities will be even larger \cite{Dolezal:2010zz, Rama:2011zz}.
That means that already now the statistical error for the 
collected samples  of $\tau \to 3\pi \nu_\tau$ is of  the 
 order of 0.03 \%. For  $\tau \to K\pi\pi \nu_\tau$ it is about 0.1 \%
and for $\tau \to KK \pi  \nu_\tau$ at the level of 0.2 \%. To exploit such valuable 
data sets, theoretical predictions need to be properly 
prepared. As typically several milions of events per channel are collected, that means that the statistical error can reach $\sim$0.03 \%.  
To match it, parametrizations 
of hadronic currents 
resulting from theoretical models must be controlled to technical precision better than 0.03 \% in 
Monte Carlo, combining theoretical aspects and full detector response.
Only then, one can be sure that the comparison of the data with theoretical 
predictions  exploits in full the statistical impact of the data, and one can concentrate
on systematic effects both for theory and experiment. 
One should stress that the
above technical precision is required 
not only for signal distribution, but for background as well. Sophisticated
techniques allowing proper comparisons of the data and models are also needed. 

A review of the status
of available tools for such studies of $\tau$ decays  can be found in 
Ref.~\cite{Actis:2010gg}.
It was concluded in Ref.~\cite{Actis:2010gg} that the appropriate
choice of hadronic current parametrization was the most essential missing  
step to perform. It was also found that, 
for the decays involving more than two pseudoscalars in the final state, the appropriate 
use of hadronic currents in fits is important too.
At present, standards of precision are at  2 \% level. This is
a factor of 100 less than what is required. For many $\tau$ decay modes
even this 2 \% precision level is far to be reached  \cite{Hayashii:2010zz}.

The original version of {\tt TAUOLA} \cite{Jadach:1993hs} uses the results of Ref.~\cite{Kuhn:1990ad, Kuhn:1992nz} and their extensions to other 
decay channels\footnote{With time, due to pressure from the experimental community,
many other parametrizations were introduced, but not in a systematic way. Some of those found its way to {\tt TAUOLA} later, Ref.~\cite{Golonka:2003xt},
and are used 
as a starting reference point for our present project as seen from the computing side.}. 
In that model each three-pseudoscalar current is constructed as a weigthed sum of products of 
Breit-Wigner functions \cite{Pich:1989pq,Kuhn:1990ad,Decker:1993ay,Decker:1992kj}. 
This approach was contested in Ref.~\cite{GomezDumm:2003ku} where it was demonstrated that the corresponding hadronic form 
factors, which
were 
written to reproduce the leading-order (LO) $\chi PT$ result~\cite{Weinberg:1978kz}, 
fail to reproduce the next-to-leading-order 
one (NLO)~\cite{Gasser:1983yg, Gasser:1984gg}. 
   The corresponding parametrization based on Breit-Wigner functions was not able to reproduce CLEO $\tau^-\to (KK\pi)^-\nu_\tau$ decays data \cite{Coan:2004ep}. This 
resulted in the CLEO collaboration reshaping the model by the introduction of two ad-hoc parameters that spoilt the QCD normalization of the Wess-Zumino part.
This shows that, although the approach of weighted products of Breit-Wigner functions 
 was sufficient and very successful twenty years ago now, with the massively increased experimental 
data samples, it is pressing to upgrade. 
As an alternative, an approach based on the Resonance Chiral Theory
\cite{Ecker:1989yg,Ecker:1988te} was proposed.
Its application to hadronic tau decays is supposed to be consistent
and theoretically well founded {(see Sect. \ref{sec:systematic} for the related discussion)}.
However, its results have  to be confronted with the experimental data before
actual improvement will be confirmed.
The hadronic currents for the two and three pseudoscalar final
states that we consider here have been calculated in the framework of
R$\chi$T
\cite{GomezDumm:2003ku,Dumm:2009va,Dumm:2009kj,Arganda:2008jj,Jamin:2008qg}
and have been prepared for  {\tt TAUOLA}.
 
Section \ref{sec:currents} is 
devoted to a general presentation of the form hadronic currents must fulfil to 
be installed into
{\tt TAUOLA} generator \cite{Jadach:1993hs}. In each subsection analytic forms of currents calculated within Resonance Chiral Theory are given channel by 
channel.
In Section \ref{sec:a1width} energy-dependent widths
as used in the parametrization of intermediate resonances are presented.
Section \ref{sec:Benchmark} is dedicated to technical tests of the channel
$\tau \to 3\pi \nu_\tau$. For channels involving kaons,
only overall benchmark distributions are collected. 
The details of technical tests are left to the project Web page \cite{web:RChL}. 
For all decay channels numerical results, which are of more 
physical interest, are collected in 
Section~\ref{sec:results}. Within it, the three-meson channels,  which have been worked out in more  depth,
are first presented  and then, the two meson channels 
 are discussed. In both cases technical aspects are worked out 
to precision better than 0.1 \%.

The organization of the hadronic currents and how they can be integrated into the {\tt TAUOLA} 
library  
is explained in  Section \ref{sec:software} and in Appendix \ref{app:B}.
Section \ref{sec:systematic} is prepared for a reader who is oriented towards the theoretical details of calculation and estimation
of theoretical uncertainties of the approach. It provides arguments necessary for discussion of the
range of parameters allowed for fits. 
 The summary in Section \ref{sec:Summary} closes the paper.
Further technical appendices are also given. 
Appendix \ref{app:a1} lists analytic functions used in the parametrization  of 
hadronic currents. Appendix \ref{app:C} provides  numerical values of the 
model parameters used all over the paper. 
It is explained which parameters and in which range  
can be modified without breaking assumptions of the model and where in 
the code they are defined.
Appendix \ref{app:D}  collects 
branching ratios of the newly prepared $\tau$ decay channels as calculated by
Monte Carlo simulation. It contains explicit references to 
the definition of the hadronic current in each decay channel, which are spread all over the 
paper. In the future, however, these definitions can be replaced by the new references. 

The implementation of final state interactions (FSI) in the two-meson $\tau$ decay modes 
is discussed in Appendix \ref{app:F}. An improvement
of the latter  and the scalar form factor 
in $\tau\to K \pi  \nu_\tau$ decays will be 
addressed in a forthcoming publication \cite{RChL:scal}.

Finally let us stress that our  paper aims at explaining how 
this new set of hadronic currents can be  
installed in {\tt TAUOLA}, independently of whether it is a standalone version, part 
of Belle/BaBar software or a different configuration, which is another purpose 
of Appendix \ref{app:B}.


\section{Hadronic current for two and three hadrons}\label{sec:currents}

Before discussing in detail the implementation of the currents 
into the program and resulting distributions, let us first collect
here all necessary formulas. In general we will follow conventions 
for normalizations as used in Ref.~\cite{Jadach:1993hs}. We will not recall
here relations between hadronic distributions and decay product distributions
though. They are rather simple and we assume that the reader is familiar with 
the necessary parts of  Ref.~\cite{Jadach:1993hs}. Let us recall the   
 matrix element $\cal M$ for the $\tau$ decay into
hadronic state  X  and a neutrino:
$\tau(P)\rightarrow X \nu_{\tau}(N)  $.
It reads
${\cal
M}=\frac{G_F}{\sqrt{2}}\bar{u}(N)\gamma^{\mu}(1-\gamma_{5})u(P)J_{\mu} .$
All dynamics of hadronic interactions is encapsulated in a current
$J_\mu$, which is a  function of  hadronic $\tau$ decay products only.

Contrary to 
{\tt TAUOLA}, as documented in Refs.~\cite{Jadach:1993hs} or \cite{Golonka:2003xt}, now hadronic currents for all
two-pseudoscalar final states are defined in separate routines and the constraint that the scalar form factor must be set to 0 is removed.

For $\tau$ decay channels with two mesons $\left[h_1(p_1)\right.$ and $\left. h_2(p_2)\right]$, the hadronic current reads
\begin{equation}
J^\mu  = N \left[ \left(p_1 - p_2-\frac{\Delta_{12}}{s}(p_1+p_2)\right)^\mu F^{V}(s) + \frac{\Delta_{12}}{s}(p_1 + p_2)^\mu F^{P}(s) \right],
\end{equation}
where $s = (p_1 +p_2)^2$ and $\Delta_{12}=m_1^2-m_2^2$. The formulas for vector, $F^{V}(s)$, and pseudoscalar, $F^{P}(s)$, form factors 
depend on the particular decay channel\footnote{The vector form factor 
of both two pions and two kaons is expected to be fixed at zero momentum transfer by gauge invariance in 
the $SU(2)$ symmetry limit \cite{Gasser:1984ux}: $F^V(0) = 1$, 
see Section~\ref{Subsect:pipi0}.\label{foot:gauge1}}
 and are given, respectively,  
for the $\pi^-\pi^0$,  
$(K \pi)^-$
 and $K^- K^0$ decay modes
 in 
subsection~\ref{Subsect:pipi0} and the following ones\footnote{Two-meson $\tau$ decays involving an $\eta$ meson, 
$\tau\to\eta^{(')} P^-\nu_\tau,\,P=\pi,K$ have a negligible branching 
fraction \cite{Nakamura:2010zzi, Pich:1987qq}.}. $SU(3)$ symmetry relates all four normalization factors by the appropriate 
Clebsch-Gordan coefficient:
\ba
N^{\pi^-\pi^0} = 1 , \, \, N^{K^-K^0} = \frac{1}{\sqrt{2}} ,\, \, N^{\pi^- \bar K^0} = \frac{1}{\sqrt{2}}, \, \, N^{\pi^0 K^-} = \frac{1}{2}\,.
\ea

For the final state of three pseudoscalars, with momenta $p_1$, $p_2$ and $p_3$, Lorentz invariance determines the decomposition 
of the hadronic current to be
\begin{eqnarray}
J^\mu &=N &\bigl\{T^\mu_\nu \bigl[ c_1 (p_2-p_3)^\nu F_1  + c_2 (p_3-p_1)^\nu
 F_2  + c_3  (p_1-p_2)^\nu F_3 \bigr]\nonumber\\
& & + c_4  q^\mu F_4  -{ i \over 4 \pi^2 F^2}      c_5
\epsilon^\mu_{.\ \nu\rho\sigma} p_1^\nu p_2^\rho p_3^\sigma F_5      \bigr\},
\label{fiveF}
\end{eqnarray}
where  $T_{\mu\nu} = g_{\mu\nu} - q_\mu q_\nu/q^2$ denotes the transverse
projector, and $q^\mu=(p_1+p_2+p_3)^\mu$ is the momentum of the hadronic system.
The decay products  are ordered and their four-momenta are denoted, 
respectively, as $p_1$, $p_2$ and $p_3$. Here and afterward in the paper $F$ stands for the pion decay constant in the chiral limit.

 Functions $F_i$   (hadronic form factors) depend in general 
on three independent invariant masses  that can be constructed from the three meson 
four-vectors. We chose $q^2=(p_1+p_2+p_3)^2$
and two invariant masses  $s_1=(p_2+p_3)^2$, $s_2=(p_1+p_3)^2$ built from 
pairs of momenta. Then
$s_3=(p_1+p_2)^2$ can be calculated from the other three invariants,
$s_3 = q^2 - s_1 - s_2 + m_1^2 + m_2^2 + m_3^2$, and $F_i$ 
written explicitly with its dependencies reads as
$F_i(q^2,s_1,s_2)$.
 This form of the hadronic current is the most general one and constrained 
only by Lorentz invariance.  For modes with an even number of kaons 
the normalization factor reads as
 $N = \mathrm{cos} \theta_{\mathrm{Cabibbo}}/F$, otherwise 
$N = \mathrm{sin} \theta_{\mathrm{Cabibbo}}/F$.

We  leave  the $F_4$ contribution  in the basis, even though 
it is of the order $\sim m_{\pi}^2/q^2$~\cite{GomezDumm:2003ku,Dumm:2009kj,Decker:1993ay} for the three-pseudoscalar channels that we consider in this work. 
It  plays a role   in the low $q^2$ region for the three-pion modes. 
We will neglect the corresponding contribution for the modes with kaons, i.e., $c_4 = 0$.  
Among the three hadronic form factors which correspond to the axial-vector part of the hadronic tensor,  ($F_1$, $F_2$, $F_3$),   
only two are independent.
We will keep the definition of
$F_1$, $F_2$ or $F_3$, exactly as shown in Eq.~(\ref{fiveF}), which is the form used in {\tt TAUOLA} since the beginning. 
However, linear combinations constructed from only two of these
 functions are in principle equally good. The decay channel dependent 
constants $c_i$  are  given in Table~\ref{table:ci}.

\begin{table}[h]
\centering
\begin{tabular}{| c | c | c | c | c | c |}
\hline
Decay mode & c$_1$ & c$_2$ & c$_3$ & c$_4$ & c$_5$ \\ 
($p_1$, $p_2$, $p_3$) &  &  &  & &  \\\hline
$\pi^-\pi^-\pi^+$ & 1 & - 1 & 0 & 1 & 0 \\ \hline
$\pi^0\pi^0\pi^-$ & 1 & - 1 & 0 & 1 & 0 \\ \hline
$K^-\pi^-K^+$ & 1  & - 1  & 0 &  0  & 1  \\ \hline
$K^0\pi^-\bar{K}^0$ & 1 & - 1  & 0  &  0  & 1  \\ \hline
$K^-\pi^0K^0$ & 0  & 1  & - 1  &  0  & - 1 \\ \hline
\end{tabular}
\caption{Coefficients for formula (\ref{fiveF}) in the isospin symmetry limit. Note, that in 
Ref.~\cite{Jadach:1993hs}
different conventions were used and coefficients
were affecting normalization too.}\label{table:ci}
\end{table}
The theoretical assumptions behind the hadronic currents that we use are discussed  in  
Section \ref{sec:systematic}.
In the model the results for all hadronic currents, with the exception of  two-pion and two-kaon modes, 
are calculated in the isospin
 limit, therefore the corresponding hadronic form factors depend only  on the average pion 
$\left[m_\pi = (m_{\pi^0}+2\cdot m_{\pi^+})/3\right]$ and kaon $\left[m_K=(m_{K^0}+m_{K^+})/2\right]$ 
masses, we relax the assumption later\footnote{At first step
we will take such assumption for our phase space generator as well. Later in the paper
 we will nonetheless return to proper masses, distinct 
for charged and neutral pseudoscalars.  
We will evaluate the numerical consequences, see Table \ref{Table:bench}.}.
For the three-pseudoscalar modes every hadronic form factor consists of 3 parts: 
a chiral contribution (direct decay, without production of any intermediate 
resonance), one-resonance  and double-resonance mediated processes. 
The results for the hadronic form factors are taken from 
Refs.~\cite{Dumm:2009va,Dumm:2009kj}. 

The  channels we will present (together with the trivial decay
-- from the Monte Carlo point of view --  into $\pi \nu_\tau$ and $K\nu_\tau $)
represent
more than 88  \% of the hadronic width of $\tau$ \cite{Nakamura:2010zzi}. 
The dominant 
missing channels $\tau \to \pi^+ \pi^- \pi^- \pi^0 \nu_\tau $ and 
$\tau \to \pi^- \pi^0 \pi^0 \pi^0 \nu_\tau $, which are together about 9.7 \% of the hadronic width of $\tau$, are more difficult to control 
theoretically\footnote{
Several older options developed for these channels  are provided for user convenience, but they 
will not be documented here. Please 
see the {\tt README} files stored in directory {\tt new-currents/other-currents} 
for details.}.
Also  attempts to describe $\tau \to \pi^+ \pi^- \pi^- \pi^0 \nu_\tau $ 
are relatively  recent.  They are  technically compatible 
with our solution for {\tt TAUOLA} currents and can be used simultaneously.
It is 
documented
 in Ref.~\cite{Bondar:2002mw}. 

Let us now describe  hadronic 
currents for 
each particular channel.


\subsection{ $\pi^-\pi^-\pi^+\nu_\tau$ and $\pi^0\pi^0\pi^-\nu_\tau$ }
\label{Subsect:pipipi}

Hadronic form factors  for the three-pion modes
have been calculated assuming the isospin symmetry%
\footnote{The inclusion of the 
complete first-order corrections to $SU(2)$ symmetry is beyond 
our present scope. In particular, electromagnetic corrections arising 
at this order 
are neglected. We restrict 
ourselves, for the moment, to the ones given by the mass splittings between members of the same $SU(2)$-multiplet.  These
enter the kinematical factors 
and phase-space integrals. 
The model-independent electromagnetic corrections can be handled with {\tt PHOTOS} \cite{Barberio:1993qi, Davidson:2010ew}, while the structure-dependent 
corrections have been 
computed only for the one- and two-meson $\tau$ decay modes (with only pions and kaons) 
in Refs. \cite{Cirigliano:2001er, Cirigliano:2002pv, FloresBaez:2006gf, Guo:2010dv}.
Its implementation in generation with {\tt PHOTOS} will follow work of Refs. \cite{Nanava:2006vv,Nanava:2009vg,Xu:2012px}. \label{foot:iso}}, as a consequence, $m_{\pi^\pm}=m_{\pi^0}$ . 
The code for the current is given in \\{\tt new-currents/RChL-currents/f3pi\_rcht.f.}

The independent set of hadronic 
form factors $F_i$, $i=1,...,5$ is chosen as: $F_1$, $F_2$ and $F_4$ ($F_3=0$ then). 
The vector form factor vanishes for the three-pion modes due to the G-parity conservation \cite{Decker:1992kj, Kramer:1984cy}: 
  $F_5 = 0$.
It is convenient to present the functions as
\begin{equation}
 F_{i} \ = \ 
 ( F_i^{\chi} \, + \, F_i^{\mbox{\tiny R}} \, + \, F_i^{\mbox{\tiny RR}} )\cdot R^{3\pi}
\ ,\qquad i=1,2,4\ ,
\end{equation}
where $F_i^{\chi}$ is the chiral contribution, $F_i^{\mbox{\tiny R}}$ is the one resonance contribution and $F_i^{\mbox{\tiny RR}}$ 
is the double-resonance part. The  $R^{3\pi} $ constant equals -1 for $\pi^0\pi^0\pi^-$ and 1 for $\pi^-\pi^-\pi^+$.

For the convention defined by Eq. (\ref{fiveF}), the form factors $F_i$ can be obtained from  
Ref.~\cite{Dumm:2009va} with the 
replacements: 
\begin{equation}
F_i(Q^2,s,t) \to F_i(q^2,s_1,s_2)/F  , \; \; \qquad i=1,2,4 ,
\end{equation}
where $F$ was defined after Eq.~(\ref{fiveF}). 
The form factors read:
\begin{eqnarray}
\label{eq:t1r}
&& F_1^{\chi}(q^2,s_1,s_2) = - \frac{2\sqrt{2}}{3 } \,,\\
&& F_1^{\mbox{\tiny R}}(q^2,s_1,s_2) =  \frac{\sqrt{2}\,F_V\,G_V}{3\,F^2} \bigg[
\, \frac{3\,s_1}{s_1-M_\rho^2 - i M_\rho \Gamma_\rho(s_1)} \, - \, \nonumber \\
&&\left( \frac{2 G_V}{F_V} - 1 \right) \, \left(
\, \frac{2 q^2-2s_1-s_3}{s_1-M_\rho^2 - i M_\rho \Gamma_\rho(s_1)} \, + \, \frac{s_3-s_1}{s_2-M_\rho^2 - i M_\rho \Gamma_\rho(s_2)} \,
\right)\bigg] \,, \nonumber \\
&& F_1^{\mbox{\tiny RR}}(q^2,s_1,s_2) =  \frac{4 \, F_A \, G_V}{3 \,F^2} \,
 \frac{q^2}{q^2-M_A^2 - i M_A \Gamma_A(q^2)} \, \bigg[- \, (\lambda' + \lambda'')
 \, \frac{3\,s_1}{s_1-M_\rho^2 - i M_\rho \Gamma_\rho(s_1)} \, \nonumber \\
&&+ H\left(\frac{s_1}{q^2},\frac{m_\pi^2}{q^2}\right)\, \frac{2 q^2 + s_1 -
s_3}{s_1-M_\rho^2 - i M_\rho \Gamma_\rho(s_1)} \, + \, H\left(\frac{s_2}{q^2},\frac{m_\pi^2}{q^2}\right) \, \frac{s_3-s_1}{s_2-M_\rho^2 - i M_\rho \Gamma_\rho(s_2)}\bigg] \ , \nonumber
\end{eqnarray} 
where
\begin{equation} \label{eq:fq2}
H(x,y)  =  - \,\lambda_0 \, y \, +  \,
\lambda'\, x \, + \,  \lambda''  \; ,
\end{equation}
 and 
\begin{eqnarray}
\lambda'& =& \frac{F^2}{2 \sqrt{2} F_A G_V} \,,\nn\\
\lambda''& =& -\left(1-2\frac{G_V^2}{F^2}\right)\lambda' \,,\nn\\
4\lambda_0& =& \lambda'+\lambda''\,.
\end{eqnarray}

  Bose symmetry implies that the form factors $F_1$ and
$F_2$ are related  $F_2(q^2,s_2,s_1) =  F_1(q^2,s_1,s_2)$
(the minus sign that comes
from the definition of the hadronic current, Eq. (\ref{fiveF}), is included in $c_2$).

The pseudoscalar form factor, $F_4=F_4^{\chi}+F_4^{\mbox{\tiny R}}$, carries the contribution from both the 
direct vertex and the one-resonance mechanism of production:
\begin{eqnarray}\label{f4_3pi}
F_4^{\chi}(q^2,s_1,s_2) & = &  \frac{2\sqrt{2}}{3} \,\frac{m_\pi^2[3(s_3-m_\pi^2)-q^2(1 + 2\kappa R^{3\pi})]}{2q^2(q^2-m_\pi^2)} \,,\nonumber \\
F_4^{\mbox{\tiny R}}(q^2,s_1,s_2) & = & - \frac{\sqrt{2}\,F_V\,G_V}{3\,F^2} \left[\alpha_2(q^2,s_2,s_1)+\alpha_2(q^2,s_1,s_2)\right] ,
\end{eqnarray}
where $\kappa = 1$ for $\tau^-\to\pi^-\pi^-\pi^+\nu_\tau$, $\kappa = 1/2$ for $\tau^-\to\pi^0\pi^0\pi^-\nu_\tau$ and  
\begin{equation}\label{eq:funct_alpha2}
\alpha_2(q^2,s_1,s_2) = \frac{3G_V}{F_V} \, \frac{s_1}{q^2}\frac{m_\pi^2}{q^2-m_\pi^2}
\frac{s_3-s_2}{s_1-M_\rho^2 - i M_\rho \Gamma_\rho(s_1)} \,.
\end{equation}
The pseudoscalar form factor $F_4$ is proportional to $m_\pi^2/q^2$~\cite{Dumm:2009va}, thus  it is 
suppressed with respect to $F_1$ and $F_2$. However, the pseudoscalar contribution can affect the $q^2$ spectrum near the threshold%
\footnote{Numerical results with and without $F_4$ are presented in Section~\ref{sec:Semi-real}.}.

Besides the pion decay constant $F$, the results for the form factors $F_i$ depend on some coupling constants of the model:
 $F_V$ (we impose  $G_V=F^2/F_V$), $F_A$  and the masses of the nonets of vector and axial-vector resonances 
 ($M_V$ and $M_A$) in the chiral 
and large-$N_C$ limits. We follow Refs.~ \cite{Dumm:2009va, Dumm:2009kj} and  
replace the masses used in the resonance Lagrangian with the masses of the corresponding
 physical states: 
$M_V \to M_{\rho}$ and\footnote{ 
In footnote $^{\ref{foot:M_A}}$ of Appendix~\ref{app:variationresparameters} we explain;  the parameter $M_A$ of
the short-distance QCD constraints should not be identified with $M_{a_1}$.
Our choice $M_A \to M_{a_1}$ is not well founded, but can be easily changed 
at the time of fits to experimental data.}
 $M_A \to M_{a_1}$.

 To include the  $\rho'$ meson we follow Ref.~\cite{Dumm:2009va}, its  Eq. (32). 
We insert  Eq.~(\ref{rhoprime}) of  the combined $\rho$ and $\rho'$ propagators 
into our Eqs. (\ref{eq:t1r}) and (\ref{eq:funct_alpha2})
\begin{equation}\label{rhoprime}
 \frac{1}{M_{\rho}^2-q^2-iM_{\rho} \Gamma_{\rho}(q^2)} \longrightarrow
\frac{1}{1+\beta_{\rho'}} \, \left[ \frac{1}{M_{\rho}^2-q^2-iM_{\rho} \Gamma_{\rho}(q^2)} \, + \,
\frac{\beta_{\rho'}}{M_{\rho'}^2-q^2-iM_{\rho'} \Gamma_{\rho'}(q^2)} \right] \, .
\end{equation}
Impact\footnote{A discussion on
the implementation of the second (third) resonance nonet to the model can be found in Section~\ref{sec:systematic}.}
 of the $\rho'$ meson on the  $d\Gamma/dq^2$ spectrum 
can be seen from Fig.~3  of Ref.~\cite{Dumm:2009va}.

In the file  {\tt new-currents/RChL-currents/f3pi\_rcht.f} the form factors $F_1$, $F_2$, $F_4$ given 
by Eqs. (\ref{eq:t1r})-(\ref{f4_3pi}) and with substitution (\ref{rhoprime}) are coded.  
For completeness, let us remember that form factors $F_3$ and $F_5$ are equal to zero.

The only eventual isospin breaking  will result from  $m_{\pi^\pm} \ne m_{\pi^0}$ used  in the phase space generator
embedded in {\tt tauola.f}. 


\subsection{ $K^-\pi^-K^+\nu_\tau$ and $K^0\pi^-\bar{K^0}\nu_\tau$}
\label{Subsect:KKpi}

Again isospin symmetry is assumed, therefore 
$m_{\pi^\pm}=m_{\pi^0}$ 
and $m_{K^0}= m_{K^\pm}$.
The code for the currents is given in {\tt new-currents/RChL-currents/fkkpi.f}.

We will neglect the contribution from
the pseudoscalar form factor $F_4$ as it is proportional again to $m_\pi^2/q^2$~\cite{Dumm:2009kj}. 
We present the result for the non-zero form factors $F_i$ 
in the same way
as before:
\begin{equation}
 F_{i} \ = \ 
  F_i^{\chi} \, + \, F_i^{\mbox{\tiny R}} \, + \, F_i^{\mbox{\tiny RR}} 
\ ,\qquad i=1,2,5\ .
\end{equation}
Taking into account the convention for the current, Eq.~(\ref{fiveF}), and the  values of the $c_i$ coefficients in Table~\ref{table:ci} 
the result for the form factor can be obtained from  
Ref.~\cite{Dumm:2009kj} with the replacements: $F_1(Q^2,s,t) \to F_2(q^2,s_2,s_1)/F$, $F_2(Q^2,s,t) \to F_1(q^2,s_2,s_1)/F$. Therefore, 
the form factor $F_1$ reads 
\begin{eqnarray}
\label{eq:f21}
 F_1^{\chi}(q^2,s_2,s_1) &= &  - \frac{\sqrt{2}}{3}  \, , \\[3.5mm]
 F_1^{\mbox{\tiny R}}(q^2,s_2,s_1) & = & - \, \frac{\sqrt{2}}{6} \, \frac{F_V\,G_V}{F^2}  \, \left[
 \, \frac{B^{\mbox{\tiny R}}(s_1,s_3,m_K^2,m_K^2)}{M_{\rho}^{2}-s_2 - i M_\rho \Gamma_\rho(s_2)}  \,  +
 \, \frac{A^{\mbox{\tiny R}}(q^2,s_1,s_3,m_K^2,m_K^2,m_{\pi}^2)}{M_{K^{*}}^{2}-s_1 - i M_{K^{*}} \Gamma_{K^{*}}(s_1)} \,\right] ,  \,  \nonumber \\[4mm]
F_1^{\mbox{\tiny RR}}(q^2,s_2,s_1) & = & \frac{2}{3} \, \frac{F_A G_V}{F^2} \, \frac{q^2}{M_{A}^2-q^2 - i M_A \Gamma_A(q^2)} \,
\, \left[ \, \frac{B^{\mbox{\tiny RR}}(q^2,s_1,s_3,s_2,m_K^2,m_K^2,m_{\pi}^2) }{M_{\rho}^2-s_2 - i M_\rho \Gamma_\rho(s_2)} \,
\right. \nonumber \\
& &  \qquad \qquad \qquad \qquad \qquad\left.
+ \, \frac{A^{\mbox{\tiny RR}}(q^2,s_1,s_3,m_K^2,m_K^2,m_{\pi}^2) }{M_{K^*}^2-s_1 - i M_{K^{*}} \Gamma_{K^{*}}(s_1)} \, \right] \, \, . \nonumber
\end{eqnarray}
where the functions $A^{\mbox{\tiny R}}$, $B^{\mbox{\tiny R}}$,
$A^{\mbox{\tiny RR}}$ and $B^{\mbox{\tiny RR}}$  are defined in
Appendix~\ref{app:a1} and $s_3$ is calculated from $q^2,s_1,s_2$ and masses.  

The form factor $F_2$ is given by
\begin{eqnarray}
\label{eq:f11}
F_2^{\chi}(q^2,s_2,s_1) &= &   F_1^{\chi} \, ,  \\
F_2^{\mbox{\tiny R}}(q^2,s_2,s_1) & = & - \, \frac{\sqrt{2}}{6} \, \frac{F_V\,G_V}{F^2}  \, \left[
\, \frac{A^{\mbox{\tiny R}}(q^2,s_2,s_3,m_K^2,m_{\pi}^2,m_K^2)}{M_{\rho}^{2}-s_2 - i M_\rho \Gamma_\rho(s_2)}  \, +
\, \frac{B^{\mbox{\tiny R}}(s_2,s_3,m_K^2,m_{\pi}^2)}{M_{K^{*}}^{2}-s_1 - i M_{K^{*}} \Gamma_{K^{*}}(s_1)} \,\right] ,  \, \nonumber  \\[4mm]
F_2^{\mbox{\tiny RR}}(q^2,s_2,s_1) & = & \frac{2}{3} \, \frac{F_A G_V}{F^2} \, \frac{q^2}{M_{A}^2-q^2 - i M_A \Gamma_A(q^2)} \,
\, \left[ \, \frac{A^{\mbox{\tiny RR}}(q^2,s_2,s_3,m_K^2,m_{\pi}^2,m_K^2) }{M_{\rho}^2-s_2 - i M_\rho \Gamma_\rho(s_2)} \,
\right. \nonumber \\
& &  \qquad \qquad \qquad \qquad \qquad\left.
+ \, \frac{B^{\mbox{\tiny RR}}(q^2,s_2,s_3,s_1,m_K^2,m_{\pi}^2,m_K^2)}{M_{K^*}^2-s_1 - i M_{K^{*}} \Gamma_{K^{*}}(s_1)} \, \right]
\, \, . \nonumber 
\end{eqnarray}

The vector form factor, $F_5$, arises from the chiral anomaly \cite{Wess:1971yu, Witten:1983tw} and the non-anomalous
odd-intrinsic-parity amplitude \cite{RuizFemenia:2003hm}. It is obtained from Ref.~\cite{Dumm:2009kj} 
with the replacement $F_3(Q^2,s,t) \to - F_5(q^2,s_2,s_1)/(4\pi^2F^3)$. It reads
\begin{eqnarray}
\label{eq:f31}
 &&F_5^{\chi}(q^2,s_2,s_1) =  \sqrt{2} \, , \\
 &&F_5^{\mbox{\tiny R}}(q^2,s_2,s_1) = \frac{16\pi^2 \, G_V}{M_V} \, \left[ \,
C^{\mbox{\tiny R}}(q^2,s_2,m_K^2,m_K^2,m_{\pi}^2) \,
\left( \sin^2 \theta_V \frac{1+ \sqrt{2}  \cot \theta_V}{M_{\omega}^2-s_2 - i M_\omega \Gamma_\omega}
\right.\right. \nonumber \\
& &+ \left. \left. \, \cos^2 \theta_V
\frac{1- \sqrt{2} \tan \theta_V }{M_{\phi}^2-s_2 - i M_\phi \Gamma_\phi} \right )
\, + \, \frac{C^{\mbox{\tiny R}}(q^2,s_1,m_K^2,m_\pi^2,m_K^2)}{M_{K^*}^2-s_1 - i M_{K^{*}} \Gamma_{K^{*}}(s_1)} \right. \nonumber \\
&& \left.
-\frac{2 \, F_V}{G_V}
\, \frac{D^{\mbox{\tiny R}}(q^2,s_2,s_1)}{M_{\rho}^2-q^2 - i M_\rho \Gamma_\rho(q^2)} \right] \, , \nonumber \\
&&F_5^{\mbox{\tiny RR}}(q^2,s_2,s_1) =  -16 \sqrt{2} \pi^2 F_V \, G_V \,  \frac{1}{M_{\rho}^2 - q^2 - i M_\rho \Gamma_\rho(q^2) } \,
\left[ 
\frac{C^{\mbox{\tiny RR}}(q^2,s_1,m_K^2)}{M_{K^*}^2-s_1 - i M_{K^{*}} \Gamma_{K^{*}}(s_1)}  + \right. \nonumber \\
&& \left. C^{\mbox{\tiny RR}}(q^2,s_2,m_{\pi}^2) \,
\left( \sin^2 \theta_V \frac{1+ \sqrt{2} \cot \theta_V}{M_{\omega}^2-s_2 - i M_\omega \Gamma_\omega}
+ \, \cos^2 \theta_V
\frac{1- \sqrt{2} \tan \theta_V }{M_{\phi}^2 -s_2 - i M_\phi \Gamma_\phi} \right) 
\right] \, ,
\nonumber
\end{eqnarray}
where $C^{\mbox{\tiny R}}$, $D^{\mbox{\tiny R}}$ and $C^{\mbox{\tiny RR}}$ 
are defined in Appendix~\ref{app:a1}.

For the widths of the narrow resonances $\omega$ and $\phi$ the PDG  \cite{Nakamura:2010zzi} values are taken and the constant 
width approximation is followed.
The parameter $\theta_V$  defines the mass eigenstates $\omega(782)$ and $\phi(1020)$~\cite{Dumm:2009kj} and is the mixing angle
between the octet and singlet vector states $\omega_8$ and $\omega_0$. In our numerical calculation we take the ideal mixing 
$\left[\theta_V=\mathrm{tan}^{-1}(1/\sqrt{2})\right]$.
In this  limit, the contribution of the $\phi(1020)$ meson in Eq. (\ref{eq:f31}) vanishes
\footnote{ \label{foot:phitau} On the other hand, one should keep in mind that the decay channel
$\tau^-\to \phi \pi^- \nu_\tau$ was observed  by BaBar  \cite{Aubert:2007mh} and 
$\tau^- \to \phi K^- \nu_\tau$  by Belle~\cite{Inami:2006vd}
.}. This will be changed in future with fits to the data.

The file  {\tt new-currents/RChL-currents/fkkpi.f} contains the form factors $F_1$, $F_2$ and $F_5$. 
The  form of Eqs. (\ref{eq:f21})-(\ref{eq:f31}) is used. As one can see, the contribution from  the excited states, 
e.g. $\rho'$, $\rho''$, is not included in the  $KK\pi$ case,  contrary to the three-pion one.
The only eventual isospin breaking assumed will result from $m_{\pi^\pm} \ne m_{\pi^0}$ and $m_{K^{\pm}}\ne m_{K^0}$ 
used in the phase space generator embedded in {\tt tauola.f}\;.

\subsection{ $K^-\pi^0K^0\nu_\tau$}
\label{Subsect:KK0pi0}

The hadronic current of the $ K^-\pi^0K^0\nu_\tau$ decay mode is again obtained with  isospin symmetry;  
$m_{\pi^\pm}=m_{\pi^0}$ and $m_{K^0}= m_{K^\pm}$. The code for the currents is given in the file\\{\tt new-currents/RChL-currents/fkk0pi0.f}\;.

If the momenta of the pseudoscalars are attributed as in Table~\ref{table:ci}, that is $K^-(p_1)\pi^0(p_2)K^0(p_3)$, then it is convenient to choose 
the independent set of hadronic 
form factors for the axial-vector part as $F_2(q^2,s_2,s_3)$ and $F_3(q^2,s_2,s_3)$. We will neglect the contribution due to 
the pseudoscalar form factor $F_4$ as it is 
again proportional to the square of pion mass over $q^2$~\cite{Dumm:2009kj}. 
Taking into account the  constants $c_i$ of Table~\ref{table:ci} and  Eq.~(\ref{fiveF}), 
the result of Ref.~\cite{Dumm:2009kj} needs the replacements
\begin{eqnarray}
F_2(Q^2,s,t) &\to& F_3(q^2,s_2,s_3)/F ,  \, \, F_1(Q^2,s,t) \to F_2(q^2,s_2,s_3)/F , \nonumber\\
F_3(Q^2,s,t) &\to& -F_5(q^2,s_2,s_3)/(4\pi^2 F^3) .
\end{eqnarray}

As before:
\begin{equation}
 F_{i} \ = \ 
  F_i^{\chi} \, + \, F_i^{\mbox{\tiny R}} \, + \, F_i^{\mbox{\tiny RR}} 
\ ,\qquad i=2,3,5\ 
\end{equation}
and  
\begin{eqnarray}
\label{eq:f12}
F_2^{\chi}(q^2,s_2,s_3)  & = & -1 \, , \\[3mm]
F_2^{\mbox{\tiny R}}(q^2,s_2,s_3) & = & - \frac{1}{6} \frac{F_V G_V}{F^2} \, \left[
\, \frac{B^{\mbox{\tiny R}}(s_2,s_1,m_K^2,m_{\pi}^2)}{M_{K^{*}}^{2}-s_3-i M_{K^*} \Gamma_{K^*}(s_3)} \, +
\, 2 \; \frac{A^{\mbox{\tiny R}}(q^2,s_2,s_1,m_K^2,m_\pi^2,m_K^2)}{M_\rho^{2}-s_2-i M_\rho \Gamma_\rho(s_2)} \,
\right. \nonumber \\
& & \left. \qquad\qquad\qquad +
\, \frac{A^{\mbox{\tiny R}}(q^2,s_1,s_2,m_\pi^2,m_K^2,m_K^2)}{M_{K^{*}}^{2}-s_1-i M_{K^*} \Gamma_{K^*}(s_1)} \, \right]
\, , \nonumber \\ [3.5mm]
F_2^{\mbox{\tiny RR}}(q^2,s_2,s_3) & = &  \frac{\sqrt{2}}{3} \frac{F_A G_V}{F^2}
\frac{q^2}{M_{A}^2-q^2-i M_{A} \Gamma_{A}(q^2)} \, \Bigg[
\, \frac{B^{\mbox{\tiny RR}}(q^2,s_2,s_1,s_3,m_K^2,m_{\pi}^2,m_K^2)}{M_{K^{*}}^{2}-s_3-i M_{K^*} \Gamma_{K^*}(s_3)} \nonumber \\
& + &\, 2 \, \frac{A^{\mbox{\tiny RR}}(q^2,s_2,s_1,m_K^2,m_\pi^2,m_K^2)}{M_\rho^{2}-s_2-i M_\rho \Gamma_\rho(s_2)} 
+ \, \frac{A^{\mbox{\tiny
 RR}}(q^2,s_1,s_2,m_\pi^2,m_K^2,m_K^2)}{M_{K^{*}}^{2} - s_1 - iM_{K^{*}} \Gamma_{K^{*}}(s_1) } \, \Bigg]
\, , \nonumber 
\end{eqnarray}
 $s_3$ is calculated from $q^2,s_1,s_2$ and masses.
The contributions to  $F_3$ read
\begin{eqnarray}
\label{eq:f22}
 F_3^{\chi}(q^2,s_2,s_3) & = & 0 \, , \\ [3mm]
F_3^{\mbox{\tiny R}}(q^2,s_2,s_3) & = & - \frac{1}{6} \frac{F_V G_V}{F^2} \, \left[
\, \frac{A^{\mbox{\tiny R}}(q^2,s_3,s_1,m_K^2,m_K^2,m_{\pi}^2)}{M_{K^{*}}^{2}-s_3-iM_{K^{*}} \Gamma_{K^{*}}(s_3)} \, +
\, 2 \; \frac{B^{\mbox{\tiny R}}(s_3,s_1,m_K^2,m_K^2)}{M_\rho^{2}-s_2-i M_\rho \Gamma_\rho(s_2)} \,
\right. \nonumber \\
& & \left. \qquad\qquad\qquad -
\, \frac{A^{\mbox{\tiny R}}(q^2,s_1,s_3,m_K^2,m_K^2,m_\pi^2)}{M_{K^{*}}^{2}-s_1-iM_{K^{*}}\Gamma_{K^{*}}(s_1)} \, \right]
\, , \nonumber 
\end{eqnarray}
\begin{eqnarray}
F_3^{\mbox{\tiny RR}}(q^2,s_2,s_3) & = &  \frac{\sqrt{2}}{3} \frac{F_A G_V}{F^2}
\frac{q^2}{M_{A}^2-q^2-i M_{A} \Gamma_{A}(q^2)} \, \Bigg[
\, \frac{A^{\mbox{\tiny RR}}(q^2,s_3,s_1,m_K^2,m_K^2,m_{\pi}^2)}{M_{K^{*}}^{2}-s_3-iM_{K^{*}}\Gamma_{K^{*}}(s_3)}
 \nonumber \\
& &
+ \, 2 \; \frac{B^{\mbox{\tiny RR}}(q^2,s_3,s_1,s_2,m_K^2,m_K^2,m_\pi^2)}{M_\rho^{2}-s_2-i M_\rho \Gamma_\rho(s_2)} \,
- \, \frac{A^{\mbox{\tiny RR}}(q^2,s_1,s_3,m_K^2,m_K^2,m_\pi^2)}{M_{K^{*}}^{2}-s_1-iM_{K^{*}}\Gamma_{K^{*}}(s_1)} \, \Bigg] \, . \nonumber 
\end{eqnarray}
\par
The form factor $F_5$ driven by the vector current is given by the sum of
\begin{eqnarray}
\label{eq:f32}
 F_5^{\chi}(q^2,s_2,s_3) & = & 0  \,,  \\ [3mm]
F_5^{\mbox{\tiny R}}(q^2,s_2,s_3) & = & \frac{8 \sqrt{2} \pi^2 \, G_V}{M_V}
\!\!\Bigg[ \frac{C^{\mbox{\tiny R}}(q^2,s_3,m_K^2,m_\pi^2,m_K^2)}{M_{K^*}^2- s_3 -iM_{K^{*}}\Gamma_{K^{*}}(s_3) } -
\frac{C^{\mbox{\tiny R}}(q^2,s_1,m_K^2,m_\pi^2,m_K^2)}{M_{K^*}^2-s_1 -iM_{K^{*}}\Gamma_{K^{*}}(s_1) }  \nonumber \\
& - & \frac{2 F_V}{G_V}
\frac{E^{\mbox{\tiny R}}(s_3,s_1)}{M_{\rho}^2-q^2-i M_\rho \Gamma_\rho(q^2)} \Bigg] \, , \nonumber \\ [3mm]
F_5^{\mbox{\tiny RR}}(q^2,s_2,s_3) & = & - 16\pi^2 F_V G_V \frac{1}{M_{\rho}^2-q^2}
\Bigg[ \frac{C^{\mbox{\tiny RR}}(q^2,s_3,m_K^2)}{M_{K^*}^2-s_3  -iM_{K^{*}}\Gamma_{K^{*}}(s_3) } \nonumber \\
& -&
\frac{C^{\mbox{\tiny RR}}(q^2,s_1,m_K^2)}{M_{K^*}^2-s_1 - iM_{K^{*}}\Gamma_{K^{*}}(s_1)} \Bigg] \, , \nonumber 
\end{eqnarray}
where the new function $E^{\mbox{\tiny R}}$, as the previous ones,  is defined in Appendix~\ref{app:a1}. 

The file  {\tt new-currents/RChL-currents/fkk0pi0.f} contains the form factors $F_2$, $F_3$ and $F_5$. 
Eqs. (\ref{eq:f12})-(\ref{eq:f32}) are used.
 As one can see, the contributions from  the excited states, e.g. $\rho'$, ${K^*}'$ are not
included in the form factors of the $\tau\to K K\pi\nu_\tau$ decay channels.
 This shall be an obvious future improvement resulting from the confrontation with the data.

The assumption that the  eventual isospin breaking will result only from $m_{\pi^\pm} \ne m_{\pi^0}$ and $m_{K^\pm}\ne m_{K^0}$ 
as used in the phase space generator embedded in {\tt tauola.f}\; is taken.

\subsection{ $\pi^-\pi^0\nu_\tau$,   $\pi^0 K^-\nu_\tau$,  $\pi^- \bar K^0\nu_\tau$  and  $K^-K^0\nu_\tau$  }
\label{Subsect:pipi0}

The two-pseudoscalar final states are simpler. They can be presented together  in one subsection. 
The code for the hadronic currents of the $\pi^-\pi^0\nu_\tau$, $K^-K^0\nu_\tau$ and $(K\pi)^-\nu_\tau$ modes is given,
respectively, in files
{\tt new-currents/RChL-currents/frho\_pi.f}, \\
 {\tt new-currents/RChL-currents/fk0k.f} and {\tt new-currents/RChL-currents/fkpipl.f}\;.

In the general case there are both vector and scalar form factors.  In the isospin symmetry limit, 
$m_{\pi^\pm} = m_{\pi^0}$, $m_{K^\pm} = m_{K^0}$, for both two-pion and two-kaon modes  the scalar form factor vanishes 
and the corresponding channel is described by the vector form factor only\footnote{The scalar form factor 
appears only at next-to-leading order in SU(2) breaking and it can 
be safely neglected (see Ref.~\cite{Cirigliano:2001er} for details).}. 
Also for the $K\pi$ mode we restrict ourselves at first to the vector form factor only,  
the scalar form factor will be properly included in Ref.~\cite{RChL:scal}.

For all three channels we use a parametrization for the vector form factor which is developed starting from the lowest-lying 
resonance contribution:
\begin{eqnarray}\label{FF_2scal}
F^V_{PQ}(s) =F^{VMD}(s)\,\,\mathrm{exp}\left[\sum_{P,Q}N_{loop}^{PQ}{\frac{-s}{96\pi^2 F^2} Re A_{PQ}(s)}\right]\,,
\end{eqnarray}
where $F^{VMD}$ is 
the contribution from the lightest vector resonance that can be exchanged in the process, and
the exponentiation resums FSI effects 
(see Appendix \ref{app:F} for a related discussion). 
The function $A_{PQ}(s)$ is a loop function for two pseudoscalars with masses $m_P$ and $m_Q$, and is presented in Appendix~\ref{app:a1}. 
$N_{loop}^{PQ}$ is a constant dictated by chiral symmetry for the different decay channels:
\begin{equation}
N_{loop}^{\pi^-\pi^0} = 1 , \, \, N_{loop}^{K^-K^0} = \frac{1}{2} , \, \,  N_{loop}^{K\pi} =  N_{loop}^{K\eta} =\frac{3}{4} . 
\end{equation}

In the case of the two-pion mode, the  theoretical calculation performed in the framework of R$\chi$T~\cite{Roig:2011iv}
gives the following result for the form factor:
\begin{eqnarray} \label{VFFpipi_rho}
F^{V}_{\pi\pi}(s) &\!\!\! =&\!\!\! 
\frac{M^2_{\rho}}{M^2_\rho-s-iM_\rho \Gamma_\rho (s)} 
\exp \Biggl\{
\frac{-s}{96\pi^2 F^2} \biggl[Re A_{\pi^- \pi^0}(s) 
+ \frac{1}{2} Re 
A_{K^- K^0}(s)  \biggr] \Biggr\} \, ,
\end{eqnarray}
 if only  $\rho$ resonance is taken into account.
The contribution of the excited resonances (both $\rho'$ and $\rho''$) modify the  form factor of formula (\ref{VFFpipi_rho}). 
Following Ref.~\cite{Roig:2011iv} it takes the form 
\begin{eqnarray} \label{VFFpipi}
F^{V}_{\pi\pi}(s) &\!\!\! =&\!\!\! 
\frac{M^2_{\rho} + s\bigl(\gamma \mathrm{e}^{i\phi_1} 
+ \delta \mathrm{e}^{i\phi_2} \bigr)}{M^2_\rho-s-iM_\rho \Gamma_\rho (s)} 
\exp \Biggl\{
\frac{-s}{96\pi^2 F^2} \biggl[Re A_{\pi^- \pi^0}(s) 
+ \frac{1}{2} Re 
A_{K^- K^0}(s)  \biggr] \Biggr\} \,
 \nn\\
& & - \frac{s\gamma \mathrm{e}^{i\phi_1}}{M^2_{\rho'}-s-iM_{\rho'} \Gamma_{\rho'} (s)} 
\exp \Biggl\{
\frac{-s \Gamma_{\rho'} }{\pi M_{\rho'}^3\sigma^3_\pi(M_{\rho'}^2)} \biggl[Re A_{\pi}(s) \biggr] \Biggr\} \, \nn \\
& & - \frac{s\delta \mathrm{e}^{i \phi_2}}{M^2_{\rho''}-s-iM_{\rho''} \Gamma_{\rho''} (s)} 
\exp \Biggl\{
\frac{-s \Gamma_{\rho''} }{\pi M_{\rho''}^3\sigma^3_\pi(M_{\rho''}^2)} \biggl[Re A_{\pi}(s) \biggr] \Biggr\} \, ,
\end{eqnarray}
where the phase--space factor 
\begin{equation}\label{eq:sigma}
\sigma_P(q^2) \equiv \sqrt{1 - 4 m_P^2/q^2}\,
\end{equation}
is used.
 The function $A_{P}$ is the same loop function $A_{PQ}$ defined in (\ref{loopfun_2pi}), but in the limit of equal masses.
It is given in Appendix~\ref{app:a1}. In our file {\tt new-currents/RChL-currents/frho\_pi.f} the form factor 
of  formula ~(\ref{VFFpipi})  is used.

The two-kaon vector form factor is written following Ref.~\cite{Arganda:2008jj} as
\ba \label{VFFKK}
F^{V}_{KK}(s) =  \frac{M^2_{\rho}}{M^2_\rho-s-iM_\rho \Gamma_\rho (s)}\exp \Biggl\{
\frac{-s}{96\pi^2 F^2}  \biggl[Re A_{\pi^- \pi^0}(s) 
+ \frac{1}{2} Re 
A_{K^- K^0}(s)  \biggr]  \Biggr\}\,.
\ea
One can see the expression for the kaon vector form factors coincides with the pion form factor from Eq.~(\ref{VFFpipi_rho}).

The excited resonances have not been taken into account in Eq.~(\ref{VFFKK}) following
 Ref.~\cite{Arganda:2008jj}. However, their implementation along the lines of Eq.~(\ref{VFFpipi}) 
is simple. The file \\ {\tt new-currents/RChL-currents/fk0k.f} contains both%
\footnote{By default, our program runs with the $K^0K^-$ vector form factor of Eq. (\ref{VFFKK}). 
However, changing the value of the parameter {\tt FFKKVEC = 0} to {\tt FFKKVEC = 1} in {\tt value\_parameter.f} allows to run the code with 
the form factor of Eq.~(\ref{VFFpipi}). Numerical effects due to the inclusion of the excited resonances 
are given in Section~\ref{sect:numer-2scal}.}
forms of the two-kaon form factor (\ref{VFFpipi}) and  (\ref{VFFKK}).

For the $K\pi$ mode we applied the result of Ref.~\cite{Jamin:2008qg} Eq.(5),
 which reads\footnote{By default, our program runs with the $K\pi$ vector form factor of Eq. (\ref{VFFKpi}). 
However, changing the value of the parameter {\tt FFKPIVEC = 1} to {\tt FFKPIVEC = 0} the code will 
run with the form factor given in Eqs. (17), (18) of Ref.~\cite{Boito:2008fq}.
For discussion, see Section~\ref{sect:numer-2scal}.}:

\begin{eqnarray} \label{VFFKpi}
F^{V}_{K\pi}(s) &\!\!\! =&\!\!\! 
\left(\frac{M^2_{K^*}+s\gamma_{K\pi}}{M^2_{K^*}-s-iM_{K^*} \Gamma_{K^*} (s)}
-  \frac{s\gamma_{K\pi}}{M^2_{K^{*\prime}}-s-iM_{K^{*\prime}} \Gamma_{K^{*\prime}} (s)} 
\right) \\
&& \exp \Biggl\{
\frac{-s}{128\pi^2 F^2} \biggl[Re A_{K\pi}(s) 
+ Re 
A_{K\eta}(s)  \biggr] \Biggr\} . \nn
\end{eqnarray}
Note that due to the FSI effects, the form factor $F^{V}_{K\pi}(0) \neq 1$.
 
The $m_{\pi^\pm} \ne m_{\pi^0}$ and $m_{K^{\pm}} \ne m_{K^0}$ used in the phase space  Monte Carlo generator, 
and discussed numerically later in the paper,
will be the only isospin breaking assumed. 
However, we plan to include electromagnetic corrections to 
$\tau\to\pi^-\pi^0\nu_\tau$ decays in the future (see $^{\ref{foot:iso}}$).

\section{Energy-dependent widths of resonances}\label{sec:a1width}
In this section we collect the formulas to calculate the energy-dependent width 
of the resonances $\rho$, $\rho'$, $\rho''$, $K^*$, $K^{*\prime}$ and $a_1$. They were used in the previous
section as ingredients for the construction of hadronic currents.
 From the technical 
side their calculation requires integration of the appropriate matrix elements over the phase space.
In this way, for example, unitarity constraints 
are taken into account \cite{Dumm:2009va, Guerrero:1997ku, GomezDumm:2000fz}.

The energy-dependent width of $\rho(770)$ resonance, calculated in the $SU(2)$ 
limit ($m_{\pi^\pm} = m_{\pi^0}$, $m_{K^\pm} = m_{K^0}$ ), is given~\cite{GomezDumm:2000fz} as
\begin{equation}\label{eq:rhowidth_su2}
\Gamma_\rho(q^2) = \frac{M_\rho q^2}{96 \pi F^2}\biggl[ \sigma_\pi^3(q^2)\theta(q^2-4m_\pi^2) 
+ \frac{1}{2}\sigma_K^3(q^2)\theta(q^2-4m_K^2) \biggr] .
\end{equation}
The phase--space factor $\sigma_\pi$ is defined in Eq.~(\ref{eq:sigma}), the
$2 \pi$ and $2 K$ loops are included. This form of the $\rho$ width is used in modes of three pseudoscalars.

In the two-pion and two-kaon modes $SU(2)$ breaking effects were taken into account in the $\rho$ off-shell width
\begin{eqnarray}\label{eq:rhowidth_nosu2}
\Gamma_\rho(q^2) = \frac{M_\rho q^2}{96 \pi F^2}\, \bigg[ \theta(q^2 - thr_{\pi\pi}) \lambda^{3/2} 
\bigg(1,\frac{m_{\pi^+}^2}{q^2},\frac{m_{\pi^0}^2}{q^2} \bigg)
  +  \frac{1}{2}\theta(q^2 - thr_{KK}) \lambda^{3/2} \bigg(1,\frac{m_{K^+}^2}{q^2},\frac{m_{K^0}^2}{q^2} \bigg) \bigg]\,,
\end{eqnarray}  
where 
 $\lambda(x,y,z) = (x - y -z)^2 - 4yz$,  $thr_{\pi\pi} = (m_{\pi^+} + m_{\pi^0})^2$ and  $thr_{KK} = (m_{K^+} + m_{K^0})^2$.

At this stage the widths of the $\rho'(1465)$ and $\rho''(1700)$ mesons are modeled as decays to two pions,
\begin{equation}\label{eq:rho1width}
\Gamma_{\rho'}(q^2) = \Gamma_{\rho'}\,\frac{q^2}{M_{\rho'}^2}\,\frac{\sigma_\pi^3\left(q^2\right)}{\sigma_\pi^3\left(M_{\rho'}^2\right)}\,\theta\left(q^2 - 4m_\pi^2\right) ,
\end{equation}
with $\Gamma_{\rho'}\equiv \Gamma_{\rho'}\left(M_{\rho'}^2\right)$ .

The energy-dependent width of the $K^*(892)$ resonance is given, in the $SU(2)$ limit ($m_{\pi^\pm} = m_{\pi^0}$, $m_{K^\pm} = m_{K^0}$), 
in Ref.~\cite{Jamin:2006tk}. It is related to $\Gamma_\rho(q^2)$ by chiral symmetry. It reads:
\begin{eqnarray}\label{eq:kstarwidth}
\Gamma_{K^*}(q^2) = \frac{M_{K^*} q^2}{128 \pi F^2}
\bigg[\lambda^{3/2}\left(1,\frac{m_K^2}{q^2},\frac{m_\pi^2}{q^2}\right)\theta(q^2 -thr_{K\pi}) +
  \lambda^{3/2}\left(1,\frac{m_K^2}{q^2},\frac{m_\eta^2}{q^2}\right)\theta(q^2 - thr_{K\eta}) \bigg] \,,
\end{eqnarray}
with $thr_{K\pi}=(m_K+m_\pi)^2$ and $thr_{K\eta}=(m_K+m_\eta)^2$.

Formula ~(\ref{eq:kstarwidth}) is used for the $KK\pi $ modes whereas for the $K\pi$ modes we use the following result
(Eq. (4) from Ref.~\cite{Jamin:2008qg})
\begin{equation}\label{eq:kstarwidth_nosu3}
\Gamma_{K^*}(q^2) = \Gamma_{K^*} \frac{q^2}{M_{K^*}^2} \frac{ \lambda^{3/2}\left(1,\frac{m_K^2}{q^2},
\frac{m_\pi^2}{q^2}\right)\theta(q^2 -thr_{K\pi})
 + \lambda^{3/2}\left(1,\frac{m_K^2}{q^2},\frac{m_\eta^2}{q^2}\right)\theta(q^2 -thr_{K\eta})}
{\lambda^{3/2}\left(1,\frac{m_K^2}{M_{K^*}^2},\frac{m_\pi^2}{M_{K^*}^2}\right)   +\lambda^{3/2}
\left(1,\frac{m_K^2}{M_{K^*}^2},\frac{m_\eta^2}{M_{K^*}^2}\right)}
\,.
\end{equation}
From Eq. (4) of Ref.~\cite{Jamin:2008qg} we have 
\begin{eqnarray}
\Gamma_{K^*}\equiv \Gamma_{K^*}(M_{K^*}^2) = \frac{G_V^2 M_{K^*}^3}{64 \pi F^4}
\left[\lambda^{3/2}\left(1,\frac{m_K^2}{M_{K^*}^2},\frac{m_\pi^2}{M_{K^*}^2}\right)
+\lambda^{3/2}\left(1,\frac{m_K^2}{M_{K^*}^2},\frac{m_\eta^2}{M_{K^*}^2}\right)\right]\,  ,
\end{eqnarray}
however, we prefer to write down Eq.~(\ref{eq:kstarwidth_nosu3}) in terms of the width ($\Gamma_{K^*}$), 
see discussion after Eq.~(17) in Ref.~\cite{Jamin:2008qg}.
The width of $K^{*\prime}(1410)$ is modelled as a decay to $K\pi$ and reads
\begin{eqnarray}
\Gamma_{K^{*\prime}}(q^2) =
\Gamma_{K^{*\prime}}\frac{q^2}{M_{K^{*\prime}}^2}
\frac{\lambda^{3/2}\left(1,\frac{m_K^2}{q^2},\frac{m_\pi^2}{q^2}\right)}
{\lambda^{3/2}\left(1,\frac{m_K^2}{M_{K^{*\prime}}^2},\frac{m_\pi^2}{M_{K^{*\prime}}^2}\right)}\theta(q^2
- thr_{K\pi}) \,  .
\end{eqnarray}

For the energy dependence of the $a_1$ resonance width%
\footnote{Calculation of $a_1$ width from $F_i$ was  already used
in Refs.~\cite{Pich:1989pq, Kuhn:1990ad,Jadach:1990mz}. }
 we use%
\footnote{There is an additional factor $1/F^2$ here and in Eq.~(\ref{eq:diff_width_a1}) compared with the 
definition, e.g. in Refs.~\cite{Dumm:2009va,Dumm:2009kj}. This is related to the normalization 
of our form factors $F_1$, $F_2$ and $F_3$. See explanation prior to the 
Eqs.~(\ref{eq:t1r}), (\ref{eq:f21}) and (\ref{eq:f12}).}
\cite{Dumm:2009va}
\begin{eqnarray}\label{eq:a1width}
\Gamma_{a_1}(q^2) &=& 2\Gamma_{a_1}^\pi(q^2)\theta\left(q^2 - 9m_\pi^2\right) \\
&+& 2\Gamma_{a_1}^{K^{\pm}}(q^2)\theta\left(q^2 - (m_\pi+2m_K)^2\right) + \Gamma_{a_1}^{K^0}(q^2)\theta\left(q^2- (m_\pi+2m_K)^2\right) , \nonumber
\end{eqnarray}
where 
\begin{eqnarray}
\Gamma_{a_1}^{\pi,K}(q^2) & = &\frac{-S}{192(2\pi)^3F_A^2 F^2 M_{a_1}}\bigg(\frac{M_{a_1}^2}{q^2} - 1 \bigg)^2  \nonumber \\
&\int & ds dt 
\left(V_1^\mu F_1 + V_2^\mu F_2 + V_3^\mu F_3 \right)^{\pi,K} \left((V_{1\mu} F_1 + V_{2\mu} F_2 +  V_{3\mu} F_3)^{\pi,K}\right)^*
\end{eqnarray}
stands for the contribution from the
individual three-pion and (two kaons - one pion) absorptive cuts. Here 
\begin{equation}
V_i^\mu = c_i T^{\mu\nu} (p_j-p_k)_\nu,  \; \; \; i\neq j \neq k = 1,2,3\,,
\end{equation}
with the coefficients $c_i$ appearing in Table 1.

In summary, $\Gamma_{a_1}^\pi(q^2)$ is the contribution of the $\pi^-\pi^-\pi^0$ and 
$\pi^0\pi^0\pi^-$ cuts, $\Gamma_{a_1}^{K^{\pm}}(q^2)$  of the $K^-\pi^-K^+$ and $K^0\pi^-\bar{K}^0$ cuts, and 
finally the $K^-\pi^0K^0$ contribution gives rise to the 
term $\Gamma_{a_1}^{K^0}(q^2)$.
 The form factors $F_i$ are presented in Sections 
\ref{Subsect:pipipi}, \ref{Subsect:KKpi} and \ref{Subsect:KK0pi0}.
The symmetry factor is defined as $S = 1/n!$, where   $n$ denotes the number of identical particles in the final state.

For reference, we include  the formula for the spectral function, the $q^2$-spectrum 
for the processes $\tau \to \text{3 pseudoscalars}\, \nu_\tau$ of this work\footnote{Our testing 
programs feature a calculation of spectral functions, including those of the formula (\ref{eq:spect_function}). 
However, we will not elaborate on this point here, even though it 
is important for future data analysis \cite{Actis:2010gg}, where results of Ref.~\cite{Kuhn:1992nz} are proposed to be used. 
We expect that a Monte Carlo sample will be used instead of semianalytical Eq.~(\ref{q2spec}).}
\begin{eqnarray}\label{eq:diff_width_a1}
\frac{d\Gamma}{dq^2} &=& \frac{G_F^2|V_{ud}|^2}{128(2\pi)^5 M_\tau F^2}\bigg(\frac{M_\tau^2}{q^2}-1\bigg)^2 
\int ds dt \bigg[ W_{SA} + \frac{1}{3} \bigg(1+2\frac{q^2}{M_\tau^2}\bigg) (W_A +W_B)\bigg] ,
\label{q2spec}
\end{eqnarray}
where
\begin{eqnarray}\label{eq:spect_function}
W_A & = & - (V_1^\mu F_1 +V_2^\mu F_2 + V_3^\mu F_3 )(V_{1\mu} F_1 + V_{2\mu} F_2 + V_{3\mu} F_3 )^* \,,\\
W_B & = & \frac{1}{64\pi^4 F^4}\left[s t u +(m_{K,\pi}^2 - m_\pi^2)(q^2 - m_{K,\pi}^2 )s 
+ m_{K,\pi}^2(2m_\pi^2-q^2)q^2 -m_{K,\pi}^2 m_\pi^4 \right] |F_5|^2 \,,\nonumber \\
W_{SA}& = & q^2 |F_4|^2 \nonumber \,.
\end{eqnarray}
The following phase space integration limits have to be used
\begin{equation}
\int ds dt = \int_{4 m_{K,\pi}^2}^{\left(\sqrt{q^2}-m_\pi\right)^2} ds \int_{t_-(s)}^{t_+(s)} dt \,,
\end{equation}
where
\begin{equation}
t_{\pm}(s) = \frac{1}{4 s}\bigg\{(q^2 - m_\pi^2)^2 - [\lambda^{1/2}(q^2,s,m_\pi^2) \mp \lambda^{1/2}(m_{K,\pi}^2,m_{K,\pi}^2,s)]^2\bigg\} .
\end{equation}

The necessary functions are  located in 
files {\tt new-currents/RChL-currents/funct\_{}rpt.f} and 
{\tt  new-currents/RChL-currents/wid\_{}a1\_{}fit.f}\;.


\section{Benchmark calculations for three-pion mode}\label{sec:Benchmark}
Since Ref.~\cite{Jadach:1993hs} has been published, numerical tests of {\tt TAUOLA} Monte Carlo functioning 
have not been repeated in a systematic way, despite the technical precision requirements
 are much higher now and reach sub-per mil level. Prior to  physics 
oriented comparisons between analytical (numerical integration)
 and Monte Carlo calculations, we need to revisit numerical 
stability of the generator and of multiple numerical integration, used in semianalytical calculations
accompanying generation and its tests as well.

 We will use the   
decay channel  $\tau\to\pi^-\pi^-\pi^+\nu_\tau$ to demonstrate our tests 
of Monte Carlo.  
For other channels, technical tests will be skipped from 
documentation\footnote{Directory {\tt new-currents/RChL-currents/cross-check} 
is devoted to such tests. The {\tt README} file explains technical details.} 
even though new issues absent in the $\tau\to\pi^-\pi^-\pi^+\nu_\tau$ case can
appear. A good example is
numerical stability at phase space edges of presamplers for relatively
narrow resonances such as  $K^*$ 
in $\tau \to K\pi  \nu_\tau$ decay.
In this case, for long runs square root of a negative number may have appeared because 
of rounding errors; an appropriate correction to the code was introduced. 
Such long  runs were never performed in the past for this channel. 
Our tests were indeed long. In some cases, for variables counting crude
events we have even run over the  
allowed maximum ($\sim2\cdot 10^9$) for {\tt FORTRAN INTEGER} type. 

To avoid problems with multidimensional integration of the $a_1$-meson propagator which is rapidly-changing as a function 
of its arguments, we first tabulated the $\Gamma_{a_1}(q^2)$ of 
 Eq.~(\ref{eq:a1width}). The
code for tabulation is located 
in {\tt new-currents/RChL-currents/tabler}~\footnote{Technical details of 
the calculations, which are quite independent from parts of 
the code loaded with {\tt TAUOLA}, are explained in {\tt README} files 
of this directory
and its subdirectories.
}.
Then we use linear interpolation to get the value of the $a_1$ width at required $q^2$.
To integrate over $s$ and $t$ variables the Gauss integration method 
has been used. The produced distribution has been checked to be numerically
stable\footnote{The source code of the integration routine {\tt gauss} has been taken from CERN program library  \cite{CERNLIB}. 
Tests  are provided  
in {\tt new-currents/RChL-currents/cross-checks}\;.
Cross-check by linear interpolation of the $q^2$ distribution from the neighbouring points demonstrates 
that the fluctuations due to numerical problems of integration are absent, the results are continuous, 
whereas the result produced with the integration method VEGAS \cite{Lepage:1977sw} had a tendency to fluctuate.}.

\subsection{Technical test}
Before we can go to the presentation of simulation results, where physical currents
are used, let us start with the simplified cases. We will begin with the calculation
for $\tau \to \pi^-  \pi^- \pi^+\nu_\tau$ where
 $F_1= F = 0.0924$ GeV, other form factors are set to zero ($F_2 = 0$, $F_4 = 0$) and 
$m_{\pi^\pm}=m_{\pi^0}=0.13804$ GeV.
It  is an 
important starting point, it helps to adjust conventions of normalization 
constants in {\tt TAUOLA} Monte Carlo and analytical calculations. 
 Phase 
space integration is free from singularities resulting from the matrix elements. 
Nonetheless, corresponding presamplers can be verified. Numerical 
integration is rather quick as 
there is
no need of invoking time consuming functions. That is why it is 
important to perform this check with a precision higher than 
for later tests. 

The total rate we obtained  from a Monte Carlo run of
$6 \cdot10^{6}$ events was 
 $(2.7414\pm 0.01 \%)\cdot 10^{-17}$ GeV.
For semianalytical numerical integration we obtain  $(2.7410 \pm0.02\%)\cdot 10^{-17}$ GeV   (with $2 \cdot 10^{-4}$ precision tag).
A difference of  $0.015 \%$ was found. This technical test performs  better than  could be achieved 
at time of work for reference \cite{Jadach:1993hs}. Statistical samples are larger by 3 orders 
of magnitude than what could have been used at that time.

For completeness, let us provide a plot of $\frac{d \Gamma}{dq^2}$  generated from Monte Carlo  divided by 
the semianalytical (numerical integration was used)  
result for this spectrum.
Reasonable agreement is found, see Fig.~\ref{Fig:fone}.

Triple Gaussian integration is used for the analytical
 calculation and double Gaussian integration enters as well into the current calculation
to be used in matrix elements of Monte Carlo generation. That is why pretabulation for the $a_1$ width, $\Gamma_{a_1}(q^2)$, is 
convenient as it speeds generation enormously. This represents another technical
feature being tested by normalization study and figures like Fig.~\ref{Fig:fone}.

Technical tests, as the one we discuss now,  belong 
to the  group of comparison booklets collected in the Web page \cite{web:RChL}. 

\subsection{Test with semirealistic parameters}\label{sec:Semi-real}
Let us now introduce the physical content of the current,  keeping
at first $m_{\pi^\pm}=m_{\pi^0}=0.13804$ GeV and dropping out statistical factor $\frac{1}{2}$ of two identical 
$\pi$'s.
Agreement between Monte Carlo and semianalytical calculation
 should  be $\sim$0.01\% again. 

\begin{figure}
\setlength{\unitlength}{0.1mm}
\begin{picture}(1600,1500)
\put(300,250){\begin{picture}( 1200,1200)
\put(0,0){\framebox( 1200,1200){ }}
\multiput(  300.00,0)(  300.00,0){   4}{\line(0,1){25}}
\multiput(    0.00,0)(   30.00,0){  41}{\line(0,1){10}}
\multiput(  300.00,1200)(  300.00,0){   4}{\line(0,-1){25}}
\multiput(    0.00,1200)(   30.00,0){  41}{\line(0,-1){10}}
\put( 300,-25){\makebox(0,0)[t]{\large $    1.0 $}}
\put( 600,-25){\makebox(0,0)[t]{\large $    2.0 $}}
\put( 900,-25){\makebox(0,0)[t]{\large $    3.0 $}}
\put(1200,-25){\makebox(0,0)[t]{\large $    4.0 $}}
\put(530,950){\makebox(0,0)[t]{\large   Ratio: MC/semianalytical, $F_1 = F$ }}
\multiput(0,    0.00)(0,  300.00){   5}{\line(1,0){25}}
\multiput(0,   30.00)(0,   30.00){  40}{\line(1,0){10}}
\multiput(1200,    0.00)(0,  300.00){   5}{\line(-1,0){25}}
\multiput(1200,   30.00)(0,   30.00){  40}{\line(-1,0){10}}
\put(-25,   0){\makebox(0,0)[r]{\large $    0.0 $}}
\put(-25, 300){\makebox(0,0)[r]{\large $    0.5 $}}
\put(-25, 600){\makebox(0,0)[r]{\large $    1.0 $}}
\put(-25, 900){\makebox(0,0)[r]{\large $    1.5 $}}
\put(-25,1200){\makebox(0,0)[r]{\large $    2.0 $}}
\end{picture}}
\put(300,250){\begin{picture}( 1200,1200)
\thinlines 
\newcommand{\x}[3]{\put(#1,#2){\line(1,0){#3}}}
\newcommand{\y}[3]{\put(#1,#2){\line(0,1){#3}}}
\newcommand{\z}[3]{\put(#1,#2){\line(0,-1){#3}}}
\newcommand{\e}[3]{\put(#1,#2){\line(0,1){#3}}}
\y{   0}{   0}{   0}\x{   0}{   0}{  11}
\y{  11}{   0}{   0}\x{  11}{   0}{  12}
\y{  23}{   0}{   0}\x{  23}{   0}{  12}
\y{  35}{   0}{   0}\x{  35}{   0}{  12}
\y{  47}{   0}{1200}\x{  47}{1200}{  12}
\z{  59}{1200}{ 410}\x{  59}{ 790}{  12}
\z{  71}{ 790}{ 110}\x{  71}{ 680}{  12}
\z{  83}{ 680}{  45}\x{  83}{ 635}{  12}
\z{  95}{ 635}{   2}\x{  95}{ 633}{  12}
\z{ 107}{ 633}{  10}\x{ 107}{ 623}{  12}
\z{ 119}{ 623}{   2}\x{ 119}{ 621}{  12}
\z{ 131}{ 621}{   8}\x{ 131}{ 613}{  12}
\y{ 143}{ 613}{   0}\x{ 143}{ 613}{  12}
\z{ 155}{ 613}{   5}\x{ 155}{ 608}{  12}
\y{ 167}{ 608}{   4}\x{ 167}{ 612}{  12}
\z{ 179}{ 612}{   4}\x{ 179}{ 608}{  12}
\y{ 191}{ 608}{   2}\x{ 191}{ 610}{  12}
\z{ 203}{ 610}{   4}\x{ 203}{ 606}{  12}
\z{ 215}{ 606}{   2}\x{ 215}{ 604}{  12}
\y{ 227}{ 604}{   4}\x{ 227}{ 608}{  12}
\z{ 239}{ 608}{   3}\x{ 239}{ 605}{  12}
\y{ 251}{ 605}{   0}\x{ 251}{ 605}{  12}
\z{ 263}{ 605}{   2}\x{ 263}{ 603}{  12}
\z{ 275}{ 603}{   1}\x{ 275}{ 602}{  12}
\y{ 287}{ 602}{   3}\x{ 287}{ 605}{  12}
\z{ 299}{ 605}{   3}\x{ 299}{ 602}{  12}
\y{ 311}{ 602}{   2}\x{ 311}{ 604}{  12}
\z{ 323}{ 604}{   2}\x{ 323}{ 602}{  12}
\z{ 335}{ 602}{   1}\x{ 335}{ 601}{  12}
\z{ 347}{ 601}{   2}\x{ 347}{ 599}{  12}
\y{ 359}{ 599}{   2}\x{ 359}{ 601}{  12}
\y{ 371}{ 601}{   1}\x{ 371}{ 602}{  12}
\y{ 383}{ 602}{   0}\x{ 383}{ 602}{  12}
\z{ 395}{ 602}{   1}\x{ 395}{ 601}{  12}
\y{ 407}{ 601}{   1}\x{ 407}{ 602}{  12}
\z{ 419}{ 602}{   3}\x{ 419}{ 599}{  12}
\y{ 431}{ 599}{   1}\x{ 431}{ 600}{  12}
\z{ 443}{ 600}{   2}\x{ 443}{ 598}{  12}
\y{ 455}{ 598}{   0}\x{ 455}{ 598}{  12}
\y{ 467}{ 598}{   3}\x{ 467}{ 601}{  12}
\z{ 479}{ 601}{   2}\x{ 479}{ 599}{  12}
\y{ 491}{ 599}{   3}\x{ 491}{ 602}{  12}
\y{ 503}{ 602}{   0}\x{ 503}{ 602}{  12}
\z{ 515}{ 602}{   4}\x{ 515}{ 598}{  12}
\y{ 527}{ 598}{   0}\x{ 527}{ 598}{  12}
\y{ 539}{ 598}{   4}\x{ 539}{ 602}{  12}
\z{ 551}{ 602}{   4}\x{ 551}{ 598}{  12}
\z{ 563}{ 598}{   1}\x{ 563}{ 597}{  12}
\y{ 575}{ 597}{   1}\x{ 575}{ 598}{  12}
\z{ 587}{ 598}{   1}\x{ 587}{ 597}{  12}
\z{ 599}{ 597}{   2}\x{ 599}{ 595}{  12}
\z{ 611}{ 595}{   1}\x{ 611}{ 594}{  12}
\y{ 623}{ 594}{   0}\x{ 623}{ 594}{  12}
\y{ 635}{ 594}{   0}\x{ 635}{ 594}{  12}
\z{ 647}{ 594}{   4}\x{ 647}{ 590}{  12}
\z{ 659}{ 590}{   1}\x{ 659}{ 589}{  12}
\y{ 671}{ 589}{   0}\x{ 671}{ 589}{  12}
\y{ 683}{ 589}{   1}\x{ 683}{ 590}{  12}
\y{ 695}{ 590}{   1}\x{ 695}{ 591}{  12}
\y{ 707}{ 591}{   3}\x{ 707}{ 594}{  12}
\z{ 719}{ 594}{   2}\x{ 719}{ 592}{  12}
\y{ 731}{ 592}{   6}\x{ 731}{ 598}{  12}
\y{ 743}{ 598}{   1}\x{ 743}{ 599}{  12}
\z{ 755}{ 599}{   6}\x{ 755}{ 593}{  12}
\y{ 767}{ 593}{   8}\x{ 767}{ 601}{  12}
\z{ 779}{ 601}{   2}\x{ 779}{ 599}{  12}
\z{ 791}{ 599}{   1}\x{ 791}{ 598}{  12}
\y{ 803}{ 598}{   1}\x{ 803}{ 599}{  12}
\z{ 815}{ 599}{   5}\x{ 815}{ 594}{  12}
\y{ 827}{ 594}{   5}\x{ 827}{ 599}{  12}
\z{ 839}{ 599}{   3}\x{ 839}{ 596}{  12}
\z{ 851}{ 596}{   5}\x{ 851}{ 591}{  12}
\y{ 863}{ 591}{  11}\x{ 863}{ 602}{  12}
\z{ 875}{ 602}{   6}\x{ 875}{ 596}{  12}
\y{ 887}{ 596}{   6}\x{ 887}{ 602}{  12}
\y{ 899}{ 602}{   3}\x{ 899}{ 605}{  12}
\y{ 911}{ 605}{   4}\x{ 911}{ 609}{  12}
\y{ 923}{ 609}{  45}\x{ 923}{ 654}{  12}
\y{ 935}{ 654}{  97}\x{ 935}{ 751}{  12}
\z{ 947}{ 751}{ 751}\x{ 947}{   0}{  12}
\y{ 959}{   0}{   0}\x{ 959}{   0}{  12}
\y{ 971}{   0}{   0}\x{ 971}{   0}{  12}
\y{ 983}{   0}{   0}\x{ 983}{   0}{  12}
\y{ 995}{   0}{   0}\x{ 995}{   0}{  12}
\y{1007}{   0}{   0}\x{1007}{   0}{  12}
\y{1019}{   0}{   0}\x{1019}{   0}{  12}
\y{1031}{   0}{   0}\x{1031}{   0}{  12}
\y{1043}{   0}{   0}\x{1043}{   0}{  12}
\y{1055}{   0}{   0}\x{1055}{   0}{  12}
\y{1067}{   0}{   0}\x{1067}{   0}{  12}
\y{1079}{   0}{   0}\x{1079}{   0}{  12}
\y{1091}{   0}{   0}\x{1091}{   0}{  12}
\y{1103}{   0}{   0}\x{1103}{   0}{  12}
\y{1115}{   0}{   0}\x{1115}{   0}{  12}
\y{1127}{   0}{   0}\x{1127}{   0}{  12}
\y{1139}{   0}{   0}\x{1139}{   0}{  12}
\y{1151}{   0}{   0}\x{1151}{   0}{  12}
\y{1163}{   0}{   0}\x{1163}{   0}{  12}
\y{1175}{   0}{   0}\x{1175}{   0}{  12}
\y{1187}{   0}{   0}\x{1187}{   0}{  12}
\end{picture}} 
\end{picture} 
\caption{An example of a figure included in tests collected in project 
web page~\cite{web:RChL}. The
ratio of the histogram and analytical formula is shown for the $\frac{d \Gamma}{dq^2}$ in $\tau \to \pi^-\pi^-\pi^+\nu_\tau$ decay mode. 
 Only $F_1= F$ is non-zero. Normalization is not adjusted.
A statistical sample of $6 \cdot 10^{6}$ events  was used and semirealistic 
initialization as explained in this section.  Agreement within statistical 
errors is found. Fluctuations at the ends of the spectra are due to 
substantially less populated bins of that region. These tests represent a 
technical test not only of Monte Carlo generation but also for semianalytical 
numerical integration.  
   \newline
 \label{Fig:fone}                                                
}
\end{figure}
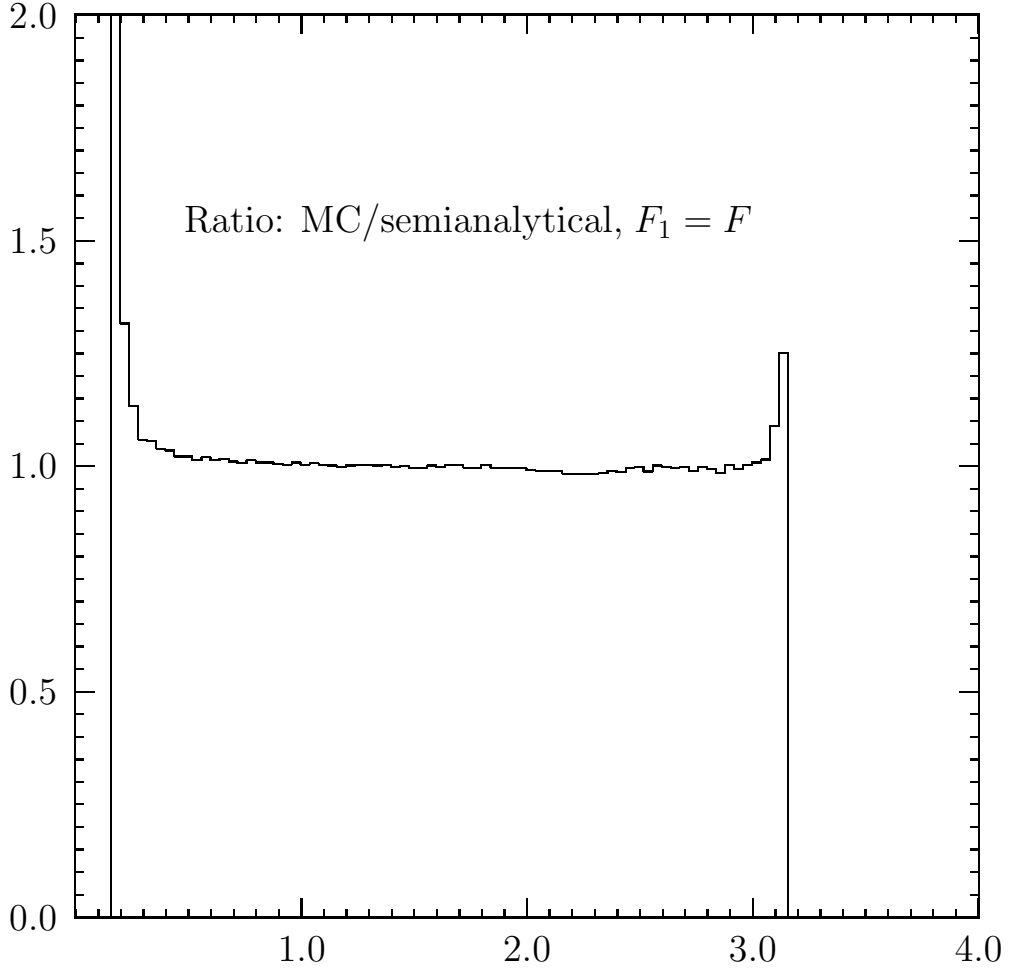

We take $F_1$ as given by Eq. (\ref{eq:t1r}), but all
other currents are set to 0. Our numerical results for the rate from  
numerical integration, $(1.8721\pm 0.02\%)\cdot 10^{-13}$ GeV, 
and from Monte Carlo generation, $(1.8722\pm 0.01 \%)\cdot 10^{-13}$ GeV (run with $6 \cdot10^{6}$ events), agree well.
The difference is only 0.005\% thus compatible with statistical error of
generated sample. Comparison of differential distributions, analogue to
Fig.~\ref{Fig:fone}, is available from the Web page~\cite{web:RChL}. For the next step we choose 
the $F_1$ and $F_2$ form factors according to Eq. (\ref{eq:t1r}), $F_4 = 0$. Numerical results are: $(4.2015\pm 0.02 \%)\cdot 10^{-13} $ GeV
for semianalytical calculation and $ (4.2023 \pm 0.01 \%)\cdot 10^{-13}$ GeV for Monte Carlo generation.
 The difference is 0.03\%, as expected. The figures are again available from Ref.~\cite{web:RChL}.

Finally, we consider the result for the total width predicted by Resonance Chiral Theory, 
given by the $F_1$, $F_2$ and $F_4$ contributions in Eqs. (\ref{eq:t1r}) and (\ref{f4_3pi}).
The $F_4$ form factor does not affect the value of the width at our precision and in this case it is  
$ (4.2025\pm 0.01 \%) \cdot 10^{-13}$ GeV (run with $6 \cdot10^{6}$ events).

The above comparisons check also that the
differential distribution  $d\Gamma/dq^2$ in  $\tau \to 3\pi \nu_\tau$ is numerically
stable. In particular that 
it is
 not affected by numerical problems 
due to double Gaussian integration. 
We have also checked, that if the  function value has been obtained
from 
interpolation of neighboring $q^2$'s,  the difference with the value
calculated directly 
was appropriately small as should be  expected\footnote{
The appropriate program for a test is available in the directory 

{\tt new-currents/RChL-currents/cross-check/check\_{}analyticity/check\_{}analyt\_{}3pi}\;.}.

\section{Numerical results for two and three-pseudoscalar channels}\label{sec:results}

In the previous section we have presented examples of
technical tests.  Let us now concentrate on numerical results, 
 corresponding to the most refined
options of the  currents included  in our distribution 
tar-ball\footnote{For the 
convenience  of updates,
 we have prepared Appendix \ref{app:D}, to be modified in the versions 
of the present paper to be included in the tar-ball. It summarizes elementary
installation benchmark results, that is the branching ratios calculated by 
the Monte Carlo.} which are of physics interest. 

In the phase space generation, 
we will take into account the differences between neutral and charged 
 pion and kaon masses, physical values will be taken. This has to be done  
to obtain proper kinematic configurations. 
On the other hand, this choice breaks constraints resulting from
isospin symmetry (see $^{\ref{foot:iso}}$) in a potentially uncontrolled way.
That is why we collect numerical results from Monte Carlo
 calculation in the form of Table~\ref{Table:bench}, where
the partial widths from Particle Data Group compilation 
\cite{Nakamura:2010zzi} are compared with our results obtained with 
isospin-averaged pseudoscalar masses and with the physical ones.

\begin{table}
\vspace{0.3cm} 
\begin{center}
{ \begin{tabular}{|r| c| c| c|} 
\toprule
Channel & \multicolumn{3}{c|}{Width, [GeV]} \\
\cline{2-4}
    & PDG  & Equal masses &   Phase space\\  
   &  &  &  with masses \\  
\midrule
{$ \pi^-\pi^0 \; \;\; \;$}  &      {($5.778 \pm 0.35\%)\cdot 10^{-13}$} &  {($5.2283 \pm 0.005\%)\cdot10^{-13}$} &  {$(5.2441\pm 0.005\%)\cdot 10^{-13}$} \\ 
{$ \pi^0 K^- \; \;\; \;$}   &       {($9.72\;\pm 3.5\%\;)\cdot 10^{-15}$} &  {($8.3981 \pm 0.005\%)\cdot10^{-15}$} &  {$(8.5810\pm 0.005\%)\cdot 10^{-15}$} \\ 
{$ \pi^-\bar K^0 \; \;\; \;$}   &  {($1.9\;\;\; \pm 5\%\;\;\;)\cdot 10^{-14}$} &  {($1.6798 \pm 0.006\%)\cdot10^{-14}$} &  {$(1.6512\pm 0.006\%)\cdot 10^{-14}$} \\ 
{$ K^-K^0 \; \;\; \;$}     &       {($3.60\; \pm 10\%\;\;)\cdot 10^{-15}$} &  {($2.0864 \pm 0.007\%)\cdot10^{-15}$} &  {$(2.0864\pm 0.007\%)\cdot 10^{-15}$} \\ 
 {$  \pi^-\pi^-\pi^+$} &           {($2.11\; \pm 0.8\%\;\;)\cdot 10^{-13}$} &  {($ 2.1013\pm 0.016\%)\cdot10^{-13}$} &   {$(2.0800\pm 0.017\%)\cdot 10^{-13}$}  \\ 
{$  \pi^0\pi^0\pi^-$}  &           {($2.10\; \pm 1.2\%\;\;)\cdot 10^{-13}$} &  {($ 2.1013\pm 0.016\%)\cdot10^{-13}$} &  {$(2.1256\pm 0.017\%)\cdot 10^{-13}$}\\ 
 {$  K^-\pi^-K^+$} &   {($3.17\; \pm 4\%\;\;\;)\cdot 10^{-15}$} &  {($3.7379 \pm 0.024\%)\cdot10^{-15}$} &   {$(3.8460\pm 0.024\%)\cdot  10^{-15}$}  \\ 
{$  K^0\pi^-\bar{K^0}$}  &         {($3.9\;\; \pm 24\%\;\;)\cdot 10^{-15}$} &  {($3.7385 \pm 0.024\%)\cdot10^{-15}$} &  {$(3.5917\pm 0.024\%)\cdot 10^{-15}$}\\
{$  K^-\pi^0 K^0$} &               {($3.60\; \pm 12.6\%\;\;)\cdot 10^{-15}$} &  {($2.7367\pm 0.025 \%)\cdot10^{-15}$} &  {$(2.7711 \pm 0.024\%)\cdot 10^{-15}$}\\
\bottomrule
\end{tabular} 
}  
\end{center}
\caption{The $\tau$ decay partial widths. For each channel,
the PDG value \cite{Nakamura:2010zzi} is compared with numerical results 
of Monte Carlo integration of our currents. The third column includes results 
with isospin averaged masses, whereas for the last column physical masses
were used.
Comparison of the last two columns enumerates the numerical effect of 
physical masses, breaking the assumption of isospin symmetry in a potentially
uncontrolled way.
Further results for individual decay channels are  given 
 in Subsections of Section \ref{sec:results}. 
} \label{Table:bench}
\end{table}


\subsection{$\pi^-\pi^-\pi^+ \nu_\tau$ and $\pi^0\pi^0\pi^- \nu_\tau$}\label{3pi_results}

 \begin{figure}[h!]
\centering
\subfigure{
\includegraphics[scale=.350]{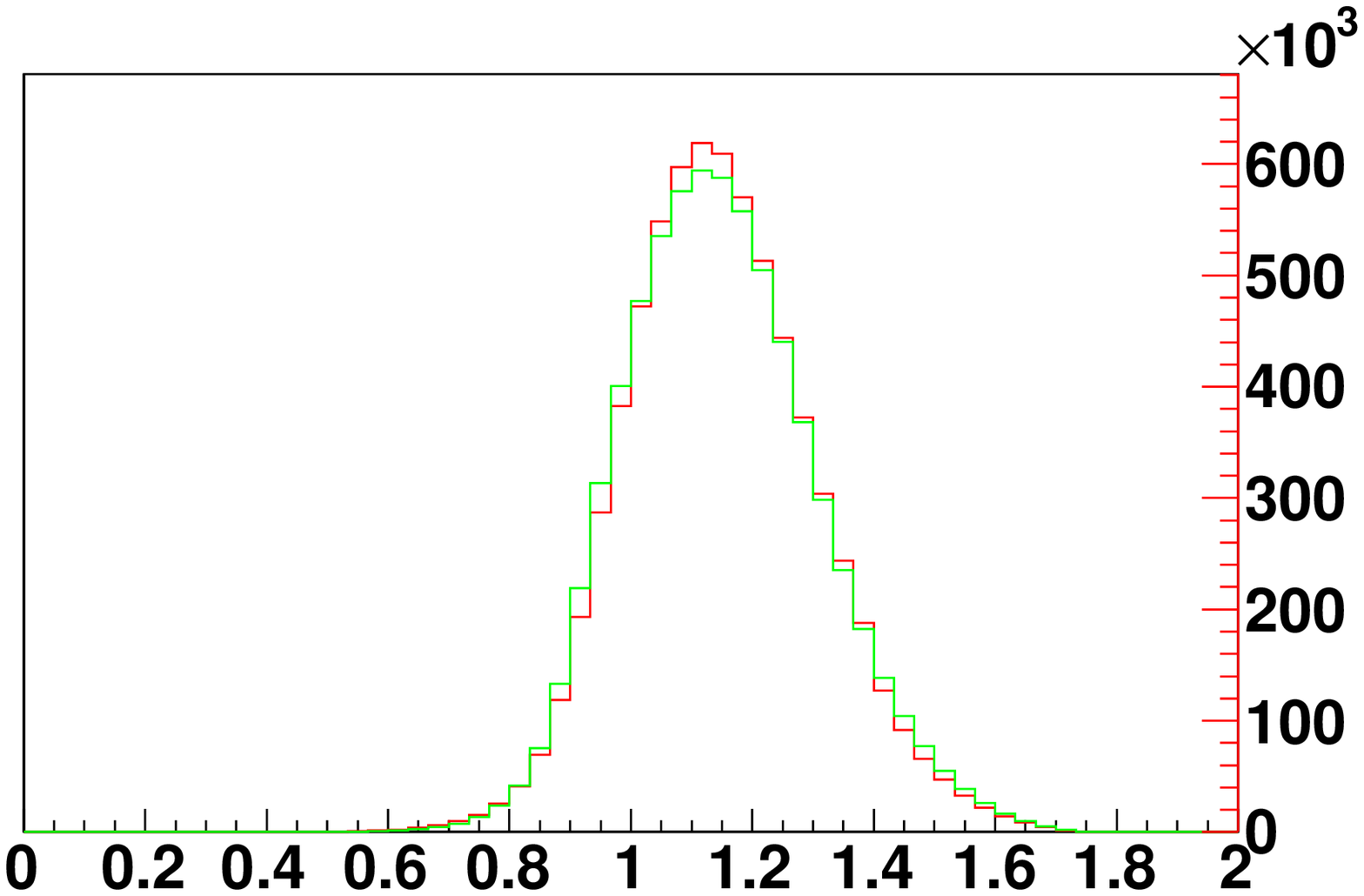}}
\subfigure{
\includegraphics[scale=.350]{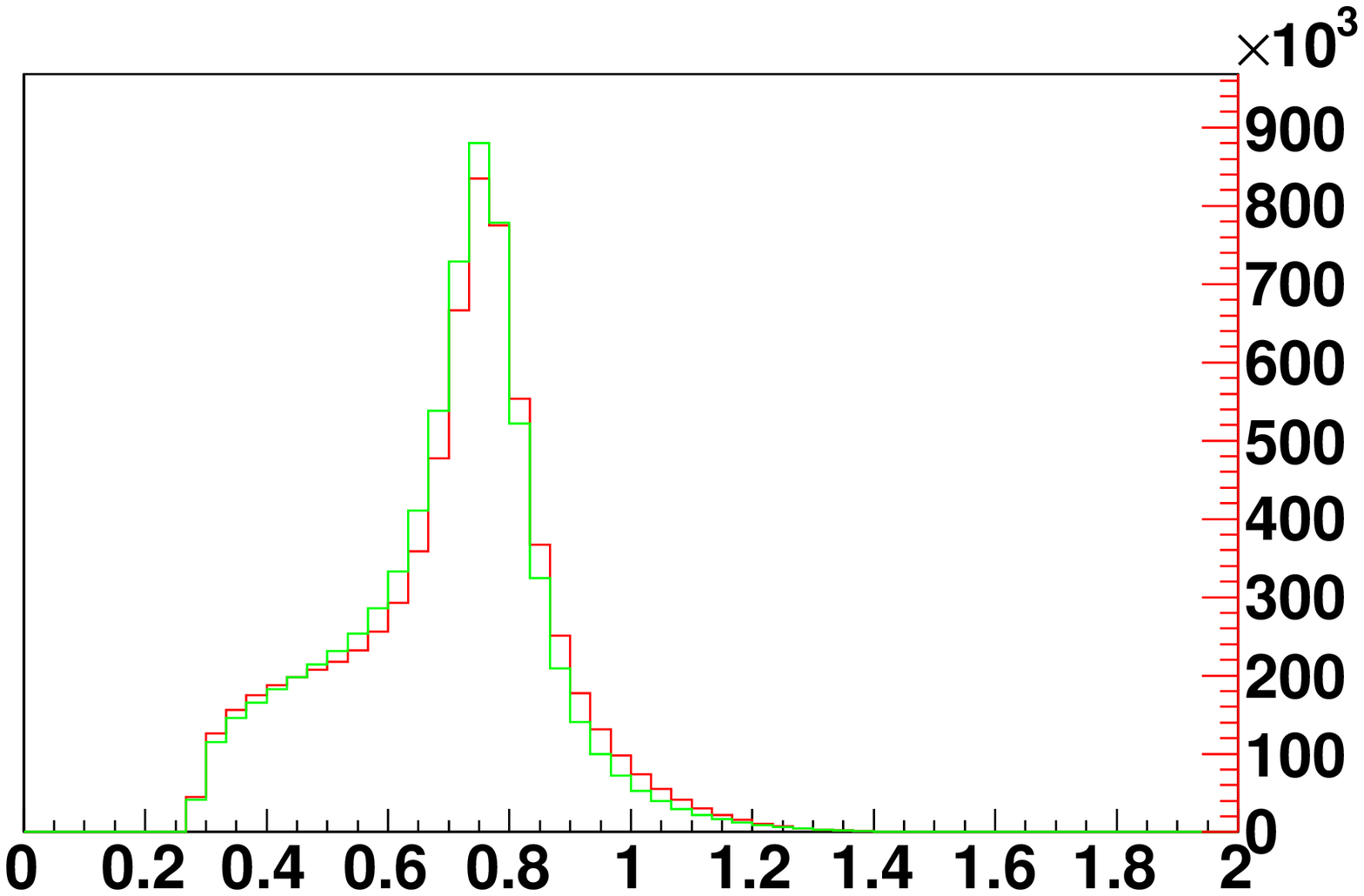}}
\caption{The   $\tau \to \pi^- \pi^-\pi^+\nu_\tau$ decay: comparison of
distributions for {\tt TAUOLA cleo} current~\cite{Golonka:2003xt}  
 and for our new current. 
On the left-hand side, the plot of $\pi^- \pi^-\pi^+$
invariant mass   is shown  and on the right-hand side  $\pi^+\pi^-$ invariant mass is given.
Green histograms (light grey) are for the new current, red (darker grey) are
for  {\tt TAUOLA cleo}. 
Distributions for the
$\tau \to  \pi^0 \pi^0 \pi^-\nu_\tau$  decay 
coincide with the ones for $\tau \to \pi^- \pi^-\pi^+\nu_\tau$. 
\label{Fig:pipipi}}
\end{figure}


Let us now turn to numerical results for the $\tau \to 3\pi \nu_\tau$ decays
obtained with our currents of Subsection \ref{Subsect:pipipi}. 
From the semianalytical calculation of partial width we get  $(2.10073\pm 0.02\%)\cdot 10^{-13}$ GeV 
for $\tau \to \pi^-\pi^-\pi^+\nu_\tau$  decay,
and  $(2.10072\pm 0.02 \%)\cdot 10^{-13}$ GeV  for $\tau \to \pi^0\pi^0\pi^-\nu_\tau$, practically the same value. The approximation of the equal masses for  $\pi^\pm$  and $\pi^0$
has been taken. In this case 
the Monte Carlo results are identical for $\pi^-\pi^-\pi^+$ and  $\pi^0\pi^0\pi^-$ final 
states\footnote{The chiral contribution $F_4^\chi(q^2,s_1,s_2)$,  Eq. (\ref{f4_3pi}), differs for  $\pi^-\pi^-\pi^+$ and  $\pi^0\pi^0\pi^-$. However,  $F_4(q^2,s_2,s_1)$ does not affect sizably the width.}; 
we have obtained $(2.1013 \pm 0.016 \%)\cdot 10^{-13}$ GeV.

For the physical, i.e., distinct  $m_{\pi^\pm}$ and $m_{\pi^0}$, masses we have 
obtained  $(2.0800 \pm 0.017 \%)\cdot 10^{-13}$ GeV for  $\pi^-\pi^-\pi^+$ and 
$(2.1256\pm 0.017 \%)\cdot10^{-13}$ GeV for  $\pi^0\pi^0\pi^-$ mode.
The difference for the distributions is too small to be seen 
and we present plots  
for the $\pi^0\pi^0\pi^-$ case. 
Only two example plots are given in Fig.~\ref{Fig:pipipi}. 
We point the reader to the web page~\cite{web:RChL}, for the booklet of 
comparisons obtained with {\tt MC-TESTER} \cite{Davidson:2008ma}. 
The figures for $d\Gamma/dq^2$ spectrum from Monte Carlo and analytical calculations
are also available from the plots of the Web page~\cite{web:RChL}. 

From the technical point of view to separate generation of the two $3\pi$ sub-channels one has to set 
the {\tt BRA1 = 0} for the  $\pi^0\pi^0\pi^-$ mode, and {\tt BRA1 = 1} for the  $\pi^-\pi^-\pi^+$,
 (e.g., in routine {\tt INITDK}, which is defined in our demonstration program \\
 {\tt new-currents/Installation/demo-standalone/taumain.f}).

An attempt on comparisons of the new model distributions and experimental data 
is also given later in the paper, in Section~\ref{sec:attempt}.


 \subsection{$K^-\pi^-K^+\nu_\tau$ and $K^0\pi^-\bar{K^0}\nu_\tau$}\label{KKpi_results}

\begin{figure}[h!]
\centering
\subfigure{
\includegraphics[scale=.350]{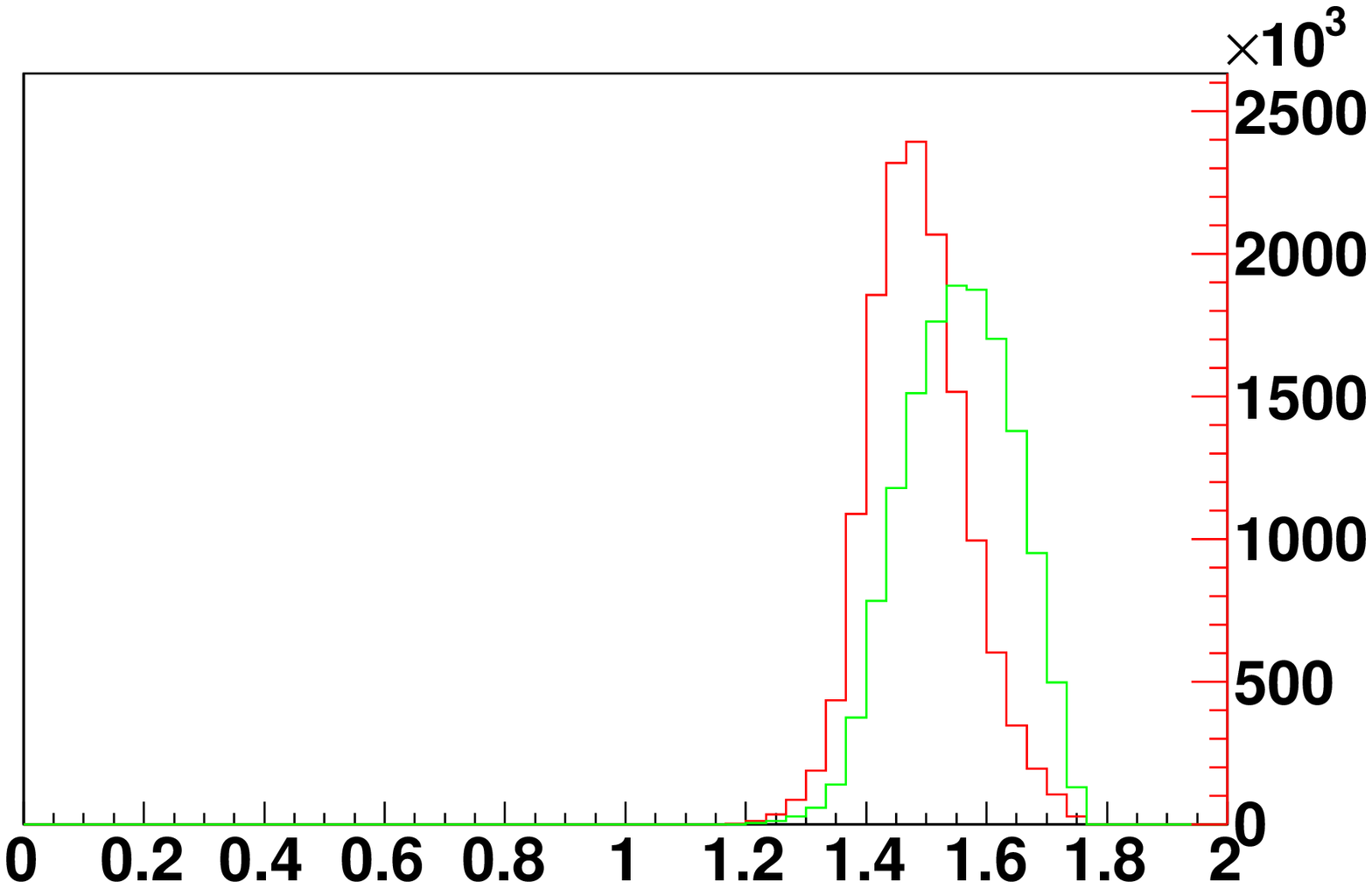}}
\subfigure{
\includegraphics[scale=.350]{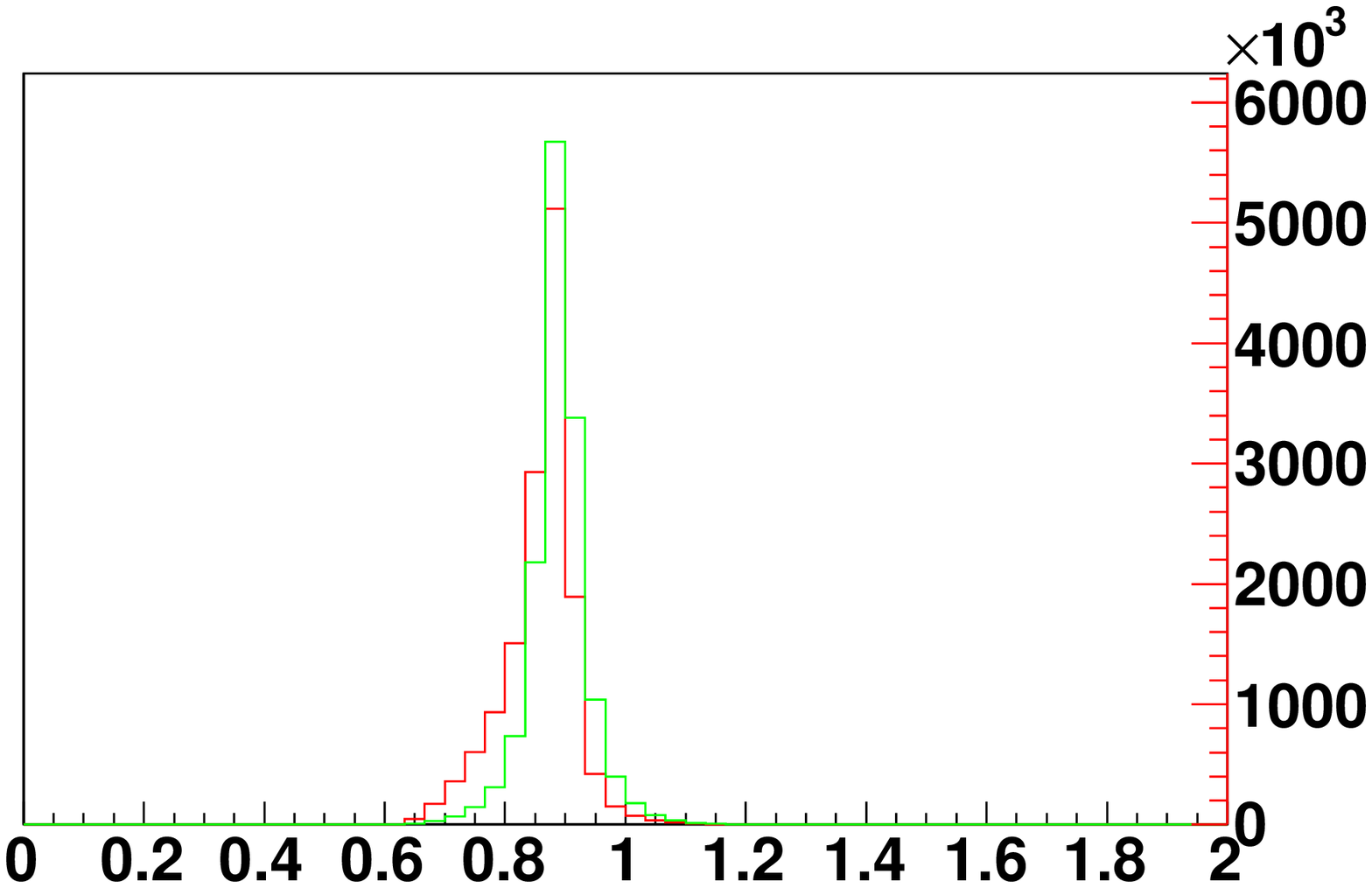}}
\caption{The   $\tau \to K^- \pi^-K^+\nu_\tau$ decay: comparison of
distributions for {\tt TAUOLA cleo} current~\cite{Golonka:2003xt}  
 and for our new current. 
On the left-hand side, plot of $K^-\pi^-K^+$ invariant mass   is shown  and on the right-hand side  $K^+\pi^-$ invariant mass is given.
Green histograms (light grey) are for the new current, red (darker grey) are
for  {\tt TAUOLA cleo}. 
\label{Fig:KKpi}}
\end{figure}

\begin{figure}[h!]
\centering
\subfigure{
\includegraphics[scale=.350]{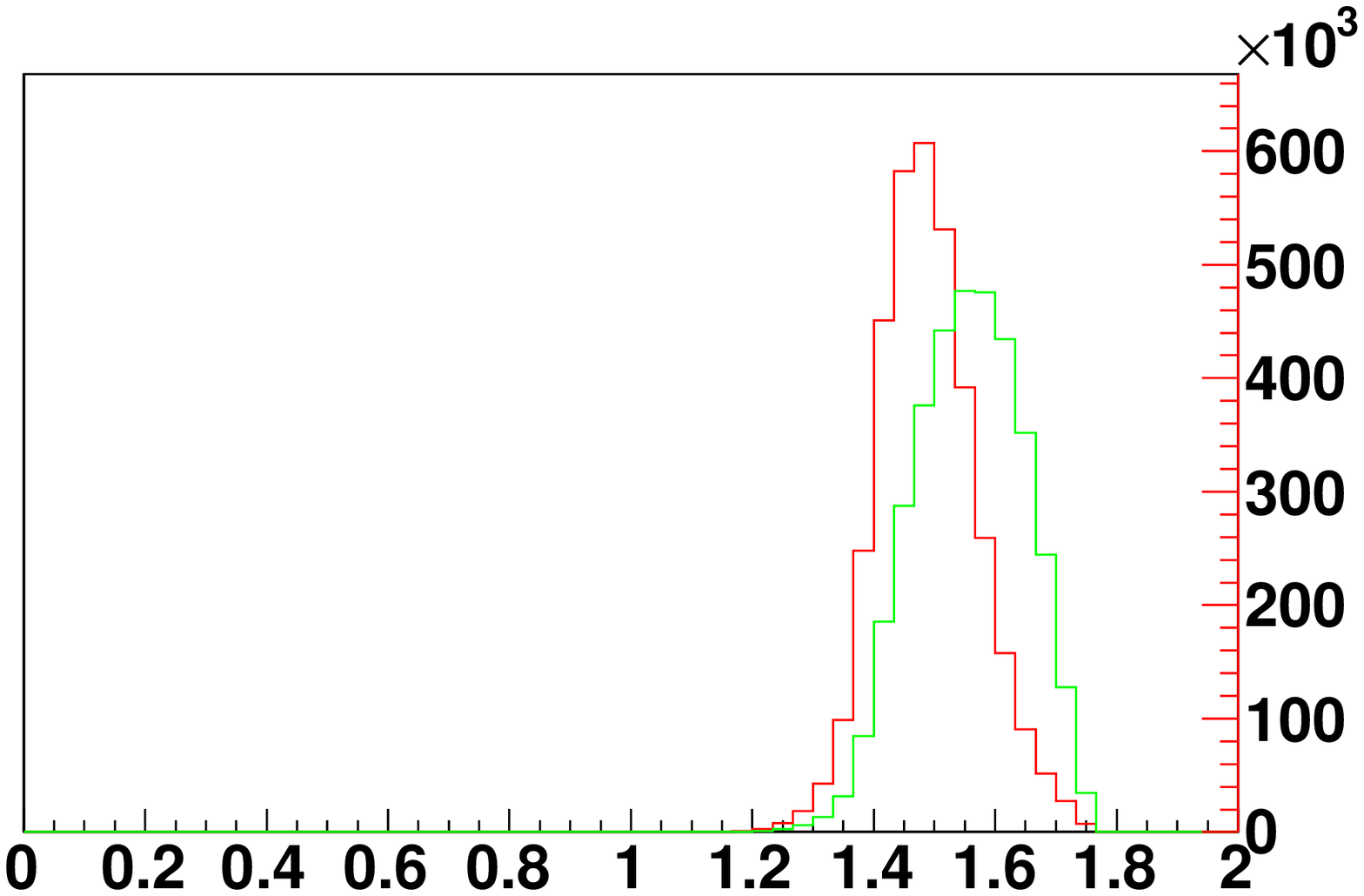}}
\subfigure{
\includegraphics[scale=.350]{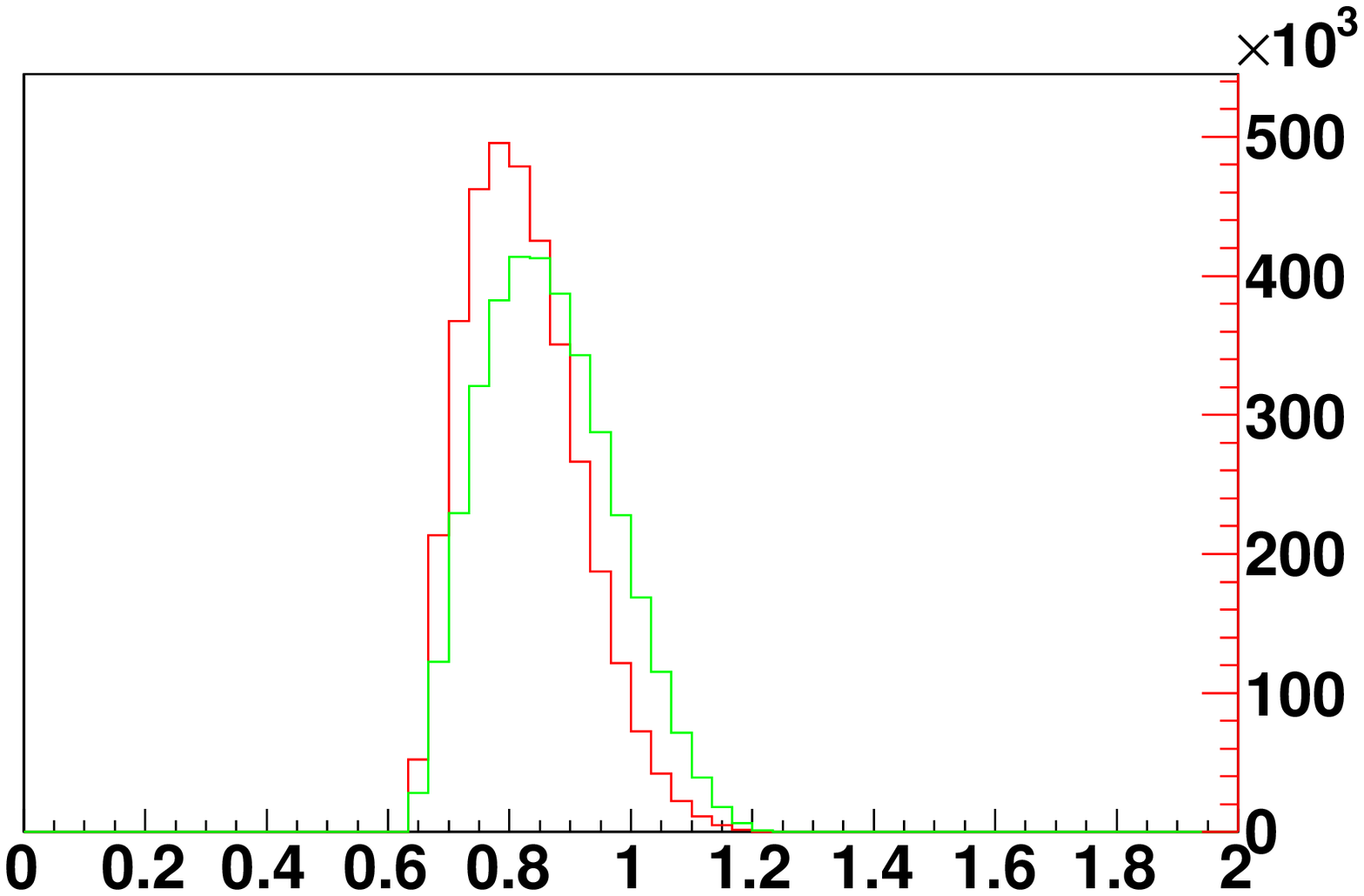}}
\caption{The   $\tau \to K^0 \pi^-\bar{K^0} \nu_\tau$ decay: comparison of
distributions for {\tt TAUOLA cleo} current~\cite{Golonka:2003xt}  
 and for our new current. 
On the left-hand side, plot of $ K_S^0 K_S^0 \pi^-$ invariant mass   is shown  and on the right-hand side $K_S^0\pi^-$  invariant mass is given.
Green histograms (light grey) are for the new current, red (darker grey) are
for  {\tt TAUOLA cleo}.   
\label{Fig:K0K0pi}}
\end{figure}

As in the case of 3$\pi$ decay modes, the Monte Carlo generated distributions
are relegated to the project web page \cite{web:RChL}.
In particular successful checks with the analytic function for $d\Gamma/dq^2$ taken 
from Ref.~\cite{Dumm:2009kj} are shown there. In the following, we present
figures \ref{Fig:KKpi} and \ref{Fig:K0K0pi} comparing the two histograms obtained with our new  and {\tt cleo}
versions of {\tt TAUOLA} currents. 
The differences are substantial.  
CLEO Collaboration \cite{Liu:2002mn, Coan:2004ep} 
introduced an ad-hoc parameter violating a property stemming directly from QCD,
the normalization 
of the vector form factor given by the chiral anomaly \cite{Portoles:2004vr}. 
Of course, in our new current we are not taking into account excited resonances. 
In the past,  two couplings could only be estimated  because of
unavailability of data to determine them from a fit. These aspects can (and should) be improved 
at the time of  confronting with the data\footnote{Whenever possible, the agreement with the
data should not be achieved by straightforward violation of the theoretical assumptions.  
Discrepancies may point to faulty background subtraction, or call for 
 improvements or replacement of the model used in currents calculation. 
In practice, this may be difficult and require significant and simultaneous effort on both theoretical and experimental sides. 
That is why one may have 
to accept temporary introduction of ad-hoc factors into the currents now as well.}.

 Let us now turn to the decay widths.
The result for $SU(2)$ symmetric masses  from {\tt TAUOLA}  with a sample of $2\cdot 10^{6}$ events is
  $\Gamma= (3.7379\pm 0.024\%)\cdot 10^{-15}$ GeV  for $K^-\pi^- K^+$  and $\Gamma=(3.7385\pm 0.024\%)\cdot10^{-15}$ GeV 
for $K^0\pi^-\bar{K^0}$. The difference for the partial width of the two channels is within statistical error. 
The analytical result is the same for both channels and  was found to  be $(3.7383\pm 0.02\%)\cdot 10^{-15}$ GeV. 
It agrees with the ones of  Monte Carlo.

For physical masses of the pseudoscalars, the Monte Carlo results for  $K^-\pi^- K^+$ and $K^0\pi^- \bar{K^0}$ 
are, respectively, $\Gamma= (3.8460\pm 0.024\%)\cdot10^{-15}$ GeV
and $\Gamma= (3.5917\pm 0.024\%)\cdot10^{-15}$ GeV. The effect of the mass adjustment in phase space gives an effect 
of the  order of 
3\%. Agreement with PDG results (see Table \ref{Table:bench}) is acceptable,
but improvements from fits are envisaged.

 \subsection{$K^-\pi^0K^0\nu_\tau$}\label{KK0pi0_results}
\begin{figure}[h!]
\centering
\subfigure{
\includegraphics[scale=.350]{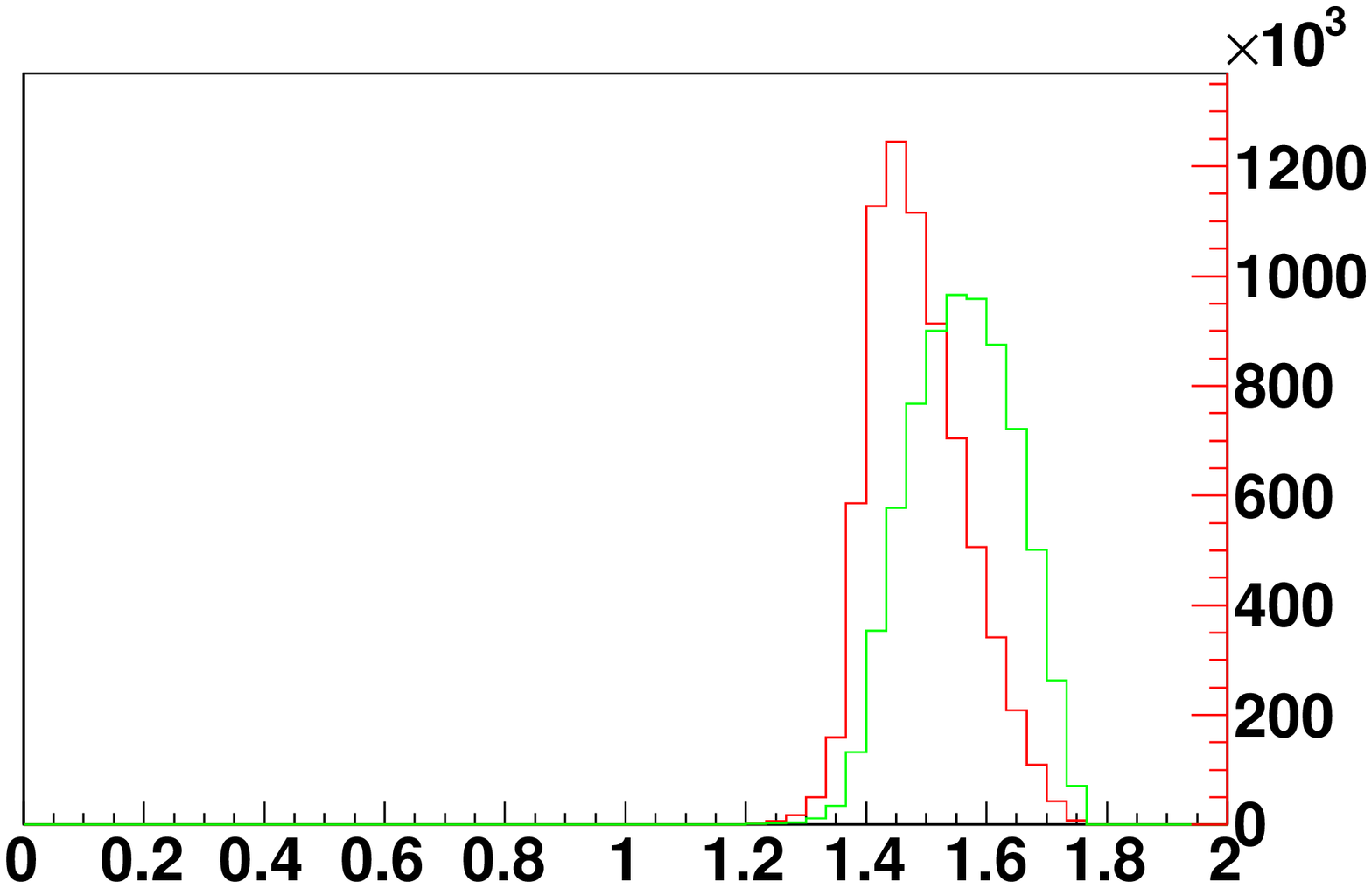}}
\subfigure{
\includegraphics[scale=.350]{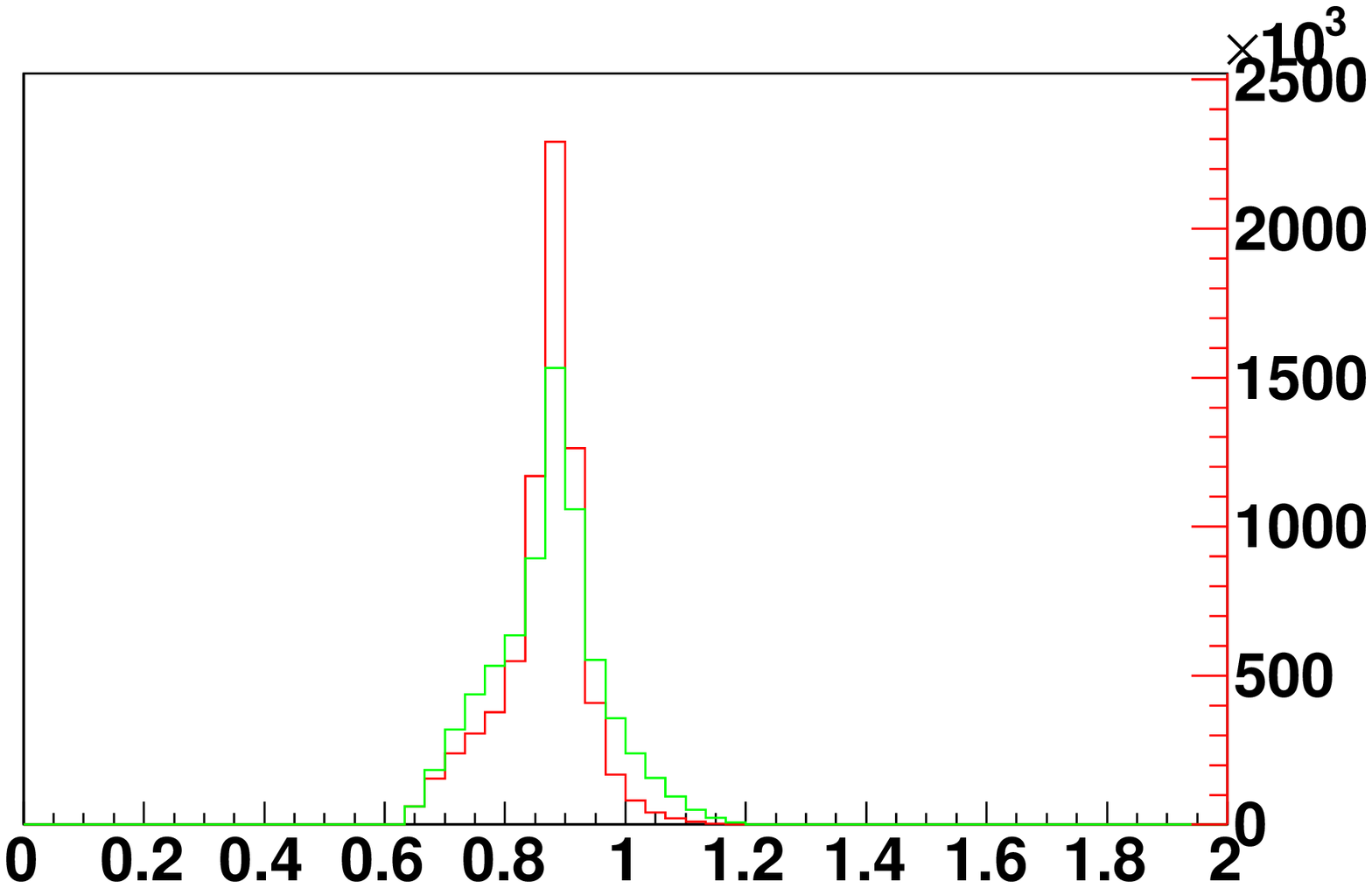}}
\caption{The   $\tau \to K^-  \pi^0 K^0\nu_\tau$ decay: comparison of
distributions for {\tt TAUOLA cleo} current~\cite{Golonka:2003xt}  
 and for our new current. 
On the left-hand side, plot of   $ K^- \pi^0 K_L^0$ invariant mass   is shown  and on the right-hand side  $K^-\pi^0$ invariant mass is given.
Green histograms (light grey) are for the new current, red (darker grey) are
for  {\tt TAUOLA cleo}.   
\label{Fig:KK0pi0}}
\end{figure}

Again, Monte Carlo generated distributions
are relegated to the project web page \cite{web:RChL}.
In particular, successful checks with analytic function for $d\Gamma/dq^2$ taken 
from Ref.~\cite{Dumm:2009kj} are shown there. Figure \ref{Fig:KK0pi0}
presents comparison between histograms obtained with present and {\tt cleo}
versions of {\tt TAUOLA} initialization. As we can see, the differences are 
substantial. Explanations of the previous subsection for the $K^-\pi^- K^+$ case 
apply.
 
Let us now turn our attention to the partial width for   $K^-\pi^0 K^0$.
The $SU(2)$ symmetric result from {\tt TAUOLA} and a sample of $2\cdot 10^{6}$ events, $ (2.7367\pm 0.025\%)\cdot 10^{-15}$ GeV, agrees well with 
the analytical result $(2.7370\pm 0.02\%) \cdot 10^{-15} $ GeV. 
The effects of realistic masses in phase space are of the order of 1 \% and
 the Monte Carlo result is $(2.7711\pm 0.024\%)\cdot 10^{-15}$ GeV. 
Agreement with the PDG result (see Table \ref{Table:bench}) is not good.

Analytical results for $K^- \pi^- K^+$ and $K^- \pi^0K^0$, which are obtained 
in the $SU(2)$ limit
 $\Gamma(\tau^- \to K^- \pi^-K^+\nu)/\Gamma(\tau^- \to  K^- \pi^0K^0\nu) = 3.7383/2.7370 \simeq 1.366\simeq 4/3$, 
compare well with the result of Ref.~\cite{Dumm:2009kj}.

With this subsection, we complete a presentation of results for $\tau$ decays 
into  three pseudoscalars. Let us now turn to the case of the decays into two 
pseudoscalars, which are simpler from technical point of view.

\subsection{ $\pi^-\pi^0\nu_\tau$,  $\pi^0 K^-\nu_\tau$,  $\pi^- \bar K^0\nu_\tau$  and $K^-K^0\nu_\tau$  }\label{sect:numer-2scal}
In this case, there is only one non-trivial invariant mass distribution, $d\Gamma/ds$, which can be constructed from the decay products. This distribution 
and its ratio to semianalytical result is given for all 
two pseudoscalar final states in the Web page \cite{web:RChL}.
 For all two-pseudoscalar modes we use samples 
of $2\cdot 10^{7}$ events%
\footnote{By default, we  include the FSI effects, the parameter {\tt FFVEC = 1}. 
FSI can be switched off if {\tt FFVEC = 0} is set
in the file {\tt new-currents/RChL-currents/value\_{}parameter.f }\;.}.

The analytical result for 
$\tau^-\to \nu_\tau \pi^-\pi^0$ equals $(5.2431\pm 0.02\%)\cdot 10^{-13} $ GeV, and for $\tau^-\to \nu_\tau K^-K^0$ 
is  $(2.0863\pm 0.02\%)\cdot10^{-15}$ GeV. The Monte Carlo 
results are respectively $(5.2441\pm 0.005 \%)\cdot 10^{-13}$ GeV and 
$(2.0864\pm 0.007 \%)\cdot 10^{-15}$ GeV.
In both channels  the physical values of pion and kaon masses  are used.
As one can see, the obtained $K^-K^0$ width is only  $\sim 58\%$ of the PDG value. 
Since the mass of the $\rho$ resonance 
is less than the two-kaon threshold, a significant contribution has to be expected from both $\rho'$ and $\rho''$. 
To check this assumption, we used the parametrization~(\ref{VFFpipi}) for the two-kaon form factor $F_{KK}^{V}$. For the moment, 
we use the same numerical value of the parameters $\gamma$ and $\delta$ as in the two-pion case, an assumption which holds in 
the $SU(3)$ symmetry limit\footnote{The parameters $\phi_1$ and $\phi_2$ are subleading and  their values 
are unsubstantial for this check.}.
The result for the partial width of $\tau^-\to \nu_\tau K^-K^0$ is $(2.6502\pm 0.008\%)\cdot10^{-15}$ GeV.
However, in the real world the parameters for pion and kaon modes are likely not to coincide and have to be fitted by the experiments. 
Corrections of order $\sim30\%$ due to $SU(3)$ breaking are expected%
\footnote{For the two-kaon mode one can  include a contribution from  excited resonances by setting {\tt FFKKVEC = 1}. 
To run with only $\rho(770)$ 
exchange, Eq.~(\ref{VFFKK}), set {\tt FFKKVEC = 0}. 
Our default is {\tt FFKKVEC = 0} and {\tt FFVEC=1.}   }.
The Monte Carlo 
result for the sum of the two channels $\pi^0K^-\nu_\tau$ and $\pi^-\bar K^0\nu_\tau$ is $(2.5197\pm 0.008 \%)\cdot 10^{-14}$ GeV
if the $SU(2)$ symmetric masses are used  and a sample of $2\cdot 10^{7}$ events is generated  with {\tt TAUOLA} and new currents. 
The corresponding analytical result is $(2.5193 \pm 0.02\%)\cdot 10^{-14}$ GeV.  
The {\tt TAUOLA} run with  physical pion and kaon masses gives $(2.5092\pm 0.008 \%)\cdot 10^{-14}$ GeV. 
Separate partial widths for  $\pi^0K^-\nu_\tau$ and  $\pi^-\bar K^0\nu_\tau$ channels  calculated from the Monte Carlo 
are given%
~\footnote{To  separate submodes one has to set
{\tt BRKS = 0} for $\pi^0 K^-$ or {\tt BRKS = 1} for $\pi^- \bar K^0$
  in routine 
{\tt INITDK} residing, e.g., in our example
{\tt new-currents/Installation/demo-standalone/taumain.f} \;.}
 in Table~\ref{Table:bench}.
This can be compared\footnote{
To run the $K\pi$ mode with the vector form factor of Eq.~(\ref{VFFKpi}), {\tt FFKPIVEC = 1} should be set
in {\tt new-currents/RChL-currents/value\_{}parameter.f }\;.
To use Eqs.~(17) and (18) 
of Ref.~\cite{Boito:2008fq} {\tt FFKPIVEC } has to be set to 0. For the default we take {\tt FFKPIVEC = 1}.
}
with the  result  $2.1829 \cdot 10^{-14}$ GeV for the $K\pi\nu_\tau$ partial width based on Eqs.~(17) and (18) of Ref.~\cite{Boito:2008fq}.
The numerical value for  parameters of our model are taken from Ref.~\cite{Boito:2008fq}, Table 4, second column. 
The difference\footnote{We are thankful to Jorge Portol\'es for the discussion on  the differences between the models.}
 between our and~\cite{Boito:2008fq}  is about\footnote{ The 
reason for this difference of 15\% is because we are not using Eq.~(19) of Ref.~\cite{Boito:2008fq}. 
The difference between the results of Refs.~\cite{Jamin:2008qg} and \cite{Boito:2008fq} for the vector 
form factor contribution is only $\sim$ 4\%. 
See Section~\ref{sec:systematic} and Appendix \ref{app:F} for a related discussion.}
 15\% for the $K\pi$ partial width.

In order to test our improvement in the treatment of FSI \footnote{Even a simple Breit-Wigner includes a crude description of FSI \cite{Jamin:2001zq}, 
where, e.g., off-shell effcts are neglected in the resummation of loops.}
we have run the program with 
the {\tt FFVEC = 0}. It corresponds to neglecting the real part of 
the loop contributions 
in $F_{\pi\pi}^{V}(s)$, $F_{KK}^{V}(s)$ and $F_{K\pi}^{V}(s)$, namely ${\mathrm Re } A_{PQ}(s) = 0$, ${\mathrm Re } A_{P}(s) = 0$ 
in Eqs.~(\ref{VFFpipi})-(\ref{VFFKpi}). 
In this case, the results for the partial widths are $\Gamma(\tau^-\to \nu_\tau \pi^-\pi^0) = (4.0642 \pm 0.005 \%)\cdot 10^{-13}$ GeV, 
$\Gamma(\tau^-\to \nu_\tau K^-K^0) = (1.2201 \pm 0.007 \%)\cdot 10^{-15}$ GeV (note that only the $\rho$ meson exchange is included, namely {\tt FFKKVEC = 0}),  
$\Gamma(\tau^-\to \nu_\tau \pi^0 K^-) = (7.4275 \pm 0.004 \%)\cdot 10^{-15}$ GeV,  
$\Gamma(\tau^-\to \nu_\tau \pi^- \bar K^0 ) = (1.4276 \pm 0.006 \%)\cdot 10^{-14}$ GeV. 

\begin{figure}[h!]
\centering
\subfigure{
\includegraphics[scale=.350]{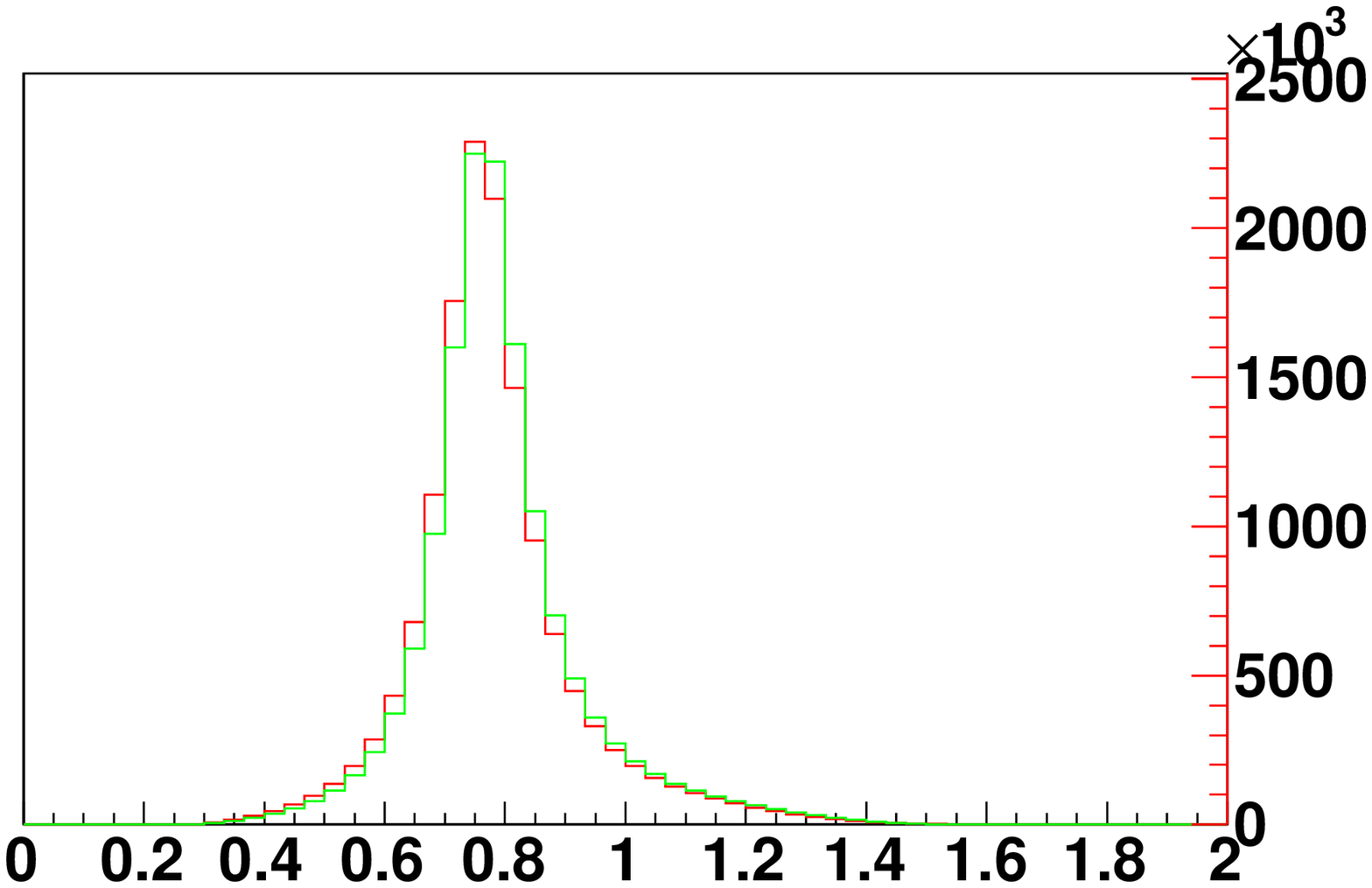}}
\subfigure{
\includegraphics[scale=.350]{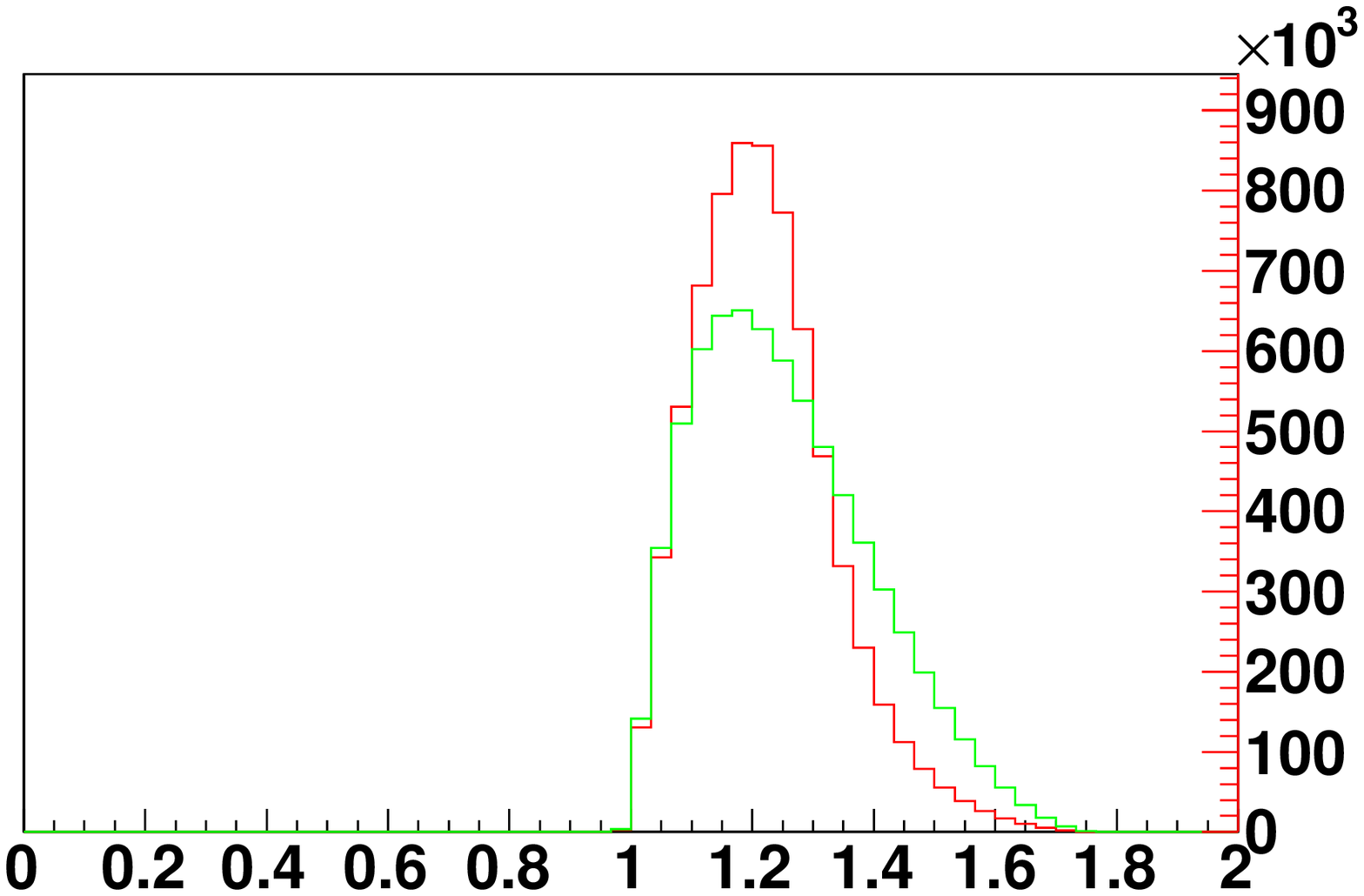}}
\caption{$\tau \to \pi^-  \pi^0\nu_\tau$ and $\tau \to K^-  K^0\nu_\tau$ decays: 
Comparison of
distributions for {\tt TAUOLA cleo} current~\cite{Golonka:2003xt}  
 and for our new current. 
On the left-hand side, plot of $ \pi^- \pi^0$ invariant mass   is shown  and on the right-hand side $ K^-K_S^0$ are for new current, red (darker grey) are
for  {\tt TAUOLA cleo}.   
\label{Fig:two-rho}}
\end{figure}

\begin{figure}[h!]
\centering
\subfigure{
\includegraphics[scale=.350]{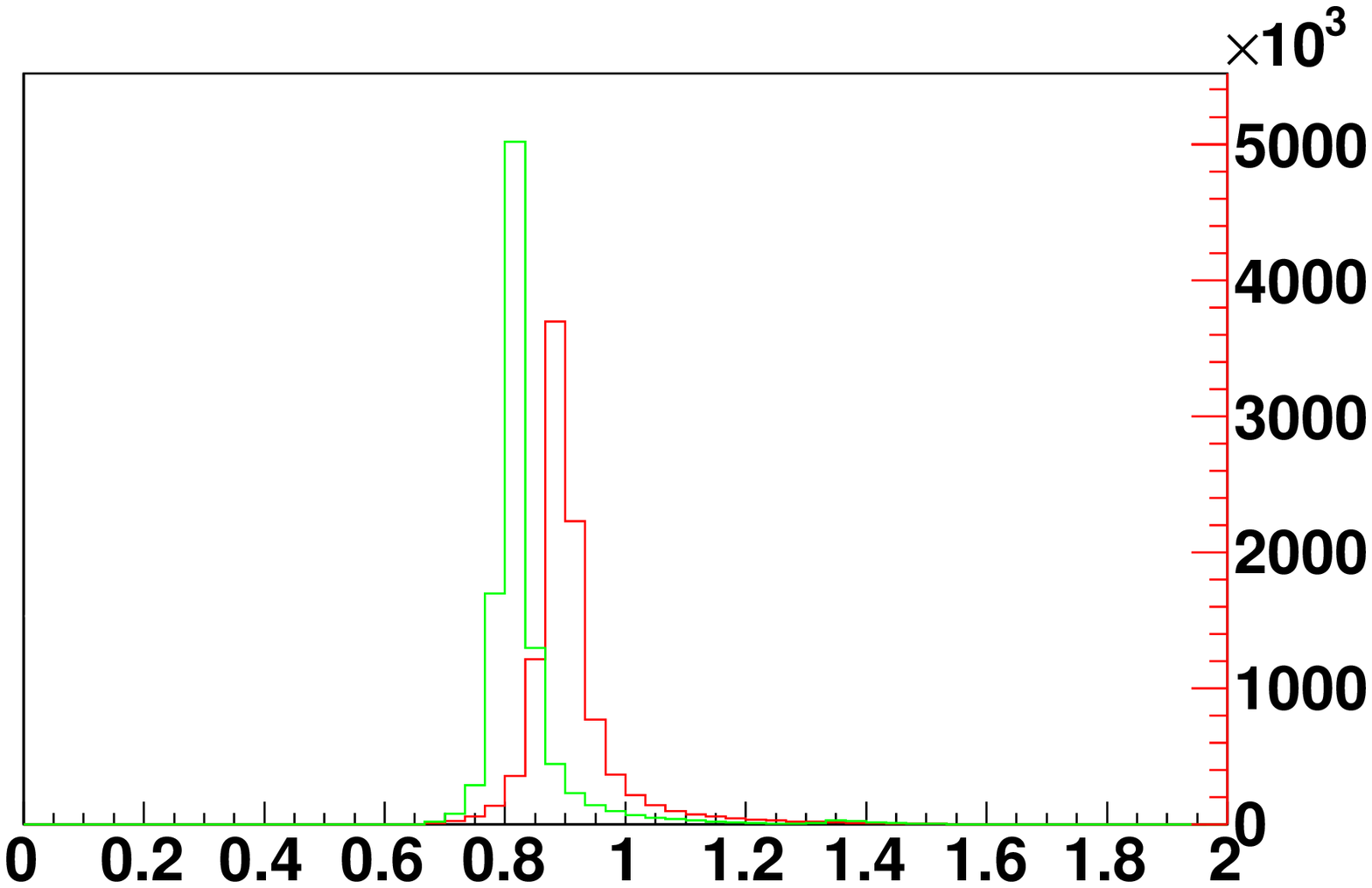}}
\subfigure{
\includegraphics[scale=.350]{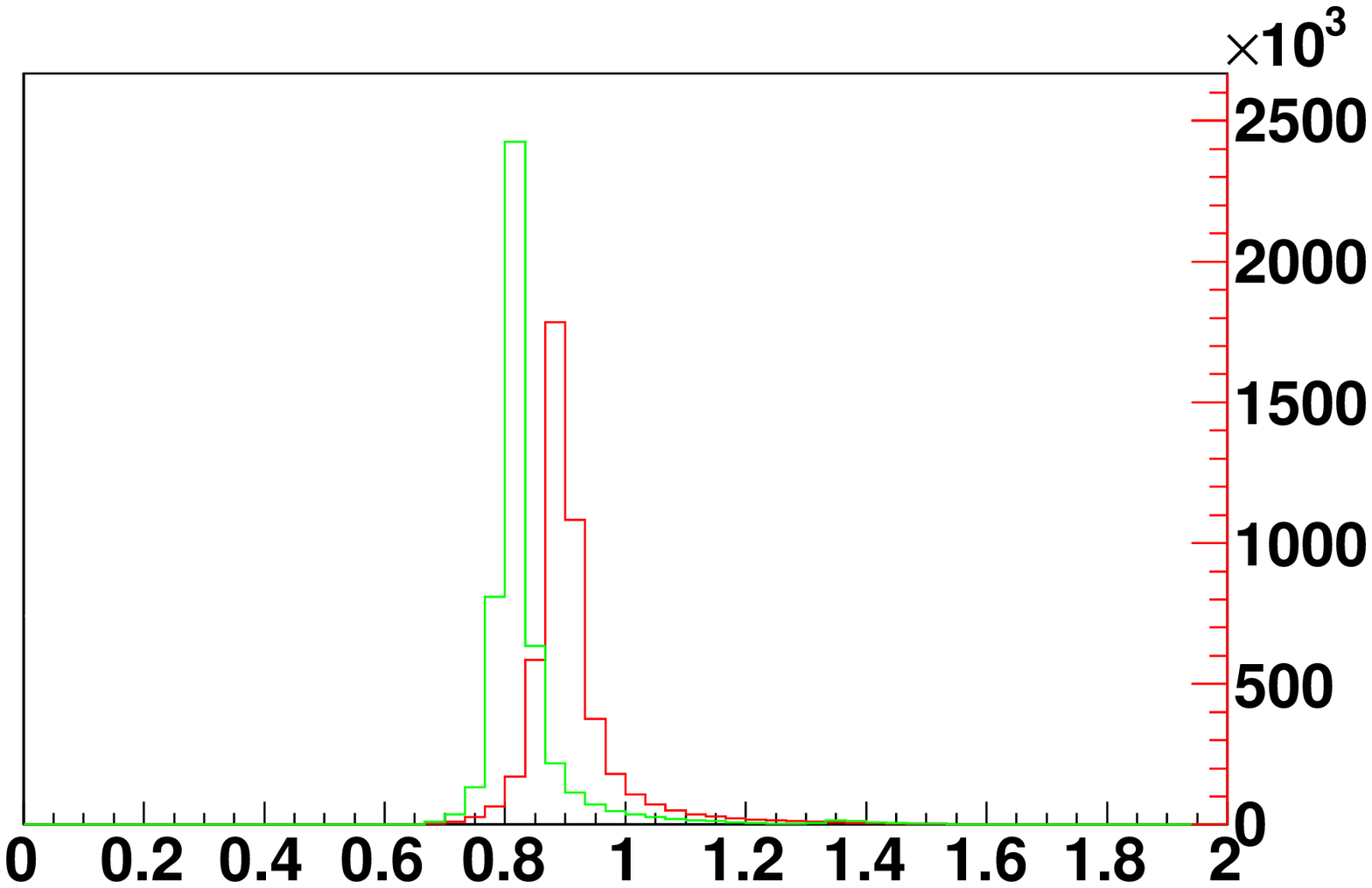}}
\caption{$\tau \to \pi^0 K^- \nu_\tau$ and $\tau \to \pi^- \bar K^0  \nu_\tau$ decays: 
Comparison of
distributions for {\tt TAUOLA cleo} current~\cite{Golonka:2003xt}  
 and for our new current. 
On the left-hand side, plot of $\pi^0 K^-$ invariant mass   is shown  and on the right-hand side  $\pi^- K_S^0$   invariant mass is given.
Green histograms (light grey) are for the new current, red (darker grey) are
for  {\tt TAUOLA cleo}.  
\label{Fig:two-Kstar}}
\end{figure}

Comparison with the data is technically simpler
and further phenomenological efforts should be delegated to that stage of the work. In fits, one has to work with single one-dimensional  $d\Gamma/ds$
spectrum and current as a function of single argument $s$ as well.
We present comparisons of $d\Gamma/ds$ spectra from {\tt cleo} current and the present parametrization of {\tt TAUOLA}, 
 for   
$\pi^-\pi^0\nu_\tau$, $K^-K_S^0\nu_\tau$ in Fig. \ref{Fig:two-rho}, 
and for $\pi^0K^-\nu_\tau$ and $\pi^-K_S^0\nu_\tau$ in Fig. 
\ref{Fig:two-Kstar}, respectively.

\subsection{Attempt at comparison with the data}\label{sec:attempt}

\begin{figure}[h!]
\centering
\subfigure{
\includegraphics[scale=.350]{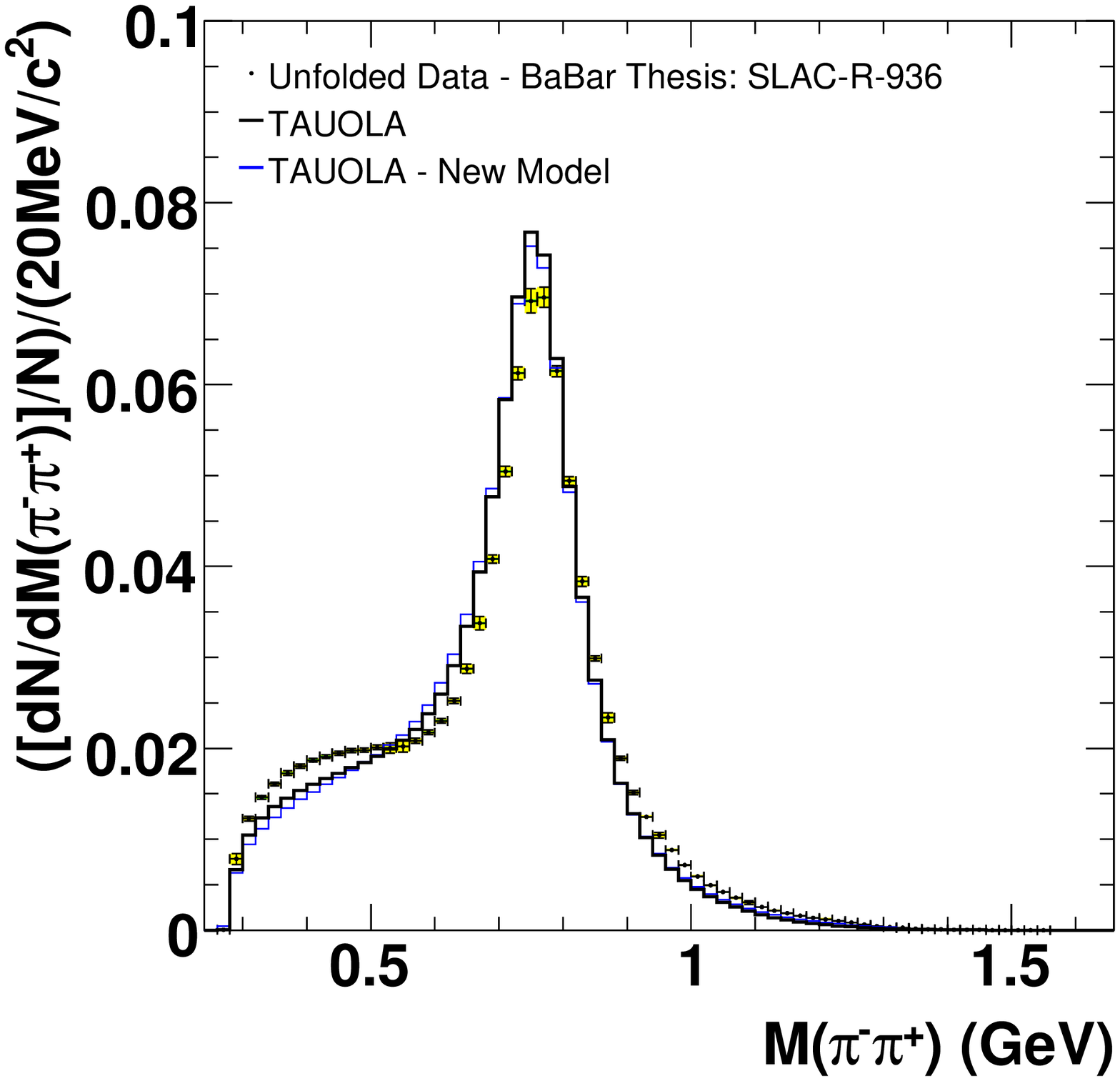}}
\subfigure{
\includegraphics[scale=.350]{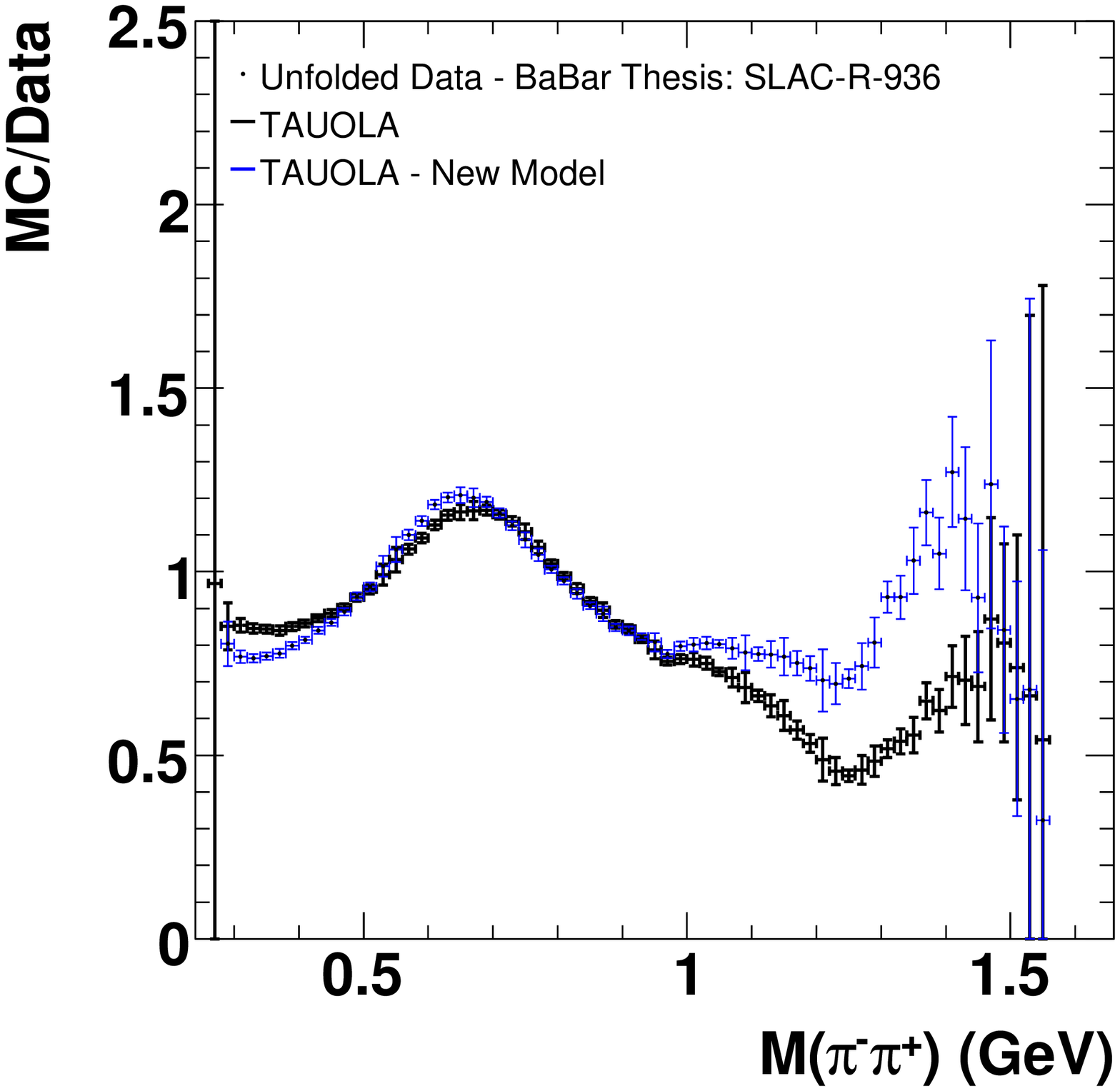}}
\caption{Invariant mass distribution of the $\pi^+\pi^-$ pair in $\tau \to \pi^- \pi^-\pi^+\nu_\tau$ decay.
Lighter grey histogram is from our model, darker grey is from default parametrization of {\tt TAUOLA cleo}.  The unfolded BaBar data are taken from Ref.~\cite{Nugent:2009zz} \label{Fig:Ian}. 
The plot on the left-hand side corresponds to the differential decay distribution, and the one on the right-hand side to plot ratios 
between Monte Carlo results and data. Courtesy of Ian Nugent.}
\end{figure}

Let us stress once again that our parametrization for all new currents is based on Resonance Chiral
Theory and is thus self-consistent. However, only minimal
attempts on adjusting to the data have  been performed. Only one dimensional 
$q^2$ distributions have been used in case of $3 \pi$~\cite{Dumm:2009va} and 
 $KK\pi$~\cite{Dumm:2009kj} decays,   to fit parameters such as $F_V$, $F_A$ etc.
 For the  $3\pi$ channel, relatively good agreement  with ALEPH data 
is shown in Fig. 3 of Ref.~\cite{Dumm:2009va}, but it represents a consistency
check  of the input.  
The proper work on fits is only to start now. The computing and theoretical 
framework is ready.
It is not
surprising that, for example, agreement with the unfolded BaBar data, 
see Fig.~\ref{Fig:Ian} of Ref.~\cite{Nugent:2009zz}, is not satisfactory. 

\section{Program organization}\label{sec:software}

The reference version of the {\tt TAUOLA} library that we  used as a starting point for the 
present work is the  {\tt TAUOLA cleo} documented in 
Ref.~\cite{Golonka:2003xt}. The choice is not accidental. This version is used as a
starting point for Belle and BaBar software as well. Also C++ implementation of 
{\tt TAUOLA} \cite{Davidson:2010rw}, installed in the library of Ref.~\cite{Kirsanov:2008zz}, uses this initialization. 
Porting into collaboration software is thus technically prepared.
This means the solution will be  convenient for many users.

Hadronic current represents a rather compact segment of the simulation package.
That is why only minor changes need to be introduced to {\tt TAUOLA cleo}
code and the makefile. Only several lines will be necessary to modify in the collaboration software.
All the rest is included in {\tt tar-ball} to be expanded in {\tt tauola} 
directory. See Appendix~\ref{app:B} for details.

The  user will be able to switch to the new current invoking simply
{\tt CALL INIRChL(1)} prior to  {\tt TAUOLA } initialization, still retaining possibility to use the old ones
with {\tt CALL INIRChL(0)} as well.

An algorithm for 
working with auxiliary weights to implement simultaneously several models of $\tau$ decays is 
a straightforward extension. 

\subsection{Weight recalculation}\label{sec:recalculation}

Present day experimental data feature very high precision over all directions of
multidimensional phase space. Nonetheless cross contamination between different
channels takes place.  This is the case, for example, if particular
decay channels differ by presence or absence of  
$\pi^0$'s. 
Figure 6 of Ref. \cite{Fujikawa:2008ma} represents such folded comparison between 
data and Monte Carlo for the $\tau^+ \to \pi^+\pi^0\bar{\nu}_\tau$ decay channel. 
This result can not be used, without additional 
information on other decay channels, for fits of hadronic current.
This takes place  even for this seemingly simple case where hadronic current
can be directly deciphered from a one-dimensional distribution.

In general,
for the precision matching of data and models it is convenient to 
simultaneously confront 
several models and take into consideration all decay channels simultaneously. 
Such a solution may be a necessary technical step if fits for unfolded
data are envisaged at the precision level better than few percents.

\begin{figure}[t!]
\begin{center}
\includegraphics[width =1.0\linewidth]{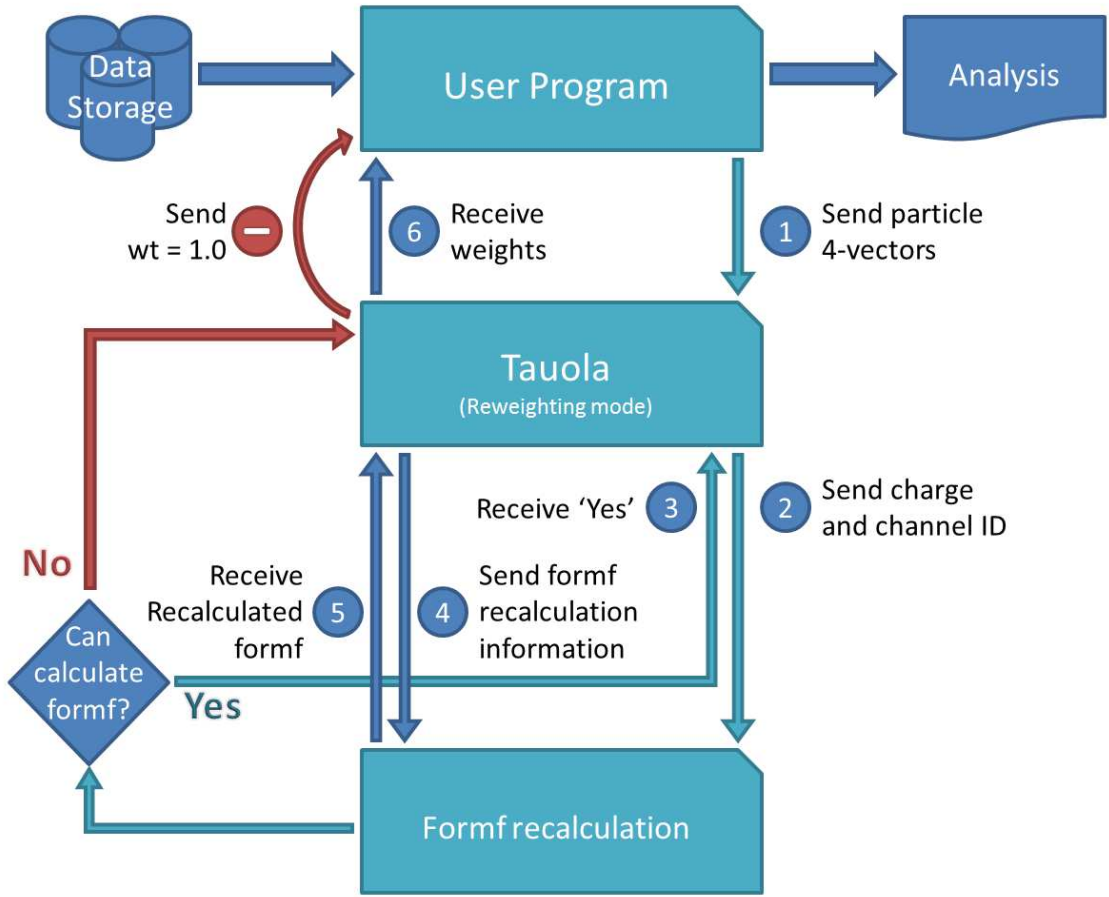}
\caption{Weight recalculation model, as discussed in Ref.~\cite{Was:2011tv}. \label{Fig:Fifo} }
\end{center}
\end{figure}

To facilitate technical tasks, a weight recalculation method is prepared, 
following discussions and recommendations given
in Ref.~\cite{Actis:2010gg}.
 It was agreed that organization of the programs should enable simulation following
the scheme of  Fig.~\ref{Fig:Fifo}.
For the single generated Monte Carlo sample (when all detector
and experimental acceptance effects are taken into account), one can  calculate 
weights enumerating change of matrix element. The procedure can be repeated as many times as needed. 
In this way the optimal choice is met when all experimental and theoretical effects can be taken into account. With this, the fit 
to multidimensional distribution of measured data can be determined.

We have prepared the necessary changes for {\tt TAUOLA}. 
The following algorithm was checked to work (its technical details are given 
in  Appendix~\ref{app:WT-recalc}):
\begin{enumerate}
\item For each  generated $\tau$  stored in a datafile, the user program reads flavours and 4-vectors
of $\tau$ and its decay products. 
\item Appropriate kinematic transformation  is  performed enabling a recalculation 
of the matrix element, exactly  as   at the generation step. 
\item An appropriate routine of {\tt TAUOLA} is chosen and the 
matrix element is calculated.
\item Another instance of {\tt TAUOLA} initialization can be activated and the matrix element recalculated
using a different physics model.
\item If one of the two models was used in the generation of the user sample, 
then the weight for model replacement can be calculated 
as a simple ratio of the two.
\end{enumerate}

Usage of this method of weight recalculation does not require any 
changes to {\tt TAUOLA} code except those related to installation of the presented upgrade, 
aiming at installation of our new {\tt RChL} currents. 
Needless to say, the method can be used several times for user  
variants of {\tt RChL} and interpolation of weights can be used in fits.
In our example, given in directory {\tt new-currents/Installation-Reweight} we provide a simple method to read generated events from file,
but it can be easily adapted to any other one. Linking libraries of 
{\tt MC-TESTER},  {\tt ROOT} or {\tt HepMC} is an option convenient also 
for an interpretation/verification of results.

Basic features which are necessary for our solution have already been
tested for Belle and BaBar software environments. Changes presented in Appendix~\ref{app:B}
are from that perspective rather straightforward.

\section{Theoretical basis of the currents}\label{sec:systematic}

\subsection{Resonance Chiral Theory framework}

The  Lorentz structures of currents are universal and a proper 
minimal set of form factors only needs to be chosen for a particular decay mode. 
One can recall the QCD symmetries 
to gain some insight in the form factors. It is particularly useful 
that the chiral symmetry of massless QCD allows one to develop an 
effective field theory description, $\chi PT$, valid for momenta much smaller 
than the $\rho$ mass~\cite{Gasser:1983yg,Gasser:1984gg}. 
However, $\chi PT$ cannot provide predictions valid all over the $\tau$ 
decay phase space \cite{Colangelo:1996hs}, it constrains nonetheless the form and normalization of the 
form factors in such limit\footnote{One should keep in mind that the effect due to 
different and nonzero masses of $u$ and $d$ quarks already now seems to 
be necessary to explain some aspects of $\tau$ data.}.

The computations done within Resonance Chiral Theory 
($R\chi T$~\cite{Ecker:1989yg,Ecker:1988te}) are able to reproduce the low-energy limit 
of $\chi PT$ up to $NLO$ \footnote{The improvement obtained when doing this can be appreciated in Fig. 1 of Ref.~\cite{Roig:2008xt}.} 
and demonstrate the right 
falloff~\cite{Brodsky:1973kr,Lepage:1980fj} in the high energy region. 
The current state-of-the-art for the hadronic $\tau$ decays form factors ($F_i$) is 
described in Refs.~\cite{Roig:2011iv, Portoles:2010yt}.

The description provided by $R\chi T$ complies with the low-energy properties of the underlying
theory,
at least up to next-to-leading order\footnote{Also the leading next-to-next-to-leading (NNLO) order terms
\cite{Bijnens:1998fm, Bijnens:2002hp} are reproduced~\cite{Guerrero:1997ku,Guerrero:1998hd}.}.
Nonetheless it is necessary to extend $\chi PT$ to the intermediate energy region, which is probed through hadronic
$\tau$ decays.
In order to do this, one relies on the large-$N_C$ expansion of QCD \cite{'tHooft:1973jz, 'tHooft:1974hx,
Witten:1979kh} which predicts that in the $N_C\to\infty$ limit there is an infinite tower of zero-width
resonances experiencing among them local effective interactions taken at tree level. We model this setting with
a spectrum that resembles the measured one\footnote{In $R\chi T$ all nine mesons [$\rho$, $K^*$, $\omega$, ($\phi$)] 
have the same mass 
in the $N_C\to\infty$ limit without taking into account $SU(3)$ breaking. The $1/N_C$ corrections make the $\phi$ heavier 
[through the axial anomaly that breaks $U(3)$ down to $SU(3)$]. $SU(3)$ breaking makes 
$M_\rho\neq M_{K^*}\neq M_\omega$.}. We introduce a $NLO$ effect (whose impact is,
however, quite sizable in phenomenological applications) in that counting, providing the resonances with an
energy-dependent width computed\footnote{The width of a
spin-one resonance is defined \cite{GomezDumm:2000fz} as the imaginary part of the pole generated by
resumming those diagrams,
with an absorptive part in the s-channel, that contribute to the two-point function of the corresponding
vector current.}  within $R\chi T$ (see Section~\ref{sec:a1width}).

The appropriate falloff at large energies \cite{Brodsky:1973kr, Lepage:1980fj} is imposed to the form factors, 
vector-vector and axial-vector---axial-vector 
correlators \cite{RuizFemenia:2003hm, Cirigliano:2004ue, Cirigliano:2005xn, Cirigliano:2006hb, Kampf:2011ty}. This 
results in a set of relations among the coupling constants of the theory, which are obtained working in the single resonance approximation 
(only the lightest multiplet of resonances 
is included per given set of quantum numbers) and in the $N_C\to\infty$ limit. Upon integration of the resonances, 
this procedure allows the saturation of the values of the 
$\chi PT$ low-energy constants at $\mathcal{O}(p^4)$ and $\mathcal{O}(p^6)$, both in the even- and odd-intrinsic parity sectors.

The results for all hadronic currents have been calculated within $R\chi T$ working, with the exception of both two-pion and two-kaon modes, in the isospin
 limit. Therefore the corresponding hadronic form factors depend only on the average pion 
 and kaon  masses:
\begin{equation}
m_\pi = (m_{\pi^0}+2\cdot m_{\pi^+})/3 \;,\;\;\; m_K=(m_{K^0}+m_{K^+})/2\;.
\label{eq:ave-mass}
\end{equation}
For the three-pseudoscalar modes every hadronic form factor consists of 3 parts: 
a chiral contribution (direct decay, without production of any intermediate 
resonance), one-resonance and double-resonance mediated processes. The precise form of the form factors for three-pion 
and (two pions - one kaon) modes is presented in Sections~\ref{Subsect:pipipi}, \ref{Subsect:KKpi} and \ref{Subsect:KK0pi0}.
For the two-meson modes the corresponding vector form factors are built from the lightest resonance contribution 
in $R\chi T$ \cite{Guerrero:1997ku}. 
FSI are resummed by means of an Omn\`es 
function \cite{Guerrero:1997ku}, see Section~\ref{Subsect:pipi0}. Both two- and three-meson $\tau$ decays are sensitive 
to the exchange of excited resonances, 
whose contribution we have to account for. 
The exchange of heavier-resonances could be computed in the same fashion as for the lightest multiplet, giving rise to new unknown 
couplings\footnote{See, for instance Refs. \cite{SanzCillero:2002bs, Mateu:2007tr}.}. This possibility has been postponed 
for the moment to prevent the increase in the number of free parameters. We believe that it is sensible to tackle this task once more 
knowledge 
on the couplings of the $R\chi T$ is achieved. In the three-meson decays, the excited resonances have been included phenomenologically, 
introducing 
an additional parameter, $\beta_{\rho'}$, in Eq.~(\ref{rhoprime}). This has been done in such a way to keep the chiral limit result 
and the QCD-ruled short-distance behaviour. 
For the two-meson processes, they have been included analogously as the lightest resonance contribution, making sure that 
the appropriate low- and high-energy limits 
are not spoiled and that unitarity and analiticity hold perturbatively%
\footnote{See subsection \ref{sect:other-s} points 1 and 3 and Appendix \ref{app:F} for an improved treatment.}. 
Following the approximation proposed in 
Ref.~\cite{Jamin:2006tk}, three new parameters, related with the $\rho(1450)$, $\rho(1700)$ and $K^{*\prime}(1410)$ couplings, appear: 
$\gamma\equiv -F_V' G_V'/F^2$ [for $\rho(1450)$], $\delta\equiv -F_V'' G_V''/F^2$ [for $\rho(1700)$] 
and $\gamma_{K\pi}\equiv -F_V' G_V'/(F F_K)$ [for $K^{*\prime}(1410)$] \footnote{Due to SU(3) breaking effects $\gamma \neq \gamma_{K\pi}$.} 
in Eqs. (\ref{VFFpipi}) and (\ref{VFFKpi}). The short-distance QCD constraint for the vector form factor will require the relation 
$F_VG_V+F_V'G_V'+F_V''G_V''+...=F^2$ to hold.

\subsection{The error associated to the $1/N_C$ expansion}
Any model based on the $1/N_C$ expansion will naturally raise the question of 
the error associated to that expansion, which can be naively estimated as $1/N_C\sim30\%$. It is natural to object to the 
convergence of a series in $1/N_C$ that ends up being $1/3$ in the real world. The associated error of the size of the expansion 
parameter would be much larger than the 
statistical error of experimental measurements or even of our precision target of a few percent. Just to give a counterexample, 
a look at the results of 
Ref.~\cite{Guerrero:1997ku} will quickly suggest that the actual error can be much smaller, which is noteworthy, 
especially taking into account that in the quoted reference, 
an impressive agreement with data in $e^+e^-\to\pi^+\pi^-$ up to $s=1$ GeV$^2$ was obtained in terms of just one parameter, $M_\rho$.

Let us recall how this can be possible. First, the large-$N_C$ expansion of QCD is not an expansion in the usual 
perturbative sense. For instance, in QED, the perturbative 
expansion means, firstly, that the diagrams with less photon couplings to fermionic lines dominate. 
After that, when we compute diagrams both at tree level, and including loops, we realize 
that the expansion parameter is $\alpha=e^2/(4\pi)^2$ and the coefficients of the series are small compared to $\alpha$:
 every order we go further in the expansion  the error reduces  by $\sim 1/\alpha\simeq$ 137.
 In the large-$N_C$ expansion of QCD, we know first, which diagrams are leading-order (planar diagrams with gluon exchanges)
and which ones are suppressed. Unfortunately, 
the expansion has a fundamental subtlety that prevents one from  determining the expansion parameter after that: there are infinite 
diagrams at any given order in the expansion, so that one cannot perform a calculation at both LO and NLO and 
compare them to know what is the expansion parameter. It is known, however, that diagrams with internal quark loops are suppressed as $1/N_C$ and 
that non-planar diagrams are suppressed as $1/N_C^2$. Moreover, if the number of quark flavours, $n_f$, 
is not considered to be smaller than $N_C$ the former diagrams can 
even scale as $n_f/N_C$. These kind of contributions would be responsible for the mixing of $q\bar{q}$ 
and $q\bar{q}q\bar{q}$ states. The fact that this effect is not observed 
in Nature and the success of the quark model classification of mesons in $q\bar{q}$ multiplets suggests 
that the coefficients of these diagrams with internal quark loops are 
tiny, in such a way that the first non-negligible correction would come from the non-planar diagrams, 
suppressed as $1/N_C^2\sim10\%$. This reasoning may explain why 
the large-$N_C$ expansion is a such a good approximation for low and intermediate-energy $QCD$, given the 
phenomenological successes of its applications in meson effective field theories \cite{Pich:2002xy} 
and the corroborated predictions given by the large-$N_C$ limit both for $\chi PT$ 
\cite{Ecker:1994gg, Pich:1995bw} and for $R \chi T$~\cite{Ecker:1988te} coupling constants. 
All these reasons seem to suggest that, quite generally, some factor comes to complement 1/$N_C$ for the value of the 
expansion parameter to be reduced and the relative accuracy to be increased.

This conclusion is supported by the investigation of some $\mathcal{O}(p^4)$ and $\mathcal{O}(p^6)$ couplings of $\chi PT$, 
which are done modelling the $NLO$ expansion in 
$1/N_C$ of $R\chi T$ \cite{Pich:2010sm, Rosell:2004mn, Rosell:2006dt, Portoles:2006nr, Pich:2008jm}. 
According to the size of the corrections, we judge that 15$\%$ can be a 
reasonable general estimate (see, however, our discussion in Appendix \ref{app:C} on the possible variations 
on the predictions of the couplings obtained in the $N_C\to\infty$ 
limit). Noticeably, the actual expansion parameter can be computed for $R\chi T$ in the study of the vector form factor 
of the pion at $NLO$ in the $1/N_C$ expansion 
\cite{SanzCillero:2009pt}, yielding
\begin{equation}
 \alpha_V\,=\,\frac{n_f}{2}\frac{2G_V^2}{F^2}\frac{M_V^2}{96\pi F^2}\,,
\end{equation}
which, at lowest order, is the ratio of the vector width and mass, $\alpha_V\sim0.2$, agreeing with the previous discussion.

Moreover, we should emphasize that our approach goes beyond the $N_c\to\infty$ limit. We supplement the lowest order in the 1/$N_C$ expansion for the theory 
in terms of mesons by the leading higher-order correction, namely by including the resonance (off-shell) widths for the wide states $\rho$, 
$K^\star$ and $a_1$. This seems to point to smaller errors than those characteristic of the $LO$ contribution in the 1/$N_C$ expansion and may be able to 
explain, altogether, an eventual fine agreement with data.
\subsection{Other sources of error} \label{sect:other-s}
Once this major concern on the reliability of the $R\chi T$ hadronic currents has been discussed, 
let us consider in turn other possible sources of error in the different 
hadronic $\tau$ decay modes considered.

\begin{enumerate}
 \item In the two-meson $\tau$ decay modes an Omn\`es type of resummation is employed for the FSI. 
The proposed expressions, Eqs.~(\ref{VFFpipi}) to 
(\ref{VFFKpi}), respect unitarity and analiticity only in a perturbative sense 
(this will be improved along the lines discussed in Appendix \ref{app:F}). The effect of these 
violations is, however, pretty small, as one can see comparing the results of Refs.~\cite{Jamin:2008qg} 
and \cite{Boito:2008fq} for the $\tau\to K\pi\nu_\tau$ decays. That is why 
we consider our current parametrizations for these decay channels a reasonable temporary approach. 
For a future update of the program we recall that Eqs.~(\ref{VFFpipi}) to (\ref{VFFKpi}) should be replaced, using the 
procedure described in Appendix \ref{app:F}. For the $K\pi$ vector form factor we are going to 
use\footnote{We note that another 
interesting approach, based on Omn\`es integral equations incorporating both chiral constraints at low energies 
and QCD short-distance relations at high energies was 
performed in Ref.~\cite{Moussallam:2007qc};  isospin-violating corrections were also studied.} 
 the parametrization 
of Ref.~\cite{Boito:2008fq} in a future upgrade of the program. 
Analogous works for the 
$\pi^-\pi^0$ and $K^-K^0$ vector form factors are under way.
 \item In the three-meson modes, scalar and pseudoscalar resonance exchange has been neglected. 
The spin-one character of the SM couplings of the hadron matrix elements in the
$\tau$ decays implies that the form factors for these processes are ruled by vector and axial-vector resonances. 
Their contribution should be minor in $\tau\to KK\pi \nu_\tau$ 
decays (see the related discussion in Section 2 of Ref.~\cite{Dumm:2009kj}) while that of the scalar resonances 
can be a bit more important in $\tau\to3\pi\nu_\tau$ decays. The lightest scalars 
are, however, suppressed in the large-$N_C$ limit (see, however, Ref. \cite{Nieves:2011gb}). In these decays, 
 we have also neglected systematically three-body FSI for the moment. They may be important in the available
 phase space and at the required precision \cite{Anisovich:1996tx, Niecknig:2012sj}, thus, should be investigated at a time of comparison to
 experimental data. The present step of our work is devoted
 predominantly to establish a technical environment. Consequently, the effort was concentrated 
 on those theoretical aspects which bring difficulties in
 program design and a careful discussion of potentially less important effects is 
 delegated to the future work.
 Finally, the computation of $\Gamma_{a_1}(q^2)$ through the optical theorem \cite{Dumm:2009va} does not give the corresponding 
 real part of the loop function, which we have disregarded. Although this approximation might be supported numerically it 
 induces a small violation of analiticity.
 \item In the $\tau\to KK \pi\nu_\tau$ decays,  for the moment, we have not included the contribution of excited resonances. 
We expect that their influence is bigger in this case 
than in the three-pion channels, since the mass of the hadronic system is larger. The impact of this error can be 
as large as the one coming from the $1/N_C$ expansion.
\end{enumerate}
\subsection{Numerical estimates of the errors in the different decay channels and distributions}
What is the numerical precision we should expect in confrontation of our currents with the data?
As already mentioned, a first crude estimation gave us an error of the order of the expansion parameter, $1/N_C$, 
thus 30\% precision tag for our results, but we already 
accumulated, during the preceding discussion, a number of indications pointing to better accuracy of the results obtained 
within this approach.
It is thus of importance \cite{Toni} to evaluate solely on the basis of theoretical considerations, for which decay 
channels 
we expect precision to be better or worse and in which regions of the phase space. This is of course a must for 
scientific theory as Resonance Chiral Theory is supposed to be. 

The answer, if scientific theory holds, always comes from the 
confrontation with the data. Results from theory have to be prepared in such a manner that agreement confirms and discrepancy
invalidates the theory under consideration. Theoretical results have to include estimation of their errors from within 
theory itself. 
That is the basic principle of science methodology  and our work has to keep this aspect 
in mind too. We have to address what is input and where such confrontation may take place.

Until now
we see that the precision can be at the level of a few percent%
\footnote{Once the parameters entering Eq.(\ref{VFFKK}) are 
fitted, we expect similar accuracy for the $KK$ modes as in the $\pi\pi$ and $K\pi$ cases.}
for the two-meson 
modes \cite{Fujikawa:2008ma,  Jamin:2008qg, Roig:2011iv, Boito:2008fq, Jamin:2006tk, :2007rf}~. 
Despite the  accuracy is at a comparable level for the 
$d\Gamma/dq^2$ distributions in the three-pion decays, the error on the distribution in the $\pi^+\pi^-$ invariant mass 
is at the level of $\sim20\%$. We expect to improve it to a few percent level once FSI are accounted for. 
In the $\tau\to KK\pi\nu_\tau$ decays the situation is, 
somehow, reversed. The largest error ($\sim30\%$) comes on the $d\Gamma/dq^2$ distribution, 
while the accuracy is much better in the $K^+\pi^-$ ($\sim 10\%$) and 
$K^+K^-$($\sim 5\%$) distributions. We expect that the errors on the $KK\pi$ modes are reduced a factor of two 
when the excited resonances contributions will be taken into account.

To get form factors with substantially better agreement with the data, one may need to introduce ad-hoc  factors,
hopefully close to unity and hopefully based on educated guesses. This may provide a valuable hint for future 
theoretical work.

\section{Summary}\label{sec:Summary}

In this paper we 
 have documented a set of currents based on 
Resonance Chiral Theory for use in hadronic $\tau$ decays into either two ($\pi^- \pi^0$, $\pi^0 K^-$, $\pi^- \bar K^0$ and $K^- K^0$) 
or three ($\pi^-\pi^-\pi^+$, $\pi^0\pi^0\pi^-$, $K^- \pi^-K^+$, $K^0  \pi^- \bar{K}^0$ and $K^- \pi^0 K^0 $) pseudoscalars. 
The set covers more 
than 88\% of total hadronic $\tau$  width. Technical tests of the installation into the 
{\tt FORTRAN} program have been documented. The set can be used as an upgrade easy to install into any version of the 
{\tt TAUOLA} $\tau$ decay library. In this way currents are ready for 
confronting the $\tau$ decay data (unfolded or not). Precision fits can be performed and arrangements
for use of model-dependent weights are ready for that purpose. 
 
On the technical side,  the $\tau$ decay algorithms themselves have been checked 
down to 0.05\% precision level. To this end, a detailed comparison between analytic and Monte Carlo 
results has been provided. Statistical samples, 2 (3) orders of magnitude larger than 
at time of reference \cite{Golonka:2003xt} (\cite{Jadach:1993hs}) were used.
This technical precision of 0.05\% is substantially better than physics precision 
of our currents which we estimate at the 5-30\% level, depending on the channel. Software environment for 
further phenomenological work is prepared.

In the present work,
we have concentrated on decay modes contributing to 88\% of the $\tau$ hadronic decay width. In particular,
the  decay modes of $\tau$ into $4\pi$'s have not been updated in our work.
The currents for these decays, constructed on the basis of low-energy 
$e^+e^- \to$ hadrons data have been  available since Ref.~\cite{Bondar:2002mw}.
 For the time being
this can be used as an alternative to {\tt TAUOLA cleo} default of Ref.~\cite{Golonka:2003xt} 
in discussion of the systematic error for background modelling. In these decay modes 
theoretical foundations are less profound, but we hope that the presented solution will not jeopardize 
analysis of other channels due to lesser control of cross contaminating channels.
Together with $4\pi$'s currents described in Ref.~\cite{Bondar:2002mw} our system covers now 97\% 
of the total hadronic $\tau$ width. 

It is straightforward to extend our work to more elaborated currents
including, e.g. scalar form factors. 
Because of different nature of benchmark distributions, technical aspects and different physics
assumptions 
we leave this task to the forthcoming work \cite{RChL:scal}. 

As a consequence of comparisons with the data,
some of the theoretical assumptions may need to be reconsidered too. 
We start from a theoretical approach common for all channels.
Empiric form factors may need to
be added later though. In this way not only agreement with the data will
be established, but eventual inefficiencies of our starting approach will
be numericaly evaluated.
That is why, it is important to ensure that model-dependent weights can be 
calculated after (and independently of)
the detector effects. There are several assumptions which lie behind our calculations
and in case of discrepancy with the data they may need to be revisited.
Let us list them in descending order of their theoretical foundation:

\begin{enumerate}
\item
We assume that Lorentz invariance will not need to be reconsidered at any step of our project.
\item With respect to the separation of the matrix element into leptonic and hadronic parts, 
we assume that electroweak corrections will not affect such separation beyond 
precision level of several permille at most, the question of an overall 
normalization factor for all hadronic channel, see e.g. Refs.~\cite{Marciano:1988vm, Braaten:1990ef, Erler:2002mv}, 
is of no practical importance for Monte Carlo.
\item
We assume that isospin symmetry should be a good guiding principle. We suppose that
it should hold more accurately for distributions than for
amplitude phases, but we do not expect large effects.
\item Some of the effective couplings can be predicted by considering the asymptotic behaviour of Green functions and form factors 
both in the effective theory ($R\chi T$) and 
in the operator product expansion of QCD \cite{Ecker:1988te}. However, these predictions are affected by different sources of errors, 
most importantly the model dependence 
on the realization of the large-$N_C$ limit of QCD \cite{Peris:1998nj, Knecht:1999gb, Peris:2000tw, Golterman:2001nk, Golterman:2001pj, Masjuan:2007ay, 
Masjuan:2008fr, Masjuan:2008fv} 
(mainly the choice of the resonance spectra, but not only). 
Special care should be taken when relating different channels, 
especially if the statistics in both of them is very different and in one decay channel only a subset of resonance parameters is used. 
One should not forget that the (formal 
or not) integration of heavy degrees of freedom out of the action affects the values of the parameters 
in the remaining lower-energy theory. Results for the individual modes should be 
analyzed consequently.
\item
Our effective couplings and interactions are based on the low-energy effective field theory of QCD ($\chi PT$), 
whose results are reproduced at NLO in the corresponding limit 
by $R \chi T$. Although the latter being formally sound, there is model dependence in any realization of 
the large-$N_C$ limit of QCD for mesons and, moreover, we are 
introducing the contribution of excited resonances only at a phenomenological level, see Eq. (\ref{rhoprime}) and the related discussion 
in Section~\ref{sec:systematic}, 
a feature that can be improved in the future. We may need to explore the limits of such 
approach and take feed back from experiments.
\item
Different solutions have been advocated for taking unitarity properties, via the propagator widths, into account. 
In particular, there is no consensus that the exponentiation 
of the real part of the resonance width is always the best solution. This point must be investigated further, 
especially in light of precision fits to high-statistics data.
\end{enumerate}

At the same time, the seeming violation of these principles may be 
a consequence of some experimental problems. 
That is why such discussion requires simultaneous participation of theorists 
and experimental 
physicists and  the proper software environment.

Such fine-tunings took place for  {\tt TAUOLA} version of 
Ref.~\cite{Jadach:1993hs}.
The improved agreement with the data of CLEO and ALEPH  has been  achieved thanks 
to the effort 
of these collaborations resulting with {\tt TAUOLA cleo } and {\tt TAUOLA aleph }
initializations. The code is available from Ref.~\cite{Golonka:2003xt}. 
We should expect similar work with our 
present parametrizations. We hope that, in this way, theoretical and experimental
constraints will be appropriately matched leading to a representation of
 experimental and theoretical achievements in common language of value for future research. 
Once such confrontation with the Belle or BaBar data is completed,
one will profit from the technical precision of the simulation
established here and thanks to prepared flexibility for 
{\tt FORTRAN} \cite{Golonka:2003xt} and C++ \cite{Davidson:2010rw} users
of high-energy experiments such as at LHC or for LC
 will benefit as well. If analytic
form of the form factors will be attained, it may help start future work for 
such research as lattice calculations.

We consider this work as a  step towards a theoretically
rigorous description of hadronic tau decay data but, at this moment,
we can still not be sure
 that the currents for all channels will be able, after fits, to describe all
data well. In this
case, detailed numerical information on the offending distributions
will be provided. Further theoretical work is  stimulated.
 Some intermediate results are presented in 
\cite{Shekhovtsova:2012yq,Roig:2012zj,TAU12:RChL}.

\section*{ Acknowledgements}
Our project is coordinated  with the effort of the Working Group on Radiative Corrections and MC Generators 
for Low Energies 
(http://www.lnf.infn.it/wg/sighad/). 
Useful discussions and help in understanding requirements our extensions must fulfil to be useful 
for BaBar and Belle go to John Michael Roney, Simon Eidelman and Hisaki Hayashii. 
We are thankful to Ian M. Nugent and Denis Epifanov for comments on the document and our program installation
procedure.
Thanks to Kenji Inami for help in organization of tests within Belle collaboration environment and to Swagato Banerjee
for discussions as well.

We would like to thank Jorge Portol\'es for fruitful discussions on the hadronic form factors and currents.
We are thankful to Matthias Jamin,
Diogo Boito and Rafel Escribano 
for  discussions of the $K\pi$ vector form factor. We are indebted to Johann Hans K\"uhn for discussions and critical concerns important 
for the present and  future 
steps of the project.

{\small 
 We acknowledge the inspiring environment of Stefano Bellucci's group at LNF, (INFN, Frascati), 
where important step of the project was achieved.
This research of OS was supported by a Marie Curie Intra European Fellowship within the 
-7th European Community Framework Programme (FP7-PEOPLE-2009-IEF) PIEF-GA-2009-253329 
and by  the Spanish Consolider Ingenio 2010 Programme 
CPAN (CSD2007-00042) as well by by MEC (Spain) under Grants FPA2007-60323, FPA2011-23778. The work of PR has been supported 
in part by MEC (Spain) under Grants FPA2007-60323, FPA2008-01430, FPA2011-23778, FPA2011-25948 and by  the Spanish Consolider Ingenio 2010 Programme 
CPAN (CSD2007-00042). The work of ZW is supported in part by the 
Polish Ministry of Science and Higher
Education grant No. 1289/B/H03/2009/37 and  of TP  
by the Polish Government grant NN202127937 (years
2009-2011) under decision DEC-2011/03/B/ST2/00107. From January 1, 2013, the affiliation of O. S. is IFJ PAN, Krakow, Poland.}
\bibliography{tau3scal}{}
\bibliographystyle{utphys_spires}

\newpage
\appendix

\setcounter{section}{0}

\section{Useful functions and notations}\label{app:a1}
To minimize repetition and to reduce the size of formulas in Section \ref{sec:currents}
the lengthy ones  were  moved to this appendix.

In the description of the three-hadron currents the following functions  were used:

\begin{eqnarray} \label{eq:cde}
A^{\mbox{\tiny R}}(q^2,x,y,m_1^2,m_2^2,m_3^2) & = &  3\, x \, + m_1^2 -m_3^2 +
\left( 1-\frac{2 G_V}{F_V} \right) \left[ 2\, q^2-2\, x-y+m_3^2-m_2^2 \right] \, , \nonumber \\[3.5mm]
B^{\mbox{\tiny R}}(x,y,m_1^2,m_2^2) & = & 2 \, \left( m_2^2-m_1^2 \right) \,
+ \, \left( 1-\frac{2 G_V}{F_V} \right) \left[ y - x + m_1^2-m_2^2\right] \, , \nonumber \\[3.5mm]
A^{\mbox{\tiny RR}}(q^2,x,y,m_1^2,m_2^2,m_3^2) & = & \left( \lambda' + \lambda'' \right) \,
(-3\, x + m_3^2-m_1^2)\, \nonumber \\ &&  + \, \left( 2\, q^2+x-y+m_1^2-m_2^2 \right)
H\left( \frac{x}{q^2}\,,\,\frac{m_2^2}{q^2} \right) \, , \nonumber \\ [3.5mm]
B^{\mbox{\tiny RR}}(q^2,x,y,z,m_1^2,m_2^2,m_3^2) & = & 2 \left(
\lambda'+\lambda'' \right) \left( m_1^2-m_2^2 \right)
 + \left( y-x+m_2^2-m_1^2 \right)
H\left( \frac{z}{q^2}\,,\,\frac{m_3^2}{q^2} \right) \, , \nonumber \\ [3.5mm]
 C^{\mbox{\tiny R}}(q^2,x,m_1^2,m_2^2,m_3^2) & = &
 (c_1-c_2+c_5) \, q^2 - ( c_1-c_2-c_5+2 c_6) \, x \nonumber \\
 & & \, + (c_1+c_2 + 8 c_3 -c_5) \, m_3^2 + 8\, c_4\, (m_1^2-m_2^2) \, , \nonumber \\[3mm]
C^{\mbox{\tiny RR}}(q^2,x,m^2) & = & d_3 \, (q^2+x)+ (d_1+8\, d_2-d_3) \, m^2 \, , \nonumber \\ [3.5mm]
D^{\mbox{\tiny R}}(q^2,x,y) & = & (g_1+2 \, g_2-g_3)\, (x+y) -2 \,g_2 \, (q^2+m_K^2)
\nonumber \\
&&  - (g_1-g_3)\, ( 3\,m_K^2+m_{\pi}^2 ) +2 \, g_4 \, (m_K^2+m_{\pi}^2) +2 \, g_5 \, m_K^2 \, , \nonumber \\[3mm]
E^{\mbox{\tiny R}}(x,y) & = & (g_1+2 \, g_2-g_3)\, (x-y) \, .
\end{eqnarray}
They follow conventions of \cite{Dumm:2009kj}.
Function $H(x,y)$ is defined in subsection \ref{Subsect:pipipi}, 
formula  (\ref{eq:fq2}).
In the description of the two pseudoscalar form factors,
following \cite{Gasser:1983yg, Gasser:1984gg},
the function $A_{PQ}(s)$ was used to describe loops involving pions, 
kaons and $\eta$ mesons 
\begin{equation}\label{loopfun_2pi}
A_{PQ}(s) \, = - \, \frac{192 \pi^2 [s M_{PQ}(s) - L_{PQ}(s)]}{s} ,
\end{equation}
where 
\begin{equation}
M_{PQ}(s) = \frac{1}{12 s}( s - 2 \Sigma_{PQ}) \bar{J}_{PQ}(s) + \frac{\Delta^2_{PQ}}{3 s^2} \tilde{J}_{PQ}(s) 
- \frac{1}{6} k_{PQ} + \frac{1}{288 \pi^2}\,,
\end{equation}
\begin{equation}
 \Sigma_{PQ} = m_{P}^2 + m_Q^2,\quad \Delta_{PQ} = m_{P}^2 - m_Q^2 , \quad k_{PQ} = \frac{F^2}{\Delta_{PQ}}(\mu_P - \mu_Q)
\end{equation}
and
\begin{equation}
L_{PQ}(s) = \frac{\Delta^2_{PQ}(s)}{4 s} \bar{J}_{PQ}(s)\,.
\end{equation}
The  $\mu_P = \mathstrut \frac{\displaystyle m_P^2}{32 \pi^2 F^2}\mathrm{ln}\left(\frac{m_P^2}{\mu^2}\right)$ 
(at present  we take $\mu = M_\rho $ for all 2 pseudo-scalars modes\footnote{Alternatively, for 
$\mu$ we may take  $M_\rho$ for $PQ=\pi\pi,\,KK$ and  $M_{K^*}$ for $PQ=K\pi$ ~\cite{RChL:scal}.}).

Finally,
\begin{eqnarray}
\bar{J}_{PQ}(s) & = &  \frac{1}{32\pi^2}\left[ 2 + 
\left( \frac{\Delta_{PQ}}{s} - \frac{\Sigma_{PQ}(s)}{\Delta_{PQ}}\right)
\ln\frac{m_Q^2}{m_P^2} - \frac{\nu}{s}\ln\frac{(s+\nu)^2-\Delta^2_{PQ}}{(s-\nu)^2-\Delta^2_{PQ}} \right] \,,\nonumber \\
\tilde{J}_{PQ}(s) & = & \bar{J}_{PQ}(s) - s\bar{J'}(0) \,,
\end{eqnarray}
where $\nu^2 = \lambda(s,m_P^2,m_Q^2)$. 
Care has to be taken to keep the imaginary part of $\nu^2$ 
in the phase-space regions where $\nu^2<0$, and where ${\mathrm Im}\, \nu$ must not be set to zero,
see the function {\tt JPQ1\_FUNCT} in the file {\tt funct\_rpt.f}\;. 
In the vector form factor for two pions, this effect shows up in the $KK$ 
contribution to the loop function from
below the threshold of  $KK$ production.

For the $\rho'$ and $\rho''$ mesons the loop function $A_\pi(s)$ 
taken from~\cite{Guerrero:1997ku} reads:
\begin{equation}
A_{\pi}(s) \, =\, 
\ln{\left( \frac{m^2_\pi}{\mu^2}\right)} + {8 \frac{m^2_\pi}{s}} -
\frac{5}{3}  + \sigma_\pi^3 \,\ln{\left(\frac{\sigma_\pi+1}{\sigma_\pi-1}\right)}\,.
\end{equation}
In the last formulas the  $SU(2)$ limit is taken and $m_{\pi^\pm} = m_{\pi^{0}}= 0.13804$ GeV.

All functions in this appendix are coded in 
file {\tt new-currents/RChL-currents/funct\_rpt.f}\;.
\appendix
\setcounter{section}{1}
\section{Installation}\label{app:B}

Our project tar-ball, even though resulting from a rather large effort,  
is not designed for independent installation. This would be of course straightforward
and we will return to that solution in the future, once the currents are optimized to improve 
agreement with the data. The parametrization will become
integrated part of the {\tt TAUOLA} distribution; for fortran or for C++ use, like in references
\cite{Golonka:2003xt,Davidson:2010rw}. At present we concentrate 
on a solution which is most convenient for the
experimental user e.g. from  Belle or BaBar collaboration aiming at combining the code with the 
version of {\tt TAUOLA} which is already being used as 
 part of the simulation set-up. 
We aim at preparing an add-up%
\footnote{As the project is developed under svn, the tar-ball is accompanied with svn label and it should be kept for reference.} for already existing set-up.
The tar-ball can be downloaded from the Web page \cite{web:RChL} of our project.

Once tar-ball is unpacked inside {\tt TAUOLA-FORTRAN/tauola} subdirectory (of user environment), the directory
{\tt tauola/new-currents} will be created,  all necessary fortran files will be found 
there. For convenience, later on, we will use the following aliases:

\begin{itemize}
\item {\tt \$\{RCHLCURRENTS\}} instead of {\tt tauola/new-currents/RChL-currents}\;.
\item {\tt \$\{OTHERCURRENTS\}} instead of {\tt tauola/new-currents/other-currents}\;.
\item {\tt \$\{INSTALLATION\}} instead of {\tt tauola/new-currents/Installation}\;.
\end{itemize}

In  {\tt tauola/new-currents} further sub-directories for more advanced use or for
documentation will be found:

\begin{itemize}
\item {\tt \$\{RCHLCURRENTS\}/tabler/a1} - programs for  pretabulations in particular of
                                           $q^2$--dependent $a_1$ width\footnote{In the directory {\tt \$\{RCHLCURRENTS\}/tabler}  
the place to calculate other pretabulated functions (as possibly the scalar form factor for $K \pi$ channel) is reserved.}.
\item {\tt \$\{RCHLCURRENTS\}/cross-check} - code for technical and numerical tests.
\item {\tt new-currents/paper} - present paper.
\item {\tt \$\{INSTALLATION\}} - instructions for modifications 
to be introduced in {\tt FORTRAN} files and {\tt makefile} residing in
 directory {\tt tauola}\;.
\item {\tt new-currents/Installation-Reweight} - instruction and example of using reweighting algorithm.
\end{itemize}

None of the directories listed above contains code which is to be loaded together with {\tt TAUOLA} library.
Code loaded with the library is located only in the main folder of {\tt \$\{RCHLCURRENTS\}} and {\tt tauola/new-currents/other-currents}\;.
Programs in \\ {\tt \$\{RCHLCURRENTS\}/tabler/a1} can  update the fortran code 
located in file \\
 {\tt \$\{RCHLCURRENTS\}/initA1Tab.f}\;.
 
Once installation is completed,
to invoke the calculation of our new currents the {\tt CALL INIRChL(1)} 
has to be invoked\footnote{For a C++ user, examples of use of {\tt inirchl\_(1) }
are given in {\tt new-currents/Installation-Reweight/} directory. } 
by user main program
prior to call on {\tt TAUOLA}
initialization. If instead {\tt CALL INIRChL(0)} is executed prior\footnote{ 
This can be done also after initialization as no initialization of tables 
is needed. Then one can revert the change again with {\tt CALL INIRChL(0)}.} 
initialization, old currents - as 
in Ref.~\cite{Golonka:2003xt} -
will be used in generation.

The {\tt CALL INIRChL(1)} may activate also new currents\footnote{Although the 
$\tau\to\eta^{(\prime)}\pi^-\pi^0\nu_\tau$ decays have been worked out within 
Resonance Chiral Theory \cite{Roig:2010jp}, the corresponding expressions for the currents 
have not been incorporated yet to the program.} e.g. 
for $\eta \pi \pi$ or $4\pi$ decay channels. At present only wrappers of currents of  
Ref.~\cite{Golonka:2003xt} and \cite{Bondar:2002mw} are prepared in 
the directory {\tt \$\{OTHERCURRENTS\}} for convenience of users and our future work. 
These  currents lead to substantially different distributions, that is why,
one  may require adjustment of phase space presampler 
used to optimize speed of generation. Anyway, as a
 default, they are turned off. 
For {\tt INIRChL(1)} the same {\tt TAUOLA cleo} currents as for 
{\tt INIRChL(1)} are used.
To turn other options,  {\tt ISWITCH} 
located in the file {\tt \$\{OTHERCURRENTS\}/ffourpi.f} has to be changed from its default value 0 to 1, 2, 3 or 4.

\subsection{Changes for host  {\tt TAUOLA} version}
\label{App:installation}

In order to use new currents, changes have to be made to the host {\tt TAUOLA} installation.
Let us document here in great detail changes to be introduced in {\tt TAUOLA cleo} version. If some 
modifications were introduced and user's host {\tt TAUOLA} installation differs from {\tt TAUOLA cleo}
of Ref.~\cite{Golonka:2003xt}, then modifications prepared in {\tt \$\{INSTALLATION\}} directory
can not be used directly and some adaptation may be necessary. In either case we advice to check if 
at least some of the numerical results from Ref.~\cite{web:RChL} are  correctly reproduced after installation.

Let us list now changes which have to be introduced to files residing in {\tt TAUOLA/tauola} directory
of the user installation .
\begin{itemize}

\item {\tt TAUOLA/tauola/makefile} \\
The list of {\tt LIB\_OBJECTS}  must be extended and additional objects
added: \\
{\tt \$\{RCHLCURRENTS\}/f3pi\_rcht.o,
     \$\{RCHLCURRENTS\}/fkkpi.o,            \\
     \$\{RCHLCURRENTS\}/fkk0pi0.o,
     \$\{RCHLCURRENTS\}/wid\_a1\_fit.o,     \\
     \$\{RCHLCURRENTS\}/frho\_pi.o,
     \$\{RCHLCURRENTS\}/funct\_rpt.o,       \\
     \$\{RCHLCURRENTS\}/value\_parameter.o,
     \$\{RCHLCURRENTS\}/initA1Tab.o,        \\
     \$\{RCHLCURRENTS\}/fkpipl.o,
     \$\{RCHLCURRENTS\}/fk0k.o,             \\
     \$\{OTHERCURRENTS\}/fetapipi.o, 
     \$\{OTHERCURRENTS\}/ffourpi.o,         \\
     \$\{OTHERCURRENTS\}/binp.o,
     \$\{OTHERCURRENTS\}/curr\_karls.o,     \\
     \$\{OTHERCURRENTS\}/curr\_karls\_extracted.o}; \\
if there are no additional dependencies the {\tt \$\{INSTALLATION\}/makefile-tauola} file can be simply copied 
into {\tt tauola/makefile}\;.

\item {\tt TAUOLA/tauola/tauola.f} \\
If the file in user's version
coincides with the one of {\tt TAUOLA cleo} distribution, the  {\tt \$\{INSTALLATION\}/tauola.f-new} file
can be simply copied into 
{\tt tauola/tauola.f}\;. To verify this, the diff file 
{\tt \$\{INSTALLATION\}/tauola.f-oldDIFFupdated} may be inspected.

\item {\tt TAUOLA/tauola/formf.f} \\
If the file in user's version
coincides with the one of {\tt TAUOLA cleo} distribution, the  {\tt \$\{INSTALLATION\}/formf.f-new} file
can be simply copied into 
{\tt tauola/formf.f}\;. To verify this, the diff file 
{\tt \$\{INSTALLATION\}/formf.f-oldDIFFupdated} may be inspected.

\end{itemize}

Once changes are introduced the new currents will be activated and the old ones will be overruled
once call to routine {\tt  INIRChL(1)} 
is invoked. Otherwise, or if  {\tt CALL INIRChL(0)}
is invoked (at any time), old currents will be then switched back on. The routine  {\tt  INIRChL(1)} has to be invoked 
by the user program
at the initialization step. For the C++ user, a definition of {\tt extern "C" void inirchl\_(int i);} has to be 
included and execution of {\tt inirchl\_(\&i);} performed.

An example has been provided in {\tt \$\{INSTALLATION\}/demo-standalone}\;.
It is based on default {\tt TAUOLA cleo} example with the only modification being the
call to {\tt INIRChL(1) } before default {\tt TAUOLA} initialization.

\subsection{ Calculating numerical tables used by form factors }
\label{App:a1-tabler}

The directory {\tt \$\{RCHLCURRENTS\}/tabler/a1} contains the 
program {\tt da1wid\_tot\_rho1\_gauss.f}; it creates a table of 
$\Gamma_{a_1}(q^2)$ according to Eq.~(\ref{eq:a1width}). The system does not use any information from {\tt TAUOLA} initialization 
except the  pion and kaon masses. 

We have also prepared a place to add tables for other functions, in the near future it will be done for the scalar form factor 
of the $K \pi$ mode.

\subsubsection{ Executing the code }

The program {\tt da1wid\_tot\_rho1\_gauss.f} produces the $q^2$ distribution of the $a_1$ off-shell width.

To compile, type {\tt make} in {\tt \$\{RCHLCURRENTS\}/tabler/a1} directory.
To run, type {\tt make run}. Each line of the produced output includes the
value of $q^2 [\mathrm{GeV}^2]$, and the value of $d\Gamma/dq^2 [\mathrm{GeV}^{-1}]$.
This table is written into   the file {\tt initA1Tab.f}  which is  the FORTRAN code ready to use.
One can shift it to {\tt \$\{RCHLCURRENTS\}} directory by {\tt make move} command.
Text format table is  written into file 
{\tt wida1\_qq\_tot\_2e5.out}\;. 

\subsubsection{ Setup }

Input parameters and common blocks are located in {\tt \$\{RCHLCURRENTS\}/parameter.inc}\;.
Other parameters are defined in {\tt \$\{RCHLCURRENTS\}/value\_parameter.f}\;.
These parameters may be changed by the user. If the parameters affect the 
$q^2$--dependent $a_1$ width (or other pretabulated functions), the tables need to be
generated anew with the help of programs residing in the directory {\tt \$\{RCHLCURRENTS\}/tabler}\;. 
A list of the parameters that affect generated tables (and thus require tables to be generated again) is 
in {\tt \$\{RCHLCURRENTS\}/value\_parameter.f}\;.
Some of the variables used in  functions from \\ 
{\tt  \$\{RCHLCURRENTS\}/funct\_rpt.f} are declared in {\tt \$\{RCHLCURRENTS\}/funct\_declar.inc}\;.

\subsection{ Tests }

Directory {\tt \$\{RCHLCURRENTS\}/cross-check} contains three subdirectories:

\begin{enumerate}
\item {\tt check\_analyticity\_and\_numer\_integr}, it includes: 
      \begin{itemize}
      \item test of numerical stability in calculations
      of 
      $\Gamma_{a_1}(q^2)$
      and for the whole $\tau$ hadronic decays as described 
      in Section~\ref{sec:currents}. For that purpose it is checked if continuity of
      results as a function of the invariant mass holds.
      \item the result for the integrated width of the $\tau \to 2\pi \nu_\tau$,
      $\tau \to K \pi  \nu_\tau$,  $\tau \to K^- K^0 \nu_\tau$,
       $\tau \to 3\pi \nu_\tau$,  $\tau \to K \pi^- K\nu_\tau$ and $\tau\to K^- \pi^0K^0\nu_\tau$.
      These results can be confronted with the result of Monte Carlo simulation 
       collected  in sub-directory {\tt tauola\_result\_modes}\;.
      \end{itemize}  
\item {\tt results\_numer\_integr\_3pion}  
      presents the results for the width of $\tau \to 3\pi \nu_\tau$
      as a function of the 3 pion invariant mass. It is calculated by numerical
      integration of the analytical formula for different choices of 
      hadronic form factors as it is described in 
      Section~\ref{sec:Benchmark}. 
\item {\tt tauola\_result\_modes} contains Monte Carlo results for both 
      differential and total width for the processes  $\tau \to 2\pi \nu_\tau$,  
       $\tau \to K \pi  \nu_\tau$,  $\tau \to K^- K^0 \nu_\tau$,
       $\tau \to 3\pi \nu_\tau$,  $\tau \to K \pi^- K \nu_\tau$ and $\tau\to K^- \pi^0K^0\nu_\tau$.
      
\end{enumerate}

\subsubsection { Numerical stability tests }

The directory {\tt \$\{RCHLCURRENTS\}/cross-check/check\_analyticity\_and\_numer\_integr}
contains six subdirectories with tests of numerical stability for hadronic $\tau$ decay modes and a subdirectory with the test 
for the $a_1$ width.
Each decay channel is located in a separate directory.
Details regarding each of these tests are described in {\tt README} files of
the directory and every subdirectory as well. That is why only the basic information is 
provided in our paper. We have checked using interpolation from neighbouring values that 
the value of $d\Gamma/dq^2$ is continuous and is not contaminated by numerical 
instability of multidimensional Gaussian integration. Also we present the 
analytical results for the partial width of every channel to be compared with 
the Monte Carlo ones. 

Content of the directory: 

\begin{itemize}
      \item {\tt check\_analyt\_3piwidth}: test of numerical 
       stability of the distribution 
       $d\Gamma(\tau \to \nu_\tau \pi\pi\pi)/dq^2$. Results are presented for 
       separate modes:
       $d\Gamma(\tau \to \nu_\tau \pi^0\pi^0\pi^-)/dq^2$ and 
       $d\Gamma(\tau \to \nu_\tau \pi^-\pi^-\pi^+)/dq^2$.  Also the value of 
       the partial widths for the channels is provided for the comparison with the
       {\tt TAUOLA} results.

\item {\tt check\_analyt\_kkpi} - test of numerical stability for 
       $d\Gamma(\tau \to \nu_\tau KK\pi)/dq^2$.
       Results are presented for separate modes: 
       $d\Gamma(\tau \to \nu_\tau K^-\pi^-K^+)/dq^2$ and 
       $d\Gamma(\tau \to \nu_\tau K^0 \pi^-\bar{K^0} )/dq^2$. The value of the 
       partial widths 
       for both channels are provided.

\item {\tt check\_analyt\_kk0pi0} - tests of numerical stability for 
      $\tau \to \nu_\tau K^-\pi^0K^0$:
      both the spectrum $d\Gamma(\tau \to \nu_\tau K^- \pi^0 K^0)/dq^2$ and the partial 
      width are provided.

\item {\tt check\_analyt\_2pi} - tests of numerical stability for 
      $\tau \to \nu_\tau \pi^- \pi^0$:
      both the spectrum $d\Gamma(\tau \to \nu_\tau \pi^- \pi^0)/dq^2$ and the partial 
      width are provided.

\item {\tt check\_analyt\_kpi} - tests of numerical stability for 
      $\tau \to \nu_\tau K \pi $:
      both the spectrum for the total width $d\Gamma(\tau \to \nu_\tau K \pi )/dq^2$ 
      and the partial width for channels $\pi^-\bar{K}^0$ and $\pi^0 K^-$ are 
      provided. The partial widths for the individual decays are checked to be 
      $2/3$ and $1/3$ of 
      the total $K\pi$ width, mass effects are negligible in this case.

\item {\tt check\_analyt\_k0k} - tests of numerical stability for
       $\tau \to \nu_\tau K^- K^0$
      both the differential distribution $d\Gamma(\tau \to \nu_\tau K^- K^0)/dq^2$ 
      and the partial width are provided.

\item {\tt check\_analyt\_a1table} - tests of numerical stability
      of 
 $\Gamma_{a_1}(q^2)$
 produced by 
      program described in Appendix \ref{App:a1-tabler}.

\end{itemize}

\subsubsection {Analytic integration test }

The results of the analytical integration test in the three-pion case are presented in the directory 
{\tt \$\{RCHLCURRENTS\}/cross-check/results\_numer\_integr\_3pion}\;. They are produced by the program {\tt totwid3pi\_qq\_table.f} 
in the directory\\
 {\tt \$\{RCHLCURRENTS\}/cross-check/check\_analyticity\_and\_numer\_integr/check\_analyt\_3pi}\;. The program
can be compiled by command {\tt make} and run with {\tt make totwid3pi\_run > output.txt}\;.

The setup file {\tt input\_f1f2f4.dat}, in\\
 {\tt \$\{RCHLCURRENTS\}/cross-check/check\_analyticity\_and\_numer\_integr/check\_analyt\_3pi} contains:
\begin{itemize}
\item {\tt eps} - defines (relative) precision of the Gaussian integration.
\item {\tt kf1} - flag for form factor $F_1$. For  {\tt kf1= } 0, 1 or 2  $F_1$
      will be set respectively  to 0, 1 or to its functional form.
\item {\tt kf2} - flag for form factor $F_2$. For  {\tt kf2= } 0, 1 or 2  $F_2$
      will be set respectively  to 0, 1 or to its functional form.
\item {\tt kf4} - flag for form factor $F_4$. For  {\tt kf4= } 0, 1 or 2  $F_4$
      will be set respectively  to 0, 1 or to its functional form.
\item {\tt chan} - flag to choose the 3 pion mode.  {\tt chan=} 1
      for $\pi^0\pi^0\pi^-$ and {\tt chan=} 2 for $\pi^-\pi^-\pi^+$.
\end{itemize}

If the functional form of the form factors is used, it will be taken from the file\\
{\tt \$\{RCHLCURRENTS\}/cross-check/check\_analyticity/check\_analyt\_3piwidth/funct\_3pi.f}\;.
If {\tt kf1} or {\tt kf2} is set to 2, pretabulated file
{\tt \$\{RCHLCURRENTS\}/initA1Tab.f} will be used for $\Gamma_{a_1}$ in the propagator of the 
$a_1$-meson\footnote{Note that the tabulated file is
 generated by the program described in Appendix \ref{App:a1-tabler}.}.

Output file contains four columns:

\begin{itemize}
\item {\tt qmin} (in $[\mathrm{GeV^2}]$) - lower boundary for the integration over 3-pion invariant mass.
\item {\tt qmax} (in $[\mathrm{GeV^2}]$) - upper boundary for the integration over 3-pion invariant mass.
\item {\tt eps} - estimate of the integration precision in the result.
\item total width (in $[\mathrm{GeV}]$).
\end{itemize}

Results for the different configurations of the form factor are presented in\\ 
{\tt tauola/RChL-currents/cross-check/results\_numer\_integr\_3pion}\;.

\subsection{{\tt TAUOLA} weight recalculation mode}\label{app:WT-recalc}

Let us present now the installation necessary for the method of weighted events, which was
 envisaged in Section~\ref{sec:recalculation}. An example of such installation code is included in our
distribution tar-ball in directory {\tt new-currents/Installation-Reweight}\;.

Before reweighting method can be used, {\tt TAUOLA} needs to be adapted to new currents as explained
in the Appendix \ref{App:installation}. Afterwards, our example program {\tt tau-reweight-test-ASCII.c}, residing in the directory
{\tt new-currents/Installation-Reweight}, can be run with the help of the simple {\tt make} command%
\footnote{Other example, {\tt tau-reweight-test-HepMC.c}, requires installation
of {\tt HepMC} and optional installation of {\tt MC-TESTER}, and their paths provided in the {\tt Makefile}\;.}.
For more details regarding the reweighting examples, refer to {\tt README} located in
{\tt new-currents/Installation-Reweight}\;.

The following subsection describes reweighting algorithm as well as initialization used in the example.
Note,  contrary to the rest of the project, reweighting algorithm, including  examples of its usage,
is written in {\tt C++}.

\subsubsection{Weight recalculation algorithm}\label{app:WT-recalc-algo}
In order to use recalculation mode, several steps have to be performed from user program:

\begin{enumerate}
\item Before {\tt TAUOLA} initialization, RChL currents have to be switched on.
      This can be done with the help of the wrapper for {\tt FORTRAN} function {\tt INIRChL(IVER)},
      by calling {\tt inirchl\_(\&i);} with {\tt i = 1;}\;.
      Two versions of currents will be used, but initialization must be done for {\tt IVER=1},
      for initialization of {\tt RChL}-specific variables and tables.
\item Initialization of {\tt TAUOLA} must be called. We are using initialization
      taken from the default {\tt TAUOLA} example, stored in wrapper function {\tt f\_interface\_tauolaInitialize}.
\item For each event, the information about $\tau$ and its decay products
      must be filled and stored in instances of {\tt SimpleParticle} class\footnote{Class {\tt SimpleParticle}
      is used only to contain  four-vector and flavour of the particle.}.
\item Once the kinematical configuration for $\tau$ decay is read from the datafile (or fifo pipe), function:
      {\tt double calculateWeight(SimpleParticle \&tau,\\ vector<SimpleParticle> \&tau\_daughters)}
      can be used to retrieve the weight.
\end{enumerate}
\vspace{1cm}

The algorithm of the function {\tt calculateWeight} is sketched in the following:

\vspace{1cm}
\begin{enumerate}
\item Particles are prepared and boosted to the appropriate frame.
\item {\tt TAUOLA} decay channel is identified.
\item {\tt TAUOLA cleo} currents are switched on with {\tt inirchl\_(\&i)}; {\tt i = 0.}
\item Call to appropriate internal {\tt TAUOLA FORTRAN} subroutine, returning weight {\tt WT1}.
\item {\tt RChL} currents are switched on with {\tt inirchl\_(\&i);}  {\tt i = 1.}
\item Call to the same routine as in step 4 is performed, returning weight {\tt WT2}.
\item Ratio of weights calculated at steps 4 and 6 gives required model replacing weight.
\item {\tt WT = WT2/WT1} is returned to the main user program.
\end{enumerate}

It is rather straightforward to extend this method to the case when more than one new version 
of physics initialization is to be used. Note that the examples are
set up so that the weight is calculated both for $\tau^-$ and $\tau^+$ and stored in
variables {\tt WT\_M} and {\tt WT\_P} respectively. In cases where  only a single
$\tau$ is present in the event, the weight corresponding to the second $\tau$ equals 1.0.

Alternatively, in cases when this approach cannot be used or is inconvenient,
variants of the method, based on fifo pipes can be useful as well. Prototypes for such solutions 
can be obtained from Ref.~\cite{Link:fifo}.

Hadronic currents for $\tau^+$ and $\tau^-$ differ due to CP 
parity. The resulting effects are taken into account in the reweight algorithm.
 
\subsection{{\tt TAUOLA++} installation}\label{app:tauola++-installation}
Thanks to the modular construction of {\tt TAUOLA C++ Interface}~\cite{Davidson:2010rw}, new
currents can be used in C++ projects in a straightforward way. It is enough
to replace the previous {\tt TAUOLA-FORTRAN} installation with the new one,
adjusting {\tt Makefile} with a list of the newly added object files.

For step-by-step instructions, we refer to {\tt \$\{INSTALLATION\}/README-TAUOLA++}\;.
Our package has already been tested to work with {\tt TAUOLA C++ 
Interface v1.0.5}, but the installation
procedure is similar for all previous versions and should remain unchanged 
for future versions as well.

\appendix
\setcounter{section}{2}
\section{Input parameters}\label{app:C}
The results collected in this paper  represent a technical test 
of program installation as well. 
Figures should be reproduced if the input parameters, 
collected in Tables \ref{Constants}, \ref{Variation_of_Parameters1},
and \ref{Variation_of_Parameters2} and defined in
routine {\tt tauola/new-currents/RChL-currents/value\_parameter.f}   
remain unmodified.
In some cases the actual numerical value of parameters depends on chosen decay channel.  
For the $KK \pi $ modes we use $M_{K^*} = (M_{K^{*\pm}} +M_{K^{* 0}})/2$. 
For the $K\pi$ modes the value of parameters depends on parametrization. It is distinct 
for the one of Ref.~\cite{Jamin:2008qg} and of Ref.~\cite{Boito:2008fq}.
In the second case the mass parameters are noticeably different from the pole values. 
The results for the latter are consistent in both approaches. 
The choice between the two parametrizations 
for $K\pi$ modes (channels 2, 3 in Table \ref{table3}) is controlled
by {\tt FFKPIVEC} again to be set in {\tt value\_parameter.f}\;. The {\tt FFKPIVEC = 0}  activates parametrization from   
Ref.~\cite{Boito:2008fq} and  {\tt FFKPIVEC = 1}  the ones of~\cite{Jamin:2008qg}. 
Numerical values of all parameters, not only the masses, 
are different for the two cases. Variables are named with big greek 
letters for {\tt FFKPIVEC = 1}  and with the small ones for  {\tt FFKPIVEC = 0}.

There are two other flags {\tt FFVEC} and {\tt FFKKVEC} in  {\tt value\_parameter.f}\;. The first one fixes run with/without 
FSI effects  ({\tt FFVEC = 1} for run with FSI effects) and the last one chooses the parametrization for two-kaon form factor 
with/without the excited $\rho$ meson states ({\tt FFKKVEC = 1} for the parametrization with $\rho'$ and $\rho''$ ). 
By default {\tt FFVEC = 1}, {\tt FFKPIVEC = 1} and {\tt FFKKVEC = 0}.

On technical side, the choice of the internal flag {\tt KAK} is made
at the start of each $\tau$ decay  generation. It depends on the  decay channel 
labelled by {\tt imode} (generated by  {\tt TAUOLA}) and the flags%
\footnote{At the moment 
 {\tt KAK} depends on {\tt FFKPIVEC} only. 
However, in future it will 
depend also on {\tt FFVEC} and {\tt FFKVEC}. }
{\tt FFVEC}, {\tt FFKKVEC} and {\tt FFKPIVEC}.
The variable {\tt KAK} is then passed into 
routine {\tt value\_parameter.f} and the appropriate choice for the parameters is made. 
The  {\tt KAK} parameter coincides with {\tt imode} for all channels except  
the $K\pi$ modes. For the $K\pi$ decay modes {\tt KAK = 70} 
if {\tt FFKPIVEC = 0} and  {\tt KAK = 71} if {\tt FFKPIVEC = 1}.

For {\tt FFKPIVEC = 0} (that is for  
{\tt KAK = 70})   
 parameters marked in Tables with $\dagger$  are used, otherwise defaults 
of Tables \ref{Variation_of_Parameters1} and \ref{Variation_of_Parameters2} are
left unmodified. 

For {\tt KAK=4}  (i.e., for $\tau \to \pi^-\pi^0\nu_\tau$)  
masses and widths of $\rho, \rho'$ and $\rho''$ result from the adjustment to the experimental 
data and do not coincide with PDG defaults.
For other channels we simply take the PDG values \cite{Nakamura:2010zzi} 
for the $\rho(\rho')$ parameters. The PDG values are also taken for the narrow width resonances $\omega$ and $\phi$, 
numerical values are collected in Table~\ref{Constants}.

The PDG value is taken for the $a_1$ mass\footnote{For discussion on the difference between the mass used in the resonance 
Lagrangian and the physical one and a possibility to substitute the first with the latter, see~\cite{Dumm:2009kj} and  footnote $^{\ref{foot:M_A}}$ in this paper.}.
The parameters of the Resonance Chiral Theory 
are given in the Table \ref{Variation_of_Parameters1} 
as well\footnote{We point out that the values for 
the parameters $\Gamma_{\rho^\prime}$, $M_{\rho^{\prime\prime}}$, $\Gamma_{\rho^{\prime\prime}}$, $\gamma$, $\delta$, $\phi_1$ and $\phi_2$ lie 
outside the educated guess for its range of variation given 
in Table \ref{Variation_of_Parameters1}. This is irrelevant for the technical check we are proposing in this Section
 but matters for the actual use of the program.}.

  The parameters $\theta_V$ and $F_K$ can be varied by the user starting
from the code version of the year 2012.
We follow Ref.~\cite{Dumm:2009kj} and the case of ideal mixing
  ($\theta_V = 35.26^\circ $).
 In this case the $\phi$ contribution to $\tau\to KK\pi \nu_\tau$
 vanishes~\cite{Dumm:2009kj}.
  However, the  absence of  intermediate $\phi$ exchange contradicts the
 results of the BaBar Collaboration \cite{Aubert:2007ym} for the isospin
 related decay $e^+e^-\to K^+\pi^0K^-$ and for $\tau$ decays themselves
 (see~$^{\ref{foot:phitau}}$).
The parameter $F_K$ is not used in our default formulas. It enters the
non-default parametrization for $K\pi$ vector form factor, i.e., for 
FFKPIVEC = 0, KAK = 70.
We follow  Ref.~\cite{Boito:2008fq} in the choice $F_K\, =\, 1.198\cdot F$.
$F_K\neq F$ is related to $SU(3)$ breaking  and higher-order chiral corrections.
The parameter $H_{t0} = -1.24004 \cdot 10^{-2}$ does not appear in the text either. It corresponds to the value 
of the $K\pi$ loop function at zero-momentum transfer, 
$\tilde{H}_{K\pi}(0)$, in Eq.(11) of Ref.~\cite{Boito:2008fq}.

\begin{table}[h!]
\begin{center}
{\begin{tabular}{|c|c|c|c|}
\toprule
Parameter& Var. name  & Default & Used in channel \\
\midrule
$m_\tau$&{\tt  MTAU}&$  1.777 $& all,$^*$\\
$m_{\nu_\tau}$&{\tt  MNUTA}&$0.001 $& all,$^*$\\
$ \mathrm{cos} \theta_{\mathrm{Cabibbo}} $&set in {\tt TAUOLA} init.&$0.975$& all \\
$G_F$& set in {\tt TAUOLA} init.&$1.166375\cdot 10^{-5}$
&  all \\
$m_{\pi^\pm}$&{\tt  mpic}&$0.13957018$& all,$^*$\\
$m_{\pi^0}$&{\tt  mpiz}&$0.1349766$& all,$^*$\\
$m_{\eta}$&{\tt  meta}&$0.547$& 2,3,5-9,$^*$\\
$m_{K^{\pm}}$&{\tt  mkc}&$0.493677$& all,$^*$\\
$m_{K^{0}}$&{\tt  mkz}&$0.497648$& all,$^*$\\
\hline
$M_\omega$&{\tt mom}&$0.78194$&7,8\\
$\Gamma_\omega$&{\tt gom}&$0.00843 $&7,8\\
$M_\phi$&{\tt mphi}&$1.019$&7,8\\
$\Gamma_\phi$&{\tt gphi}&$0.0042 $&7,8\\
\bottomrule
  \end{tabular}
}
\end{center}
\caption{
Initialization  parameters defined in {\tt TAUOLA} main code or in file
{\tt new-currents/RChL-currents/value\_parameter.f}: constants and defaults. 
In this table our defaults used for plots or parameters
 not requiring to be changed in fits are collected. 
Channels identification numbers are defined in Table \ref{table3}.
 Energy units are powers of GeV.
 Variables requiring
rerun of pretabulation {\tt new-currents/RChL-currents/tabler/a1/da1wid\_tot\_rho1\_gauss.f} are marked with $^*$. 
}
\label{Constants}
\end{table}

Let us stress that in practice  the  
parameters may need to be varied. The defaults and the expected variation ranges 
are
given in Tables \ref{Variation_of_Parameters1} and \ref{Variation_of_Parameters2}.

\subsection{Range of variation of the non-resonance input parameters} \label{app:variationnonresparameters}
As long as the PDG values do not change, the values listed in Table~\ref{Constants}
should remain unchanged%
\footnote{The PDG limit on $m_{\nu_\tau}$ ($18.2$ MeV) is not used in the program. 
If one wants to play with this limit \cite{GomezCadenas:1989ag, GomezCadenas:1990uj}, 
our test results shall change in a rather negligible way.
We use $m_{\nu_\tau}=0.01$ GeV.}.

In order to account for the uncertainty given by higher-order chiral 
corrections, we suggest to vary 
 $F_K$ as indicated in Table \ref{Variation_of_Parameters1}.

\subsection{Range of variation of the resonance input parameters}\label{app:variationresparameters}
The resonance parameters are of different nature in this respect. 
Apart from the safe identification $M_V\equiv M_\rho$ \cite{Cirigliano:2003yq}, there is more uncertainty and 
model dependence on them. For the program user this is translated in a relative freedom to change the values 
of $M_\rho$, $M_{a_1}$, $M_{\rho'}$, $\Gamma_{\rho'}$, $M_{\rho''}$, $\Gamma_{\rho''}$, $\gamma$, $\delta$, $\phi_1$, $\phi_2$,
$M_{K^{*\pm}}$, $M_{K^{*0}}$, $M_{K^{*}}$, $M_{K^{*\prime}}$, $m_{K^{*}}$, $m_{K^{*\prime}}$, $\gamma_{{K^*}}$,
 $\gamma_{K^{*'}}$, $\Gamma_{K^{*'}}$,
$F_V$, $G_V=F^2/F_V$  
(although some deviations to this relation -below 20$\%$- 
may be expected due to the effect of excited resonances), $F_A$, $\beta_\rho$ and $\gamma_{K\pi}$. 
The changes of these parameters 
can be guided by the educated guesses on their range%
\footnote{Keep in mind, however, the warning concerning the 
relation $G_V=F^2/F_V$ and the one affecting Eq.(\ref{Weinberg sum rules}) for the range for $G_V$ and $F_A$, respectively.}, displayed in 
Table \ref{Variation_of_Parameters1}.

The  warning is that the  $F_V$ and $F_A$ 
cannot be changed independently since they should satisfy, to a reasonable accuracy\footnote{\label{foot:M_A}Checking 
the first of Eqs. (\ref{Weinberg sum rules}) is straightforward; for the second one, it should be observed that 
in the different relations among couplings which can be 
obtained from short-distance QCD constraints \cite{Brodsky:1973kr, Lepage:1980fj, Floratos:1978jb, Lepage:1979zb} and 
involving $M_A$, the identification 
$M_A\sim M_{a_1}$ is not appropriate \cite{Cirigliano:2003yq}. There is some tension on the value of $M_A$: $998(49)$ MeV 
in Ref.~\cite{Mateu:2007tr} versus 
$920(20)$ MeV in Ref.~\cite{Pich:2010sm}. The range $\left[900,1050\right]$ MeV should accommodate reasonable variations 
of this parameter in order to estimate the possible 
violations of the second Weinberg sum rule. The interval given for $M_{a_1}$ is only marginally consistent with the PDG value 
\cite{Nakamura:2010zzi}. However, this is not an issue, since it depends strongly on the precise definition of the resonance 
mass used to extract it;  the PDG one and the one in Ref.~\cite{Dumm:2009va} are different.},
 the first and second Weinberg 
sum rules taken in the single resonance approximation~\cite{Weinberg:1967kj}:
\begin{equation}\label{Weinberg sum rules}
 F_V^2-F_A^2\,=\,F^2\,,\quad F_V^2 M_V^2 \,=\, F_A^2 M_A^2\,.
\end{equation}
Violations of these relations can be due to the modelization of the resonance spectrum in the large-$N_C$ limit but should remain 
below 20$\%$.

\begin{table}[h!]
\begin{center}
{\begin{tabular}{|c|c|c|c|c|}
\toprule
Parameter& Var. name  & Default & [suggested range] & Used in channel \\
\midrule
$M_\rho$              & {\tt mro}         & $0.77554$ & $\left[0.770,0.777\right]$   & 1\\
$M_\rho$              & {\tt mro}         & $0.775  $ & $\left[0.770,0.777\right]$   & 4-9,$^*$\\
$M_{a_1}$             & {\tt mma1 }       & $1.12   $ & $\left[1.00,1.24\right]$     & 5-9,$^*$\\
$M_{\rho'}$           & {\tt mrho1 }      & $1.453  $ & $\left[1.44,1.48\right]$     & 1\\
$M_{\rho'}$           & {\tt mrho1 }      & $1.465  $ & $\left[1.44,1.48\right]$     & 4,5,6,$^*$\\
$\Gamma_{\rho'}$      & {\tt grho1 }      & $0.50155$ & $\left[0.32,0.39\right]$     & 1\\
$\Gamma_{\rho'}$      & {\tt grho1 }      & $0.4    $ & $\left[0.32,0.39\right]$     & 4,5,6,$^*$\\
$M_{\rho''}$          & {\tt mrho2 }      & $1.8105 $ & $\left[1.68,1.78\right]$     & 1, 4\\
$\Gamma_{\rho''}$     & {\tt grho2 }      & $0.4178 $ & $\left[0.08,0.20\right]$     & 1, 4\\
$\gamma$              & {\tt coef\_ga}    & $0.14199$ & $\left[0.077,0.099\right]$   & 1, 4 \\
$\delta$              & {\tt coef\_de}    & $-0.12623$& $\left[-0.035,-0.012\right]$ & 1, 4 \\
$\phi_1$              & {\tt phi\_1 }     & $-0.17377$& $\left[0.5,0.7\right]$       & 1, 4\\
$\phi_2$              & {\tt phi\_2 }     & $0.27632$ & $\left[0.5,1.1\right]$       & 1, 4\\
$M_{K^{*\pm}}$        & {\tt mksp}        & $0.89166$ & $\left[0.891,0.892\right]$   & 2,3,7-9,$^*$ \\
$M_{K^{*0}}$          & {\tt mks0}        & $0.8961 $ & $\left[0.895,0.897\right]$   & 2,3,7-9,$^*$\\
$M_{K^*}$             & {\tt mkst}        & $0.8953 $ & $\left[0.8951,0.8955\right]$ & 2,3\\
$M_{K^*}$             & {\tt mkst}        & $\left(M_{K^{*\pm}}+M_{K^{*0}}\right)/2$&&7-9,$^*$\\
$m_{K^{*}}$           & {\tt mkst}        & $0.94341$ & $\left[0.9427,0.9442\right]$ & 2$^\dagger${},3$^\dagger${} \\ 
$\Gamma_{K^*}$        & {\tt gamma\_kst } & $0.0475 $ & $\left[0.047,0.048\right]$ & 2,3\\
$\gamma_{K^*}$        & {\tt gamma\_kst } & $0.06672$ & $\left[0.0655,0.0677\right]$ & 2$^\dagger${},3$^\dagger${}\\
$\Gamma_{K^*\prime}$  & {\tt gamma\_kstpr}& $0.206  $ & $\left[0.155,0.255\right]$ & 2,3\\
$\gamma_{K^{*\prime}}$& {\tt gamma\_kstpr}& $0.240  $ & $\left[0.120,0.380\right]$   & 2$^\dagger${},3$^\dagger${}\\
$M_{K^{*\prime}}$     & {\tt mkstpr}      & $1.307  $ & $\left[1.270,1.350\right]$   & 2,3\\
$m_{K^{*\prime}}$     & {\tt mkstpr}      & $1.374  $ & $\left[1.330,1.450\right]$   & 2$^\dagger${},3$^\dagger${}\\
\midrule
$F$                   & {\tt fpi\_rpt}    & $0.0924 $ & $\left[0.0920,0.0924\right]$ & all,$^*$ \\
$F_K$                 & {\tt  fk\_rpt       }    & $1.198$F        & $\left[0.94F,1.2F\right]$  &    3,4       \\
$F_V$                 & {\tt fv\_rpt}     & $0.18   $ & $\left[0.12,0.24\right]$     & 5-9,$^*$ \\
$G_V$                 & {\tt gv\_rpt}     & $F^2/F_V$ & $ \left[0.xxF^2/F_V,1.xxF^2/F_V      \right] $              & 5-9,$^*$\\
$F_A$                 & {\tt fa\_rpt}     & $0.149  $ & $\left[0.10,0.20\right]$     & 5-9,$^*$\\
$\beta_\rho$          & {\tt beta\_rho}   & $-0.25  $ & $\left[-0.36,-0.18\right]$   & 5,6$^*$\\
$\gamma_{K\pi}$       & {\tt gamma\_rcht} & $-0.043 $ & $\left[-0.033,-0.053\right]$                           & 2,3\\
$\gamma_{K\pi}$       & {\tt gamma\_rcht} & $-0.039 $ & $\left[-0.023,-0.055\right]$ & 2$^\dagger${},3$^\dagger${}\\
$\theta_V$&{\tt THETA}&$35.26^\circ$& $\left[15^o,50^o\right]$ &7,8\\
\bottomrule
  \end{tabular}
}
\end{center}
\caption{ \small
Initialization  parameters defined in file
{\tt new-currents/RChL-currents/value\_parameter.f}: part 2, fit parameters. 
An educated guess for the variation of some of the resonance parameters is given. Energy units are powers of GeV.
Channels identification numbers are defined in Table \ref{table3}.
 Variables requiring
rerun of pretabulation {\tt new-currents/RChL-currents/tabler/a1/da1wid\_tot\_rho1\_gauss.f} are marked with $^*$. 
The parameters coresponding to non-default currents of $K\pi$ modes ({\tt FFKPIVEC=0})  are marked with $^\dagger${}.
}
\label{Variation_of_Parameters1}
\end{table}

There is more uncertainty on the couplings belonging to the odd-intrinsic parity sector, namely the $c_i$, $d_i$ and $g_i$ values and variation ranges are given in Table~\ref{Variation_of_Parameters2}. 
Some remarks on the relations are in place:
\begin{itemize}
 \item $c_1-c_2+c_5 \neq 0$ would violate maximally the short-distance QCD-ruled behaviour for the vector-vector 
correlator 
$\Pi_V(q^2)$ \cite{Dumm:2009kj},  this condition must not be changed.
 \item $2g_4+g_5$ comes from $\Gamma(\omega\to\pi^+\pi^-\pi^0)$. Both the direct production mechanism \cite{Dumm:2009kj} 
and the one-resonance exchange \cite{RuizFemenia:2003hm} were taken into account consistently 
and the error is under control: $2g_4+g_5=-0.60\pm0.02$. If the PDG value for this decay width does not change, 
the value of this combination of couplings should be changed within the quoted error only.
 \item The high-energy large-$N_C$ predictions for the set $\left\lbrace g_2,\,g_1-g_3,\,c_1-c_2-c_5+2c_6,\,d_3\right\rbrace$
 come at the same order in the 
expansion of $\Pi_V(q^2)$ in powers of $1/q^2$ \cite{Dumm:2009kj}. Therefore, changes on the values of these parameters shall be expected 
from subleading corrections in 1/$N_C$ and, moreover, 
they will be highly correlated. Variations of 1/3 with respect to the values of any of them, see
Table~\ref{Variation_of_Parameters2}, may occur.
 \item The predictions for $d_1+8d_2-d_3$ and $c_1+c_2-8c_3-c_5$ were not obtained in hadronic $\tau$ decays but 
in the study of the $\left\langle VVP\right\rangle$ octet 
\cite{RuizFemenia:2003hm} and singlet \cite{Chen:2012vw} Green function. Therefore, an educated conservative guess 
yields a deviation up to some 50\% of the former. 
For the latter, 
a non-zero value may arise provided it has a minor effect in the 
observables, since its contribution vanishes in the chiral limit.
 \item Our current understanding seems to point to different values of $c_4$ and $g_4$ than those shown 
in Table \ref{Variation_of_Parameters2}. We would suggest that they are 
allowed to vary freely in the fits (keeping control on the branching ratio if it is not included as a data point in the fit).
\end{itemize}
 In addition to what is explained above, the values of the couplings can be affected by the introduction 
of the second multiplet of resonances in 
$\tau\to KK\pi \nu_\tau$ decays.
\begin{table}[h!] 
\begin{center}
{\begin{tabular}{|c|c|c|c|c|}
\toprule
Parameter & Var. name & Default& [suggested range] & Used in channel  \\
\midrule
$c_1-c_2+c_5$                           &{\tt c125}             &$0.0$&                        & 7-9 \\
$2g_4+g_5$                              &{\tt  g4; g5}           &$-0.6 $&$\left[-0.64,-0.56\right]$&7,8\\
$g_2$                                   &{\tt g2}       & $\frac{M_V}{192\pi^2 \sqrt{2} F_V}$ &$\left[-33\%,+33\%\right]$ & 7-9\\
$g_1-g_3$                               &{\tt  g13}     & $\frac{-2 M_V}{192 \pi^2 \sqrt{2} F_V}$ &$\left[-33\%,+33\%\right]$ & 7-9\\
$c_1-c_2-c_5+2c_6$                      &{\tt  c1256}   & $-\frac{3 F_V M_V}{96 \pi^2 \sqrt{2} F^2}$ &$\left[-33\%,+33\%\right]$ & 7-9\\
$d_3$                                   &{\tt  d3}      & $-\frac{M_V^2}{64 \pi^2 F^2}$   &$\left[-33\%,+33\%\right]$ & 7-9\\
$d_1+8d_2-d_3$                          &{\tt d123 }             &$0.05$& $\left[-50\%,+50\%\right]$ & 7-9\\
$c_1+c_2-8c_3-c_5$                      &{\tt  c1235 }           &$0.0$&$\left[-0.25,+0.25 \right]$ & 7-9\\
$c_4$                                   &{\tt c4}                &$-0.07$ & free                       & 7-9\\
$g_4$                                   &{\tt g4}                &$-0.72$ &free                       & 7,8\\
\bottomrule
  \end{tabular}
}
\end{center}
\caption{Initialization part 3: 
odd-intrinsic parity sector. File
{\tt new-currents/RChL-currents/value\_parameter.f} .
Channels identification numbers are defined in Table \ref{table3}.
The defaults, which follow Refs.~\cite{Dumm:2009kj, Roig:2007yp},   are needed
to reproduce our figures.
 \label{Variation_of_Parameters2}}
\end{table}

Finally there are parameters in {\tt  value\_parameter.f} which are already
prepared for the future update, but not yet used in the program:
{\tt  gro, mf2, gf2, mf0, gf0, msg, gsg}. 
\appendix
\setcounter{section}{3}
\section{Benchmark results}\label{app:D}
In Table \ref{table3} we collect results as coming from our default setting. 
It documents directly the distribution tar-ball. Results have to be 
checked once tar-ball is unpacked and installed in the particular environment.
It will check if sufficiently large samples are installed, correctness 
of coupling constant setting, etc.  This table will be updated in the future,
once currents are modified, and included in the paper; version of the distribution tar-ball. 

\begin{table}
\vspace{0.3cm} 
\begin{center}
{ \begin{tabular}{|c| c| c| c| c|} 
\toprule 
No. &Channel &  Width  [GeV] & Reference & In {\tt new-currents/RChL-currents}  \\  
    &        &               &           &directory  channel's current:  \\ 
    &        &               &           & {\tt file} $\to$ {\tt routine}  \\  
\midrule
1.& {$ \pi^-\pi^0 \; \;\; \;$}  &  {$5.2441\cdot10^{-13} \pm 0.005\%$ }  &  Subs. \ref{Subsect:pipi0} &  {\tt frho\_{}pi.f} $\to$ {\tt  CURR\_{}PIPI0} \\ 
2.& {$ \pi^0 K^- \; \;\; \;$}   &  {$8.5810 \cdot10^{-15}\pm 0.005\%$ }  &   Subs. \ref{Subsect:pipi0}& {\tt fkpipl.f} $\to$ {\tt  CURR\_{}KPI0 }\\ 
3.& {$ \pi^-\bar K^0 \; \;\; \;$}   &  {$1.6512 \cdot 10^{-14}\pm 0.006\%$ }  &  Subs. \ref{Subsect:pipi0}& {\tt fkpipl.f} $\to$ {\tt CURR\_{}PIK0 } \\ 
4.& {$ K^-K^0 \; \;\; \;$}     &  {$ 2.0864 \cdot 10^{-15}\pm 0.007\%$ }  &  Subs. \ref{Subsect:pipi0}& {\tt  fk0k.f}$\to$ {\tt CURR\_KK0} \\ 
5.&  {$  \pi^-\pi^-\pi^+$}     &  {$2.0800 \cdot 10^{-13}\pm 0.017 \%$ }  &    Subs. \ref{Subsect:pipipi}&  {\tt f3pi\_r{}cht.f }$\to$ {\tt  F3PI\_{}RCHT}$^*$ \\ 
6.& {$  \pi^0\pi^0\pi^-$}      &   {$2.1256\cdot 10^{-13}\pm 0.017 \%$ } &   Subs. \ref{Subsect:pipipi}& {\tt f3pi\_{}rcht.f}$\to$ {\tt F3PI\_{}RCHT}$^*$ \\ 
7.& {$  K^-\pi^-K^+$}         &   {$3.8460\cdot 10^{-15}\pm 0.024\% $ }  &    Subs. \ref{Subsect:KKpi} & {\tt fkkpi.f}$\to$ {\tt FKKPI}$^*$ \\ 
8.& {$  K^0\pi^-\bar{K^0}$}    &   {$3.5917 \cdot 10^{-15} \pm 0.024\% $ }    & Subs. \ref{Subsect:KKpi} & {\tt  fkkpi.f}$\to$ {\tt FKKPI}$^*$ \\
9.& {$  K^-\pi^0 K^0$}         &   {$2.7711\cdot 10^{-15}\pm 0.024 \% $  }     &  Subs. \ref{Subsect:KK0pi0}& {\tt  fkk0pi0.f}$\to$ {\tt FKK0PI0}$^*$ \\
\midrule
 &   &   &    & $^*${The $F_i$ of formula (\ref{fiveF}).} \\ 
\midrule
1.& {$ \pi^-\pi^0 \; \;\; \;$}   &  {$4.0642 \cdot 10^{-13}\pm 0.005\%$}  &   Subs. \ref{Subsect:pipi0}& {\tt frho\_{}pi.f} $\to$ {\tt  CURR\_{}PIPI0$^{**}$ }\\ 
2.& {$ \pi^0 K^- \; \;\; \;$}   &  {$7.4275 \cdot 10^{-15}\pm 0.005\%$}  &   Subs. \ref{Subsect:pipi0}& {\tt fkpipl.f} $\to$ {\tt  CURR\_{}KPI0$^{**}$ }\\ 
3.& {$ \pi^-\bar K^0 \; \;\; \;$}   &  {$1.4276 \cdot 10^{-14}\pm 0.006\%$}  &  Subs. \ref{Subsect:pipi0} & {\tt fkpipl.f} $\to$ {\tt CURR\_{}PIK0$^{**}$ } \\ 
4.& {$ K^-K^0 \; \;\; \;$}   &  {$1.2201 \cdot 10^{-15}\pm 0.007\%$}  &   Subs. \ref{Subsect:pipi0} & {\tt fkpipl.f} $\to$ {\tt  CURR\_{}KK0$^{**}$ }\\ 
\midrule 
& &  &    & $^{**}$FSI off \\ 
\bottomrule
\end{tabular} 
}  
\end{center}
\caption{Collection of numerical results to be obtained from the {\tt demo-standalone}. Possible future extensions
going beyond the published paper will be documented in this table in the paper version included with the distribution tar-ball. 
References to subsections may be replaced in the future 
by references to the forthcoming papers. 
Last column includes references to routines of the currents code.
 This table is complementary to Table~\ref{Table:bench}. Results for the case when Final State Interactions (FSI)  switched 
off with the help of
{\tt FFVEC = 0} in file {\tt new-currents/RChL-currents/value$\_${}parameter.f} are also given, then  {\tt FFKKVEC = 1} was used.
}
\label{table3}
\end{table}

A wealth of data is available from the project Web page \cite{web:RChL}. 
Many of the results collected there were obtained using 
{\tt MC-TESTER} \cite{Davidson:2008ma}. The distributions can be thus easily used 
for benchmarking any other Monte Carlo program independently if it is used 
for simulations coded in C++ or {\tt FORTRAN}, provided that
event record such as {\tt HEPEVT} or {\tt HepMC} \cite{Dobbs:2001ck} is used.

\appendix
\setcounter{section}{4}
\section{Final state interactions}\label{app:F}

Final state interactions have been taken into account for the two-meson $\tau$ 
decays but not for the three-meson  channels. 
It is of great importance to include 
them in the two-meson decays, because otherwise the right normalization 
at the peak is clearly lost and the curve would go systematically below 
the data in the resonance 
region\footnote{This reduction of the $|F_V^{PQ}(s)|^2$ peak value amounts to $13\%\leftrightarrow30\%$ for 
$PQ=\pi\pi$, $KK$, and $K\pi$, depending on the decay channel and the exchanged resonances accounted for. 
See also the analogue comparison for the decay width 
in Section~\ref{sect:numer-2scal}.}. This effect is certainly much 
smaller for the three-meson $\tau$ decays, where nonetheless it can show up in 
the two-particle invariant mass distributions in particular regions of phase 
space, where our calculation is particularly far from experimental data, see 
Fig.~\ref{Fig:Ian}. 
For the two-meson modes ($\tau\to P Q \nu_\tau$), FSI 
are taken into account through the Omn\`es resummation \cite{Omnes:1958hv} 
provided by the 
exponents in Eqs. (\ref{VFFpipi}) and (\ref{VFFKpi}). In this approximation \cite{Guerrero:1997ku}, 
the imaginary part of the loop function is kept in the denominator, 
providing the width of the exchanged resonance. The real part is resummed in the exponential --while 
the exponential is a common factor in Eq. (\ref{VFFKpi}), it 
is different in every term in Eq. (\ref{VFFpipi}). If we had good theoretical knowledge of the energy-dependent meson widths 
(also for excited resonances) the latter 
procedure would be preferable. For the time being, both approaches are equivalent within the theoretical uncertainties.
This resummation respects analyticity and unitarity only in a perturbative sense, a feature that can be 
improved~\cite{Boito:2008fq} 
to hold to all 
orders as follows:
\begin{enumerate}
 \item The relevant phase-shift is obtained as the ratio between the imaginary 
and real parts of the vector form factor 
$\delta^{PQ}(s)=Im\left[F_V^{PQ}(s)\right]/Re\left[F_V^{PQ}(s)\right]$. 
$F_V^{PQ}(s)$ stands for a form factor which can be obtained from
 Eqs. (\ref{VFFpipi}), (\ref{VFFKK}) and (\ref{VFFKpi}) by taking out the exponentials 
and placing back the real part of 
the relevant loop functions in the denominator 
 (see Refs. \cite{RChL:scal} and \cite{GomezDumm:2000fz} for details).
 \item A three-times subtracted dispersion relation is used to resum FSI effects and the 
final form factor reads
\begin{equation}
 F_V^{PQ}(s)=
\mathrm{exp}\left\lbrace\alpha_1 s+\alpha_2 s^2+\frac{s^3}{\pi}\int^{s_{cut}}_{s_{thr}}
 \mathrm{d}s' \frac{\delta^{PQ}(s')}{s'^3(s'-s-i\epsilon)}\right\rbrace\,,
\end{equation}
with $s_{cut}\sim4$ GeV$^2$ \cite{Boito:2008fq, Boito:2010me, Pich:2001pj} and $s_{thr}=(m_P+m_Q)^2$. 
The subtraction constants $\alpha_1$ and $\alpha_2$ are to be fitted to data.
\end{enumerate}

This procedure will be followed in a future update of the program.
Technically the method requires calculation of Cauchy Principal Value integrations which can be rather time-consuming and  necessitates  numerical stability checks.

\begin{figure}[t!]
\begin{center}
\includegraphics[width =1.0\linewidth]{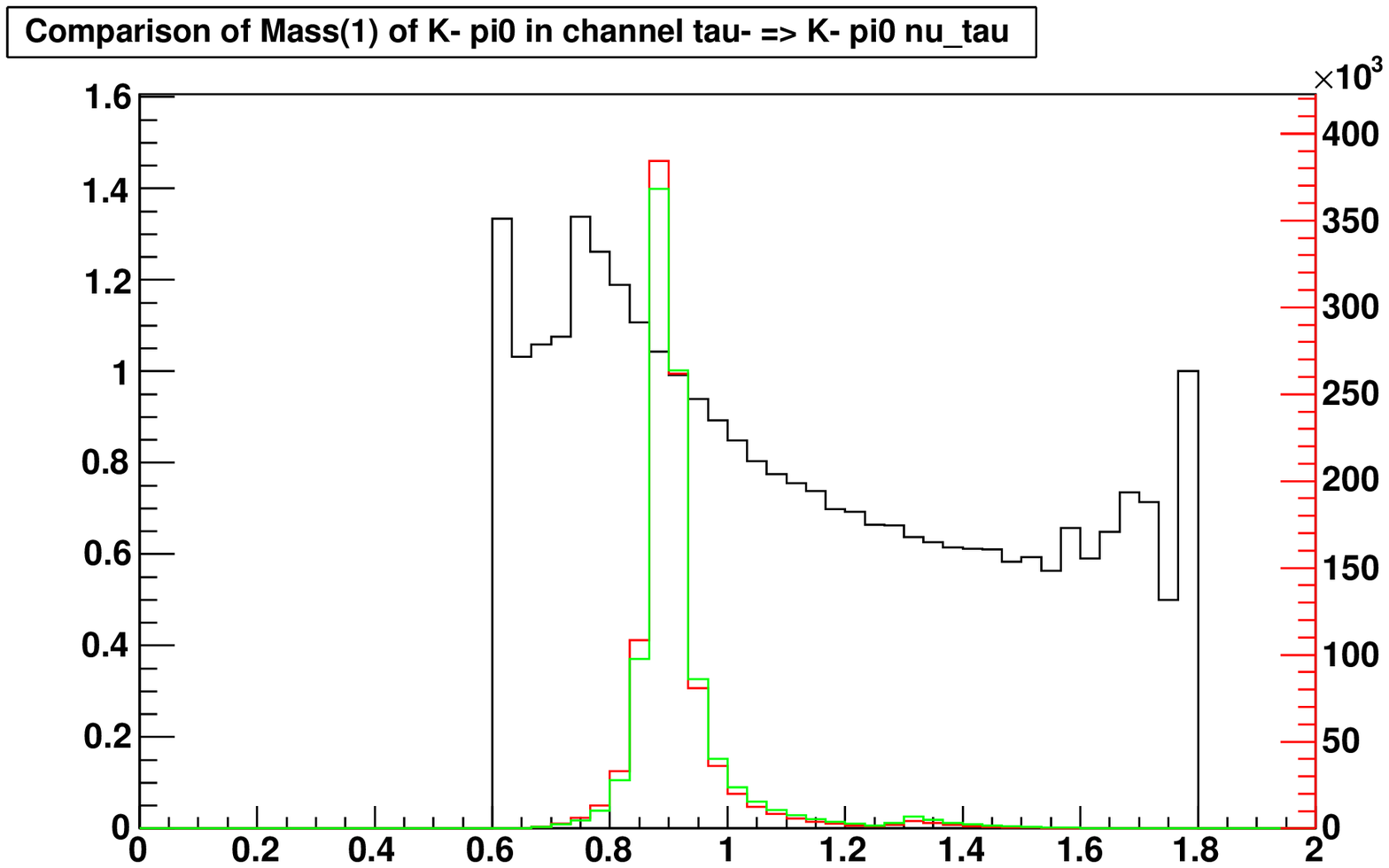}
\caption{Normalized events distribution vs. invariant mass of $\pi^0 K^-$ pair: no FSI case corresponds to 
the red (darker grey) line, 
for FSI  the green (lighter grey) line was used. 
Adjustment of numerical parameters is taken into account (Shift of $K^*$ mass, etc.).
Ratio of the two histograms is given by the black one (only in this case the left side scale should be used).  We have used
{\tt MC-TESTER},  Ref.~\cite{Davidson:2008ma}, for preparation of the plot. 
Some of  automatically generated
  markings  are explicitly left on the plot in this case.
The program, {\tt MC-TESTER}, is used also for  plots
of our Web page~\cite{web:RChL}. 
\label{fig:FSI-Kpi0} }
\end{center}
\end{figure}

For the moment we consider FSI effects as they are given in Chapter~\ref{Subsect:pipi0}, Eqs.
(\ref{VFFpipi_rho})-(\ref{VFFKpi}).
To run the code with the FSI effects one has to fix {\tt FFVEC = 1} in {\tt value\_parameter.f}\;.
Effects of FSI are presented in Table~\ref{table3}, last four lines are provided for comparison, and in Fig.~\ref{fig:FSI-Kpi0}, they 
change by  $14\%-  32\%$  the decay width, 
depending on the channel. 
Further results are collected in our Web page~\cite{web:RChL}.

In the three-meson modes these interactions are neglected at the moment. 
We plan to include them in the future, at least for 
the $\tau\to\pi\pi\pi\nu_\tau$ decays.

\end{document}